\documentclass[12pt,twoside,english]{article}
\usepackage[T1]{fontenc}
\usepackage{units}
\usepackage{amsmath}
\usepackage{amssymb}
\usepackage{graphicx}

\makeatletter
\newcommand{\lyxaddress}[1]{
\par {\raggedright #1
\vspace{1.4em}
\noindent\par}
}

\usepackage{a4wide}
\usepackage{graphicx}
\usepackage{fancyhdr}
\usepackage{amsmath}
\usepackage{amssymb}
\usepackage{subfigure}
\usepackage{setspace}
\usepackage{verbatim} 
\usepackage{mathtools} 
\usepackage{color}
\usepackage{hyperref}
\usepackage{caption} 
\usepackage{lmodern}

\definecolor{HyperRefColor}{gray}{0.3} 

\hypersetup{
    hidelinks = true
}

\numberwithin{equation}{section}

\newcommand{\vek}[1]{\mathchoice{\displaystyle\boldsymbol#1}
{\textstyle\boldsymbol#1}{\scriptstyle\boldsymbol#1}
{\scriptscriptstyle\boldsymbol#1}}
\newcommand{\mat}[1]{\mathchoice{\displaystyle\mathbf#1}
{\textstyle\mathbf#1}{\scriptstyle\mathbf#1}
{\scriptscriptstyle\mathbf#1}}

\makeatother

\usepackage{babel}
\begin{document}

\title{Higher-order surface FEM for incompressible\\Navier-Stokes flows
on manifolds}

\author{T.P. Fries}

\maketitle

\lyxaddress{\begin{center}
Institute of Structural Analysis\\
Graz University of Technology\\
Lessingstr. 25/II, 8010 Graz, Austria\\
\texttt{www.ifb.tugraz.at}\\
\texttt{fries@tugraz.at}
\end{center}}
\begin{abstract}
Stationary and instationary Stokes and Navier-Stokes flows are considered
on two-dimensional manifolds, i.e., on curved surfaces in three dimensions.
The higher-order surface FEM is used for the approximation of the
geometry, velocities, pressure, and Lagrange multiplier to enforce
tangential velocities. Individual element orders are employed for
these various fields. Stream-line upwind stabilization is employed
for flows at high Reynolds numbers. Applications are presented which
extend classical benchmark test cases from flat domains to general
manifolds. Highly accurate solutions are obtained and higher-order
convergence rates are confirmed.

\end{abstract}
\newpage{}\tableofcontents{}\newpage{}

\section{Introduction\label{sec:Introduction-1}}

The solution of boundary value problems on curved surfaces has many
practical applications in mathematics, physics, and engineering. For
example, there are transport\emph{ }processes on interfaces, e.g.,
in foams, biomembranes and bubble surfaces \cite{Edwards_1991a,Gross_2011a,Slattery_2007a},
or structure-related phenomena such as in membranes and shells \cite{Blaauwendraad_2014a,Chapelle_2011a}.
Herein, Stokes and incompressible Navier-Stokes flows on curved, two-dimensional
manifolds are considered. The governing equations for flows on \emph{moving}
surfaces are discussed in \cite{Bothe_2010a,Jankuhn_2017a} based
on fundamental surface continuum mechanics and conservation laws and
in \cite{Koba_2017a}, an energetic approach is presented. Earlier
works in a similar context may be traced back to \cite{Ebin_1970a,Gurtin_1975a,Scriven_1960a,Temam_1988a}.
For an excellent overview, the reader is refered to \cite{Jankuhn_2017a}.
The references given above often focus on mathematical properties
such as the existence and uniqueness of the solutions or stabilitity
analyses. Applications are often two-phase flows where the fluid field
in the bulk and on the moving interface are coupled. However, it is
also worthwhile to consider the situation for \emph{fixed} manifolds,
e.g., related to meterology and oceanography where the flows take
place on (part of) a sphere. Special geometries such as hyperbolic
planes and spheres are discussed in \cite{Chan_2013a,Khesin_2012a,Kobayashi_2008a}.

Herein, the focus is on the approximation of stationary and instationary
(Navier-)Stokes flows on fixed manifolds based on the surface finite
element method as outlined in \cite{Demlow_2009a,Dziuk_1988a,Dziuk_2013a}.
The governing equations resemble the three-dimensional (Navier-)Stokes
equations where the classical gradient and divergence operators are
replaced by their tangential counterparts derived from tangential
differential calculus \cite{Jankuhn_2017a}. The equations are formulated
in the classical stress-divergence form, contrasted to the approach
in \cite{Nitschke_2012a}. An additional constraint is required to
enforce that the velocities remain in the tangent space of the manifold;
it is labelled ``tangential velocity constraint''. The models are
first given in strong form and are then transformed to the weak form
to enable a numerical solution based on the surface FEM. Finite element
spaces of different orders are employed for the approximation of the
geometry and of the involved physical fields, i.e., the velocities,
pressure and the Lagrange multiplier field required to enforce the
tangential velocity constraint. It is found that the balance of these
element orders is critical for the accuracy and conditioning of the
system of equations. In particular, the well-known Babu\v ska-Brezzi
condition applies \cite{Babuska_1971a,Brezzi_1974,Franca_1988a} as
both, the incompressibility constraint and the tangential velocity
constraint are enforced using Lagrange multipliers. For the case of
the instationary Navier-Stokes equations, the Crank-Nicolson time
stepping scheme is employed for the semi-discrete sytem of equations
resulting from using the surface FEM in space. Surface FEM based on
linear elements is used in the recent work \cite{Reuther_2018a},
where the penalty method is employed to enforce tangential velocities
and a projection method rather than a monolithic approach is suggested
to solve for the different physical fields. Alternatives for the surface
FEM are the TraceFEM \cite{Deckelnick_2014a,Grande_2016a,Reusken_2014a}
and CutFEM \cite{Hansbo_2004a,Hansbo_2014a}, where the basis functions
are generated from a background mesh in the bulk surrounding the manifold
of interest.

Using the FEM for the Navier-Stokes flows at large Reynolds numbers
requires stabilization. Herein, the streamline-upwind Petrov-Galerkin
(SUPG) approach is used \cite{Brooks_1982a,Tezduyar_2000a}. Alternatively,
other variants such as the Galerkin least squares stabilization \cite{Hughes_1989a}
and variational multiscale approaches \cite{Hughes_1998a,Gravemeier_2006a}
may also be employed. Stabilization for advection-diffusion applications
on manifolds are considered in \cite{Olshanskii_2014a}.

The numerical results show that higher-order convergence rates are
achieved provided that the finite element spaces are properly chosen.
Also the conditioning of the system of equations depends on the element
orders employed for the approximation of the individual physical fields.
The presented results are based on well-known benchmark test cases
in two dimensions such as driven cavity flows and cylinder flows with
vertex shedding which, herein, are extended to curved surfaces. Due
to the higher-order elements, the results are highly accurate and
may serve as future benchmarks in the context of (Navier-)Stokes flows
on manifolds. Most test cases are carried out on parametrized surfaces,
however, also the situation of flows on zero-isosurfaces is covered
herein.

To the best of our knowledge, this is the first time, where (i) general
higher-order surface FEM is used for the (in)stationary (Navier-)Stokes
equations on manifolds including stabilization, (ii) numerical convergence
studies are presented confirming higher-order convergence rates, and
(iii) benchmark test cases are proposed and solutions presented. Furthermore
the notation employed is closely related to the typical engineering
literature and aims to provide a bridge from the mathematical to the
engineering community.

The paper is organized as follows: In Section \ref{X_Preliminaries},
some requirements and properties of surfaces are described and tangential
differential operators are defined based on \cite{Delfour_2011a,Dziuk_2013a}.
Section \ref{X_GoverningEquations} covers the governing equations
for (i) Stokes flow, (ii) stationary, and (iii) instationary Navier-Stokes
flows on two-dimensional manifolds. They are given in strong form,
weak form, and discretized weak form according to the surface FEM.
Numerical results are presented in Section \ref{X_NumericalResults}.
Convergence studies are performed for a test case for which an analytic
solution is available and it is shown that higher-order convergence
rates can be achieved. For the other test cases where no analytic
solutions are available it is confirmed that in the flat two-dimensional
case, well-known reference solutions are reproduced. Various meshes
with different orders and resolutions have been employed to obtain
highly accurate results on curved surfaces. Finally, a summary and
outlook are given in Section \ref{X_Conclusions}.

\section{Preliminaries\label{X_Preliminaries}}

\subsection{Surfaces\label{XX_Surfaces}}

\begin{figure}
\centering

\subfigure[]{\includegraphics[height=0.2\textwidth]{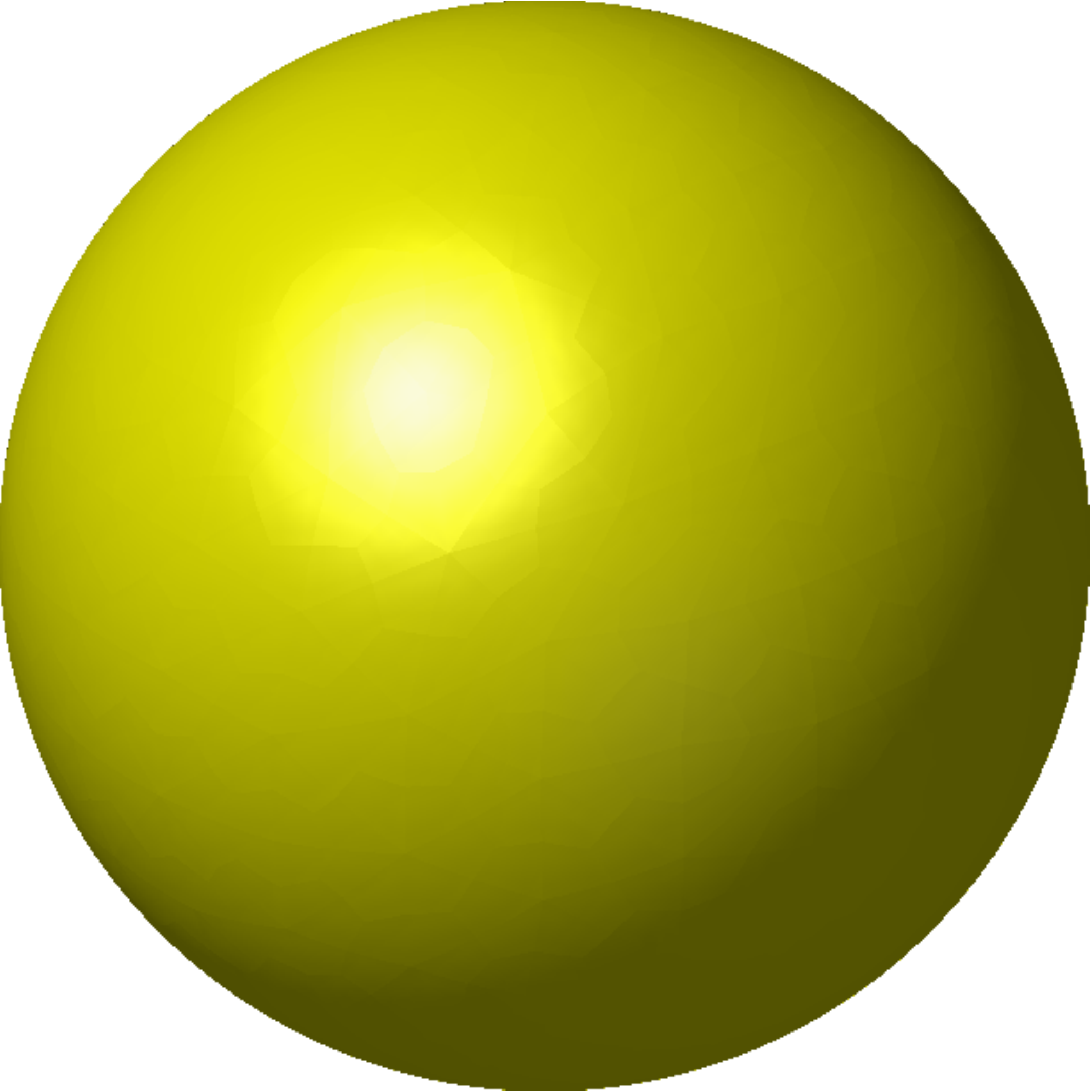}}\quad\subfigure[]{\includegraphics[height=0.2\textwidth]{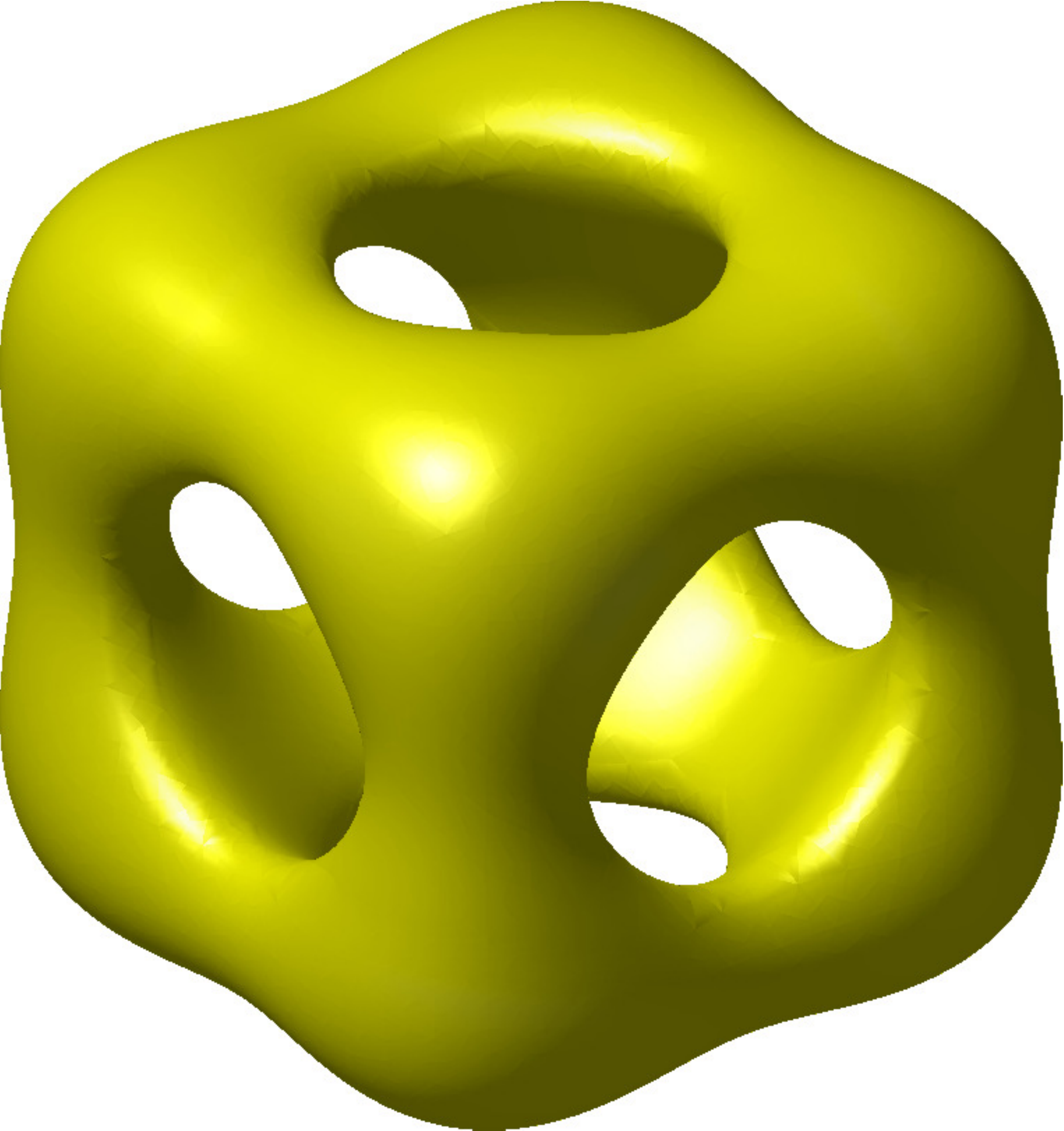}}\quad\subfigure[]{\includegraphics[height=0.2\textwidth]{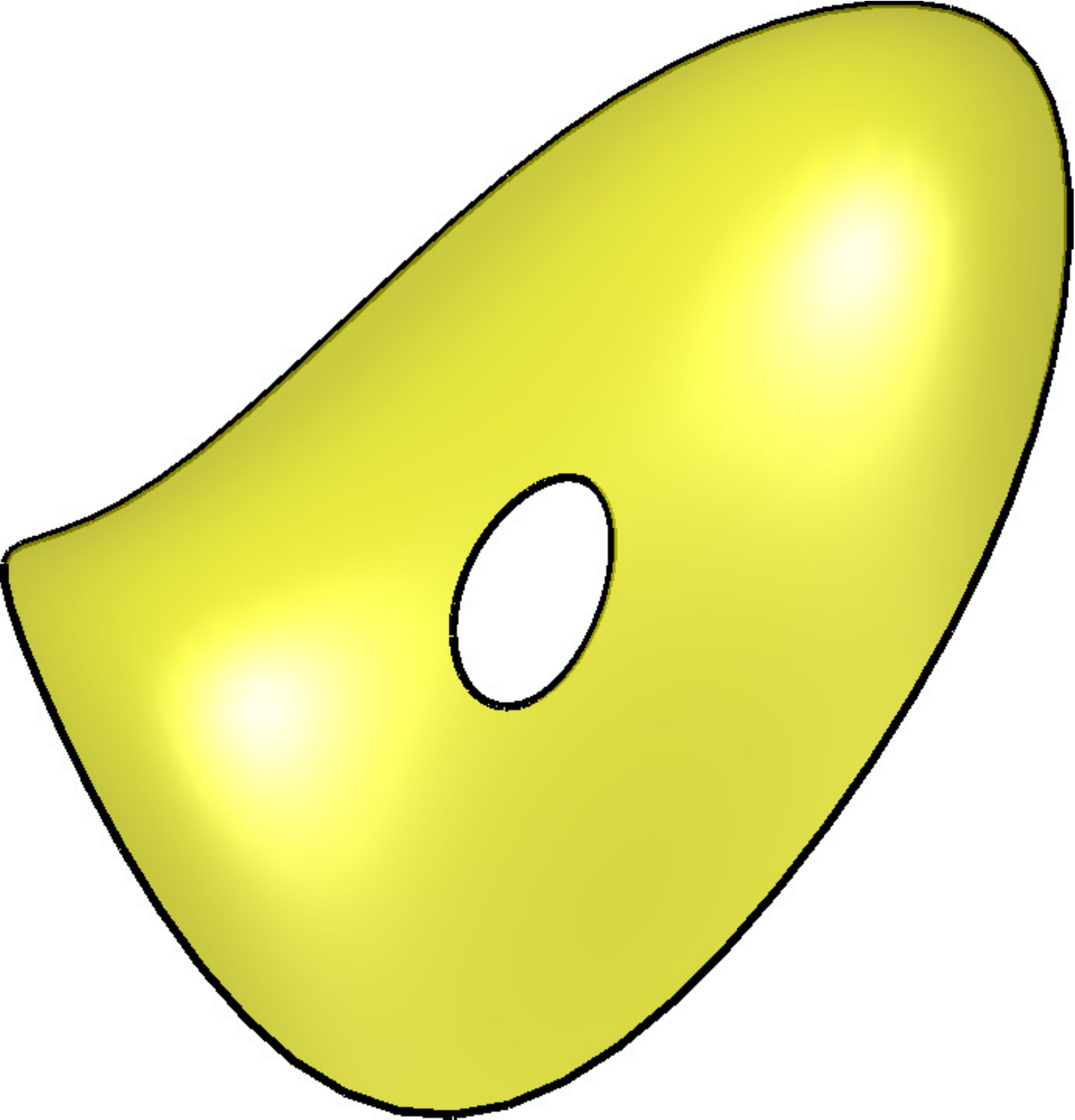}}\quad\subfigure[]{\includegraphics[height=0.2\textwidth]{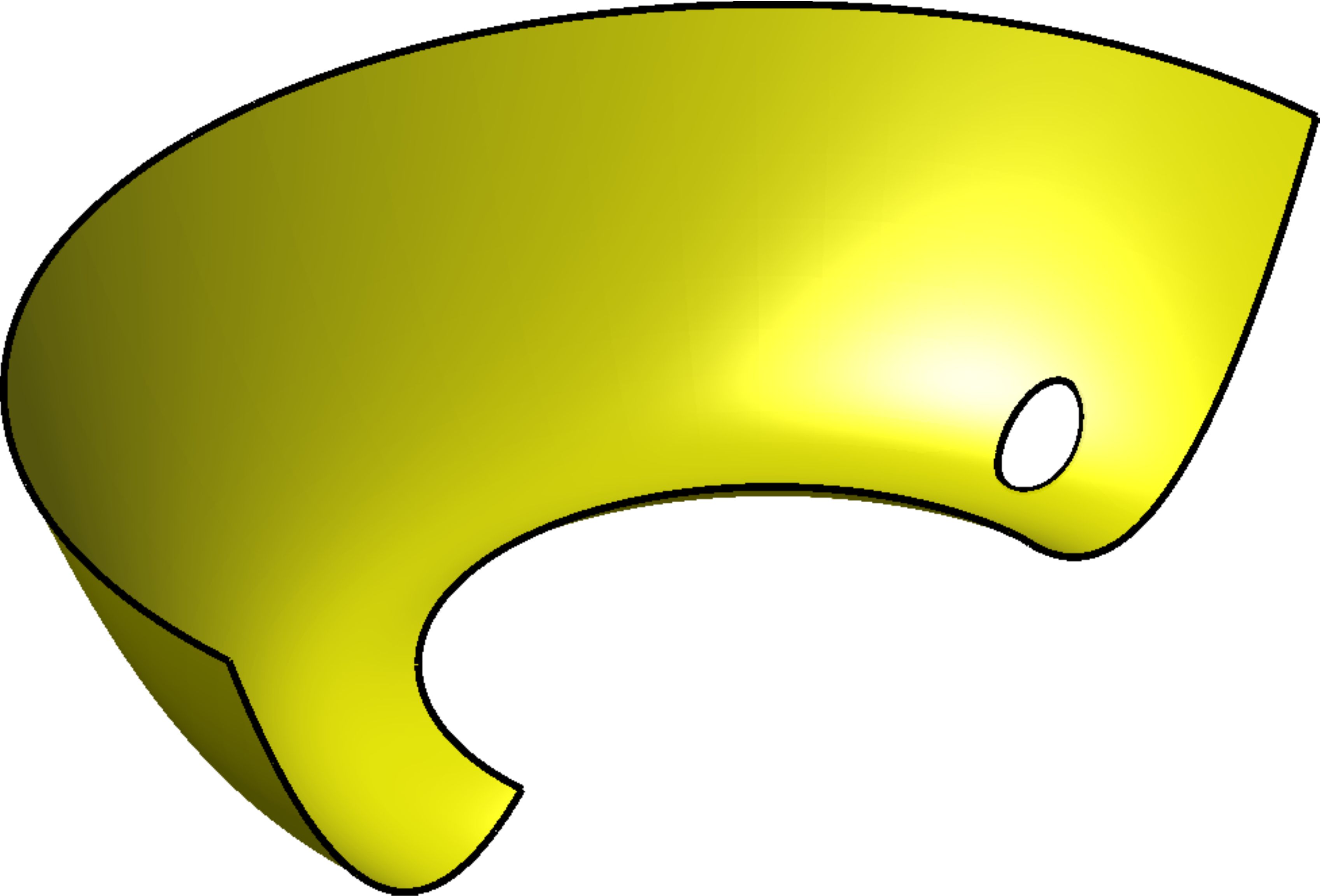}}

\caption{\label{fig:Manifolds}Some examples of (a, b) compact manifolds with
$\partial\Gamma=\emptyset$ and (c, d) manifolds with boundary $\partial\Gamma$.}
\end{figure}

The task is to solve a boundary value problem (BVP) on an arbitrary
surface $\Gamma$ in three dimensions. Let the surface be fixed in
space over time, possibly curved, sufficiently smooth, orientable,
connected (so that there is only \emph{one} surface), and feature
a finite area. There is a unit normal vector $\vek n_{\Gamma}\in\mathbb{R}^{3}$
on $\Gamma$. The surface may be compact, i.e., without a boundary,
$\partial\Gamma=\emptyset$, see Figs.~\ref{fig:Manifolds}(a) and
(b) for examples. Otherwise, it may be bounded by $\partial\Gamma$
as shown in Figs.~\ref{fig:Manifolds}(c) and (d). Then, associated
with $\partial\Gamma$, there is a tangential vector $\vek t_{\partial\Gamma}$
pointing in direction of $\partial\Gamma$ and a co-normal vector
$\vek n_{\partial\Gamma}=\vek n_{\Gamma}\times\vek t_{\partial\Gamma}$
pointing ``outwards'' and being normal to $\partial\Gamma$ and
tangent to $\Gamma$. The surface may be given in parametrized form
or implied, e.g., based on the level-set method; both situations are
considered herein. For the equivalence of these two cases and more
mathematical details, see, e.g., \cite{Dziuk_2013a}.

\subsection{Surface operators\label{XX_SurfaceOperators}}

\subsubsection{The tangential projector}

On the manifold $\Gamma$, the tangential projector $\mat P\left(\vek x\right)\in\mathbb{R}^{3\times3}$
is defined by the normal vector as 
\[
\mat P\left(\vek x\right)=\mat I-\vek n_{\Gamma}\left(\vek x\right)\otimes\vek n_{\Gamma}\left(\vek x\right).
\]
Some important properties are: (i) $\mat P\cdot\vek n_{\Gamma}=\vek0$,
(ii) $\mat P=\mat P^{\mathrm{T}}$, and (iii) $\mat P\cdot\mat P=\mat P$.

\subsubsection{Surface gradient of scalar quantities}

The tangential gradient operator $\nabla_{\Gamma}$ of a differentiable
scalar function $u:\Gamma\to\mathbb{R}$ on the manifold is given
by
\begin{eqnarray}
\nabla_{\Gamma}u\left(\vek x\right) & = & \mat P\left(\vek x\right)\cdot\nabla\tilde{u}\left(\vek x\right),\quad\vek x\in\Gamma,\label{eq:TangGradImplicit}
\end{eqnarray}
where $\nabla$ is the standard gradient operator, and $\tilde{u}$
is a smooth extension of $u$ in a neighborhood $\mathcal{U}$ of
the manifold $\Gamma$. Of course, $\tilde{u}$ may also be some \emph{given}
function (rather than an arbitrary extension) in global coordinates,
i.e., $\tilde{u}\left(\vek x\right):\mathbb{R}^{3}\to\mathbb{R}$.
For the case of parametrized surfaces defined by the map $\vek x\left(\vek r\right):\mathbb{R}^{2}\to\mathbb{R}^{3}$,
and a given scalar function $u\left(\vek r\right):\mathbb{R}^{2}\to\mathbb{R}$,
the tangential gradient may be determined without explicitly computing
an extension $\tilde{u}$ using
\begin{equation}
\nabla_{\Gamma}u\left(\vek x\left(\vek r\right)\right)=\mat J\left(\vek r\right)\cdot\mat G^{-1}\left(\vek r\right)\cdot\nabla_{\vek r}u\left(\vek r\right),\label{eq:TangGradExplicit}
\end{equation}
with $\mat J=\nicefrac{\partial\vek x}{\partial\vek r}$ being the
($3\times2$)-Jacobi matrix and $\mat G=\mat J^{T}\cdot\mat J$ being
the metric tensor (first fundamental form). Equation (\ref{eq:TangGradExplicit})
shall be used later in the context of the FEM to determine tangential
gradients of shape functions. It is noteworthy that $\nabla_{\Gamma}u$
is in the tangent space of $\Gamma$ and, thus, $\mat P\cdot\nabla_{\Gamma}u=\nabla_{\Gamma}u$
and $\nabla_{\Gamma}u\cdot\vek n_{\Gamma}=0$. The components of the
tangential gradient are denoted by
\[
\nabla_{\Gamma}u\left(\vek x\right)=\left(\partial_{x}^{\Gamma}u,\partial_{y}^{\Gamma}u,\partial_{z}^{\Gamma}u\right)^{\mathrm{T}},
\]
representing the first-order partial derivatives on $\Gamma$. Second-order
partial derivatives may be denoted by 
\[
\mathrm{He}_{ij}\left(u\left(\vek x\right)\right)=\partial_{x_{i}\,x_{j}}^{\Gamma}u\left(\vek x\right)=\partial_{x_{i}}^{\Gamma}\left(\partial_{x_{j}}^{\Gamma}u\left(\vek x\right)\right),
\]
where $\mathrm{He}_{ij}\left(u\left(\vek x\right)\right)$ is the
tangential Hessian matrix. In the context of manifolds, this matrix
is not symmetric \cite{Delfour_2011a}, i.e., for mixed second derivatives
$\partial_{x_{i}\,x_{j}}^{\Gamma}u\neq\partial_{x_{j}\,x_{i}}^{\Gamma}u$
for $i\neq j$.

\subsubsection{Surface gradient of vector quantities}

Next, operators for vector quantitites $\vek u\left(\vek x\right):\Gamma\to\mathbb{R}^{3}$
are considered. The ``directional gradient'' of $\vek u$ is the
tensor of tangential derivatives and defined as
\[
\nabla_{\Gamma}^{\mathrm{dir}}\vek u\left(\vek x\right)=\nabla_{\Gamma}^{\mathrm{dir}}\left[\begin{array}{c}
u\left(\vek x\right)\\
v\left(\vek x\right)\\
w\left(\vek x\right)
\end{array}\right]=\left[\begin{array}{ccc}
\partial_{x}^{\Gamma}u & \partial_{y}^{\Gamma}u & \partial_{z}^{\Gamma}u\\
\partial_{x}^{\Gamma}v & \partial_{y}^{\Gamma}v & \partial_{z}^{\Gamma}v\\
\partial_{x}^{\Gamma}w & \partial_{y}^{\Gamma}w & \partial_{z}^{\Gamma}w
\end{array}\right]=\nabla\tilde{\vek u}\cdot\mat P.
\]
In contrast, the \emph{covariant} derivatives are
\[
\nabla_{\Gamma}^{\mathrm{cov}}\vek u\left(\vek x\right)=\mat P\cdot\nabla_{\Gamma}^{\mathrm{dir}}\vek u\left(\vek x\right)=\mat P\cdot\nabla\tilde{\vek u}\cdot\mat P.
\]
One has to carefully distinguish these two different gradient operators.
It is noted that $\nabla_{\Gamma}^{\mathrm{cov}}\vek u$ appears frequently
in the modeling of physical phenomena on manifolds, i.e., in the governing
equations. On the other hand, $\nabla_{\Gamma}^{\mathrm{dir}}\vek u$
is often used in straightforward extensions of identities such as
product rules and divergence theorems. For example, we have for a
scalar function $f\left(\vek x\right)$ and vector functions $\vek u\left(\vek x\right)$,
$\vek v\left(\vek x\right)$ 
\begin{eqnarray*}
\nabla_{\Gamma}^{\mathrm{dir}}\left(f\cdot\vek u\right) & = & \nabla_{\Gamma}f\otimes\vek u+f\cdot\nabla_{\Gamma}^{\mathrm{dir}}\vek u,\\
\vek v^{\mathrm{T}}\cdot\nabla_{\Gamma}^{\mathrm{dir}}\vek u & = & -\vek u^{\mathrm{T}}\cdot\nabla_{\Gamma}^{\mathrm{dir}}\vek v+\nabla_{\Gamma}^{\mathrm{T}}\left(\vek u\cdot\vek v\right),
\end{eqnarray*}
however, the relations are less straightforward for the covariant
counterparts $\nabla_{\Gamma}^{\mathrm{cov}}\left(f\cdot\vek u\right)$
and $\vek v^{\mathrm{T}}\cdot\nabla_{\Gamma}^{\mathrm{cov}}\vek u$,
respectively. Later on, in the context of FEM implementations, it
proves useful to transform covariant derivatives systematically to
directional ones. This allows the computation of directional derivatives
of FE shape functions with respect to $\vek x\in\mathbb{R}^{3}$ independent
of the integration of the weak form of the governing equations.

\subsubsection{Divergence operators and divergence theorem}

The divergence of a vector function $\vek u\left(\vek x\right):\Gamma\to\mathbb{R}^{3}$
is given as
\[
\mathrm{div}_{\Gamma}\vek u\left(\vek x\right)=\mathrm{tr}\left(\nabla_{\Gamma}^{\mathrm{dir}}\vek u\right)=\mathrm{tr}\left(\nabla_{\Gamma}^{\mathrm{cov}}\vek u\right)\eqqcolon\nabla_{\Gamma}\cdot\vek u.
\]
For a tensor function $\mat A\left(\vek x\right):\Gamma\to\mathbb{R}^{3\times3}$,
there holds
\[
\mathrm{div}_{\Gamma}\mat A\left(\vek x\right)=\left[\begin{array}{c}
\mathrm{div}_{\Gamma}\left(A_{11},A_{12},A_{13}\right)\\
\mathrm{div}_{\Gamma}\left(A_{21},A_{22},A_{23}\right)\\
\mathrm{div}_{\Gamma}\left(A_{31},A_{32},A_{33}\right)
\end{array}\right]\eqqcolon\nabla_{\Gamma}\cdot\mat A.
\]
It may be shown that $\mathrm{div}_{\Gamma}\mat P=-\varkappa\cdot\vek n_{\Gamma}$
with $\varkappa=\mathrm{tr}\left(\mat H\right)$ being the mean curvature
and $\mat H=\nabla_{\Gamma}^{\mathrm{cov}}\vek n_{\Gamma}$ being
the second fundamental form.

The following divergence theorem on manifolds is later needed for
deriving the weak forms \cite{Delfour_1996a,Delfour_2011a},
\begin{equation}
\int_{\Gamma}\vek u\cdot\mathrm{div}_{\Gamma}\mat A\,\mathrm{d}A=-\int_{\Gamma}\nabla_{\Gamma}^{\mathrm{dir}}\vek u:\mat A\,\mathrm{d}A+\int_{\Gamma}\varkappa\cdot\vek u\cdot\mat A\cdot\vek n_{\Gamma}\,\mathrm{d}A+\int_{\partial\Gamma}\vek u\cdot\mat A\cdot\vek n_{\partial\Gamma}\,\mathrm{d}s,\label{eq:DivTheorem}
\end{equation}
where $\nabla_{\Gamma}^{\mathrm{dir}}\vek u:\mat A=\mathrm{tr}\left(\nabla_{\Gamma}^{\mathrm{dir}}\vek u\cdot\mat A^{\mathrm{T}}\right)$.
For \emph{tangential} tensor functions with $\mat A=\mat P\cdot\mat A\cdot\mat P$,
the term involving the curvature $\varkappa$ vanishes because then
$\mat A\cdot\vek n_{\Gamma}=\vek0$. In this case, we also have $\nabla_{\Gamma}^{\mathrm{dir}}\vek u:\mat A=\nabla_{\Gamma}^{\mathrm{cov}}\vek u:\mat A$.

\section{Governing equations\label{X_GoverningEquations}}

In the following, we consider (i) stationary Stokes flow, (ii) stationary
Navier-Stokes flow, and (iii) instationary Navier-Stokes flow on fixed
manifolds. The governing equations are first given in strong and weak
forms. The surface FEM is then applied for the discretization of the
weak forms. As mentioned above, these models are also considered,
e.g., in \cite{Bothe_2010a,Jankuhn_2017a,Koba_2017a} among others.

\subsection{Flow models in strong form\label{XX_StrongForms}}

\subsubsection{Stationary Stokes flow}

Starting point is stationary Stokes flow on a manifold. Let $\vek u\left(\vek x\right)\in C^{2}\left(\Gamma\right)$
be the three-dimensional velocity field on the surface $\Gamma$,
$p\left(\vek x\right)\in C^{1}\left(\Gamma\right)$ a pressure field,
and $\vek f_{\!t}\left(\vek x\right)$ a tangential body force, e.g.,
with unit $\unitfrac{N}{m^{2}}$. The governing field equations (in
stress-divergence-form \cite{Donea_2003a}) to be fulfilled $\forall\vek x\in\Gamma$
are
\begin{eqnarray}
-\mat P\cdot\mathrm{div}_{\Gamma}\,\vek\sigma\left(\vek u,p\right) & = & \vek f_{\!t},\label{eq:MomEqtStatStokes}\\
\mathrm{div}_{\Gamma}\,\vek u & = & 0,\label{eq:ContinuityConstraint}\\
\vek u\cdot\vek n_{\Gamma} & = & 0.\label{eq:VelocityConstraint}
\end{eqnarray}
Equation (\ref{eq:MomEqtStatStokes}) expands to three momentum equations,
equation (\ref{eq:ContinuityConstraint}) is the incompressibility
constraint and equation (\ref{eq:VelocityConstraint}) represents
the tangential velocity constraint that restricts the velocities to
the tangent space of $\Gamma$. Two different strain tensors are introduced,
\begin{eqnarray}
\vek\varepsilon^{\mathrm{dir}}\left(\vek u\right) & = & \frac{1}{2}\cdot\left(\nabla_{\Gamma}^{\mathrm{dir}}\vek u+\left(\nabla_{\Gamma}^{\mathrm{dir}}\vek u\right)^{\mathrm{T}}\right),\label{eq:StrainTensorDir}\\
\vek\varepsilon^{\mathrm{cov}}\left(\vek u\right) & = & \frac{1}{2}\cdot\left(\nabla_{\Gamma}^{\mathrm{cov}}\vek u+\left(\nabla_{\Gamma}^{\mathrm{cov}}\vek u\right)^{\mathrm{T}}\right),\label{eq:StrainTensorCov}
\end{eqnarray}
which are related to each other as $\vek\varepsilon^{\mathrm{cov}}\left(\vek u\right)=\mat P\cdot\vek\varepsilon^{\mathrm{dir}}\left(\vek u\right)\cdot\mat P$.
The stress tensor is then defined as
\[
\vek\sigma\left(\vek u,p\right)=-p\cdot\mat P+2\mu\cdot\vek\varepsilon^{\mathrm{cov}}\left(\vek u\right)
\]
where $\mu\in\mathbb{R}^{+}$ is the (constant) dynamic viscosity.
It is easily shown that
\[
-\mat P\cdot\mathrm{div}_{\Gamma}\,\vek\sigma\left(\vek u,p\right)=\nabla_{\Gamma}p-2\mu\mat P\cdot\mathrm{div}_{\Gamma}\,\vek\varepsilon^{\mathrm{cov}}\left(\vek u\right)
\]

Suppose there exists a boundary $\partial\Gamma$ of the manifold
that consists of two non-overlapping parts, the Dirichlet boundary,
$\partial\Gamma_{\mathrm{D}}$, and the Neumann boundary, $\partial\Gamma_{\mathrm{N}}$.
The corresponding boundary conditions are given as 
\begin{equation}
\begin{array}{ccccc}
\vek u\left(\vek x\right) & = & \hat{\vek u}\left(\vek x\right) & \qquad & \text{on }\partial\Gamma_{\mathrm{D}},\\
\vek\sigma\left(\vek x\right)\cdot\vek n_{\partial\Gamma}\left(\vek x\right) & = & \hat{\vek t}\left(\vek x\right) & \qquad & \text{on }\partial\Gamma_{\mathrm{N}},
\end{array}\label{eq:BoundaryConditions}
\end{equation}
where the prescribed velocities $\hat{\vek u}$ and tractions $\hat{\vek t}$
are in the tangent space of $\Gamma$, i.e., $\hat{\vek u}\cdot\vek n_{\Gamma}=\hat{\vek t}\cdot\vek n_{\Gamma}=0.$

Note that, in general, there are no explicit boundary conditions needed
for the pressure $p$. In cases where no Neumann boundary is present,
i.e., $\partial\Gamma_{\mathrm{N}}=\emptyset$ and $\partial\Gamma_{\mathrm{D}}=\partial\Gamma$,
the pressure is defined up to a constant \cite{Donea_2003a,Gresho_2000a}.
This includes compact manifolds where $\partial\Gamma=\emptyset$.
In such situations, the pressure may be prescribed at a given point
on $\Gamma$ or it is imposed by a constraint in the form of $\int_{\Gamma}p\;\mathrm{d}A=0$.

\paragraph{Vorticity on manifolds.}

The vorticity $\vek\omega$ is a physical quantity frequently computed
in flow problems. In the context of manifolds, we shall define
\begin{equation}
\vek\omega=\nabla_{\Gamma}^{\mathrm{cov}}\times\vek u.\label{eq:RegularVorticity}
\end{equation}
Note that $\vek\omega$ is co-linear to the normal vector $\vek n_{\Gamma}$,
hence, $\mat P\cdot\vek\omega=\vek0$. Therefore, it is useful to
determine the signed magnitude of $\vek\omega$, that is, the scalar
function 
\begin{equation}
\omega^{\star}\left(\vek x\right)=\vek\omega\cdot\vek n_{\Gamma}=\pm\left\Vert \vek\omega\right\Vert \qquad\forall\vek x\in\Gamma.\label{eq:SpecialVorticity}
\end{equation}
This scalar quantity may also be obtained using directional derivatives,
i.e., $\omega^{\star}=\left(\nabla_{\Gamma}^{\mathrm{dir}}\times\vek u\right)\cdot\vek n_{\Gamma}$.

\subsubsection{Stationary Navier-Stokes flow}

For stationary \emph{Navier-Stokes} flow, a non-linear advection term
is added to equation (\ref{eq:MomEqtStatStokes}) resulting into
\begin{equation}
\varrho\cdot\left(\vek u\cdot\nabla_{\Gamma}^{\mathrm{cov}}\right)\vek u-\mat P\cdot\mathrm{div}_{\Gamma}\,\vek\sigma\left(\vek x\right)=\vek f_{\!t}\left(\vek x\right),\label{eq:MomentumEqtNSstat}
\end{equation}
where $\varrho\in\mathbb{R}^{+}$ is the (constant) fluid density
with unit $\unitfrac{kg}{m^{2}}$ and $\left(\vek u\cdot\nabla_{\Gamma}^{\mathrm{cov}}\right)\vek u\coloneqq\left(\nabla_{\Gamma}^{\mathrm{cov}}\vek u\right)\cdot\vek u$.
It is quite common to express the body force in the form $\vek f_{\!t}\left(\vek x\right)=\varrho\cdot\vek g_{t}\left(\vek x\right)$
where $\vek g_{t}$ may consider gravity as $\vek g_{t}=\mat P\cdot\left[0,0,-9.81\right]^{\mathrm{T}}\unitfrac{m}{s^{2}}$
for instance. The remaining equations (\ref{eq:ContinuityConstraint})
and (\ref{eq:VelocityConstraint}) and the boundary conditions (\ref{eq:BoundaryConditions})
remain unchanged. The solution of the non-linear governing equations
can be obtained iteratively based on the Newton-Raphson method or
other fixed-point iterations such as Picard iterations. Because the
advection operator is not self-adjoint, well-known stability issues
may arise for large Reynolds numbers in a numerical context.

\subsubsection{Instationary Navier-Stokes flow}

For \emph{instationary} Navier-Stokes flow, the momentum equation
(\ref{eq:MomEqtStatStokes}) changes to
\begin{equation}
\varrho\cdot\left(\partial_{t}\vek u\left(\vek x,t\right)+\left(\vek u\cdot\nabla_{\Gamma}^{\mathrm{cov}}\right)\vek u-\vek g_{t}\left(\vek x,t\right)\right)-\mat P\cdot\mathrm{div}_{\Gamma}\,\vek\sigma\left(\vek x,t\right)=\vek0.\label{eq:MomentumEqtNSinstat}
\end{equation}
The functions representing the physical fields live in space (on $\Gamma$)
\emph{and} time, i.e., in the time interval $\tau=\left[0,T\right]$.
Therefore, Eqs.~(\ref{eq:MomentumEqtNSinstat}), (\ref{eq:ContinuityConstraint}),
and (\ref{eq:VelocityConstraint}) have to be solved in the space-time
domain $\Gamma\times\tau$. Herein, we restrict ourselves to spatially
fixed manifolds $\Gamma$.

The boundary conditions (\ref{eq:BoundaryConditions}) also extend
in time dimension, hence, there are prescribed velocities $\hat{\vek u}\left(\vek x,t\right)$
along $\partial\Gamma_{\mathrm{D}}\times\tau$ and tractions $\hat{\vek t}\left(\vek x,t\right)$
along $\partial\Gamma_{\mathrm{N}}\times\tau$. Furthermore, an initial
condition is needed,
\begin{equation}
\vek u\left(\vek x,0\right)=\vek u_{0}\left(\vek x\right),\;\text{with}\;\mathrm{div}_{\Gamma}\,\vek u_{0}=0\;\text{and}\;\vek u_{0}\cdot\vek n_{\Gamma}=0\quad\forall\vek x\in\Gamma\;\text{at}\;t=0.\label{eq:InitialCond}
\end{equation}

\subsection{Flow models in weak form\label{XX_WeakForms}}

The following trial and test function spaces are introduced, 
\begin{eqnarray}
\mathcal{S}_{\vek u} & = & \left\{ \vek u\in\mathcal{H}^{1}\left(\Gamma\right)^{3},\:\vek u=\hat{\vek u}\:\textrm{on}\:\partial\Gamma_{\mathrm{D}}\right\} ,\label{eq:TrialU}\\
\mathcal{V}_{\vek u} & = & \left\{ \vek w_{\vek u}\in\mathcal{H}^{1}\left(\Gamma\right)^{3},\:\vek w_{\vek u}=\vek0\:\textrm{on}\:\partial\Gamma_{\mathrm{D}}\right\} ,\label{eq:TestU}\\
\mathcal{S}_{p}=\mathcal{V}_{p} & = & \mathcal{L}_{2}\left(\Gamma\right),\label{eq:TrialAndTestP}\\
\mathcal{S}_{\lambda}=\mathcal{V}_{\lambda} & = & \mathcal{L}_{2}\left(\Gamma\right).\label{eq:TrialAndTestLM}
\end{eqnarray}
As mentioned previously, if no Neumann boundary exists, i.e., $\partial\Gamma_{\mathrm{N}}=\emptyset$,
the pressure is defined up to a constant and one may replace $\mathcal{S}_{p}$
by 
\[
\mathcal{S}_{p}^{0}=\big\{ p\in\mathcal{L}_{2}\left(\Gamma\right),\:\int_{\Gamma}p\;\mathrm{d}A=0\big\}.
\]

\subsubsection{Stationary Stokes flow}

The weak form of the Stokes problem becomes: Given viscosity $\mu\in\mathbb{R}^{+}$,
body force $\vek f\left(\vek x\right)$ in $\Gamma$, and traction
$\hat{\vek t}\left(\vek x\right)$ on $\partial\Gamma_{\mathrm{N}}$,
find the velocity field $\vek u\left(\vek x\right)\in\mathcal{S}_{\vek u}$,
pressure field $p\left(\vek x\right)\in\mathcal{S}_{p}$, and Lagrange
multiplier field $\lambda\left(\vek x\right)\in\mathcal{S}_{\lambda}$
such that for all test functions $\left(\vek w_{\vek u},w_{p},w_{\lambda}\right)\in\mathcal{V}_{\vek u}\times\mathcal{V}_{p}\times\mathcal{V}_{\lambda}$,
there holds in $\Gamma$ 
\begin{eqnarray}
\int_{\Gamma}\nabla_{\Gamma}^{\mathrm{dir}}\vek w_{\vek u}:\vek\sigma\left(\vek u,p\right)\mathrm{d}A+\int_{\Gamma}\lambda\cdot\left(\vek w_{\vek u}\cdot\vek n_{\Gamma}\right)\mathrm{d}A & = & \int_{\Gamma}\vek w_{\vek u}\cdot\vek f\mathrm{d}A+\int_{\partial\Gamma_{\mathrm{N}}}\!\!\!\vek w_{\vek u}\cdot\hat{\vek t}\,\mathrm{d}s,\label{eq:WeakFormMomentum}\\
\int_{\Gamma}w_{p}\cdot\mathrm{div}_{\Gamma}\,\vek u\:\mathrm{d}A & = & 0,\label{eq:WeakFormContinuity}\\
\int_{\Gamma}w_{\lambda}\cdot\left(\vek u\cdot\vek n_{\Gamma}\right)\mathrm{d}A & = & 0.\label{eq:WeakFormTangVelConstraint}
\end{eqnarray}
In order to obtain Eq.~(\ref{eq:WeakFormMomentum}), the divergence
theorem (\ref{eq:DivTheorem}) was applied to $-\int_{\Gamma}\vek w_{\vek u}\cdot\mathrm{div}_{\Gamma}\,\vek\sigma\,\mathrm{d}A$
where the curvature term vanishes due to $\vek\sigma\cdot\vek n_{\Gamma}=\vek0$.
Using the definition of the stress tensor, we get
\[
\int_{\Gamma}\nabla_{\Gamma}^{\mathrm{dir}}\vek w_{\vek u}:\vek\sigma\left(\vek u,p\right)\mathrm{d}A=-\int_{\Gamma}\nabla_{\Gamma}^{\mathrm{dir}}\vek w_{\vek u}:\left(p\cdot\mat P\right)\mathrm{d}A+2\mu\cdot\int_{\Gamma}\nabla_{\Gamma}^{\mathrm{dir}}\vek w_{\vek u}:\vek\varepsilon^{\mathrm{cov}}\left(\vek u\right)\mathrm{d}A
\]
The following relations are easily derived:
\begin{eqnarray}
\nabla_{\Gamma}^{\mathrm{dir}}\vek w_{\vek u}:\left(p\cdot\mat P\right) & = & p\cdot\mathrm{tr}\left(\nabla_{\Gamma}^{\mathrm{dir}}\vek w_{\vek u}\cdot\mat P\right)\nonumber \\
 & = & p\cdot\mathrm{div}_{\Gamma}\,\vek w_{\vek u}\nonumber \\
\nabla_{\Gamma}^{\mathrm{dir}}\vek w_{\vek u}:\vek\varepsilon^{\mathrm{cov}}\left(\vek u\right) & = & \mathrm{tr}\left(\nabla_{\Gamma}^{\mathrm{dir}}\vek w_{\vek u}\cdot\vek\varepsilon^{\mathrm{cov}}\left(\vek u\right)\right)\nonumber \\
 & = & \mathrm{tr}\left(\vek\varepsilon^{\mathrm{cov}}\left(\vek w_{\vek u}\right)\cdot\vek\varepsilon^{\mathrm{cov}}\left(\vek u\right)\right)\nonumber \\
 & = & \mathrm{tr}\left(\mat P\cdot\nabla_{\Gamma}^{\mathrm{dir}}\vek w_{\vek u}\cdot\vek\varepsilon^{\mathrm{dir}}\left(\vek u\right)\cdot\mat P\right)\label{eq:DiffContractionVariantC}
\end{eqnarray}

It is readily verified that solutions of the strong form also fulfill
the weak form from above. This is obvious for Eqs.~(\ref{eq:WeakFormContinuity})
and (\ref{eq:WeakFormTangVelConstraint}) due to Eqs.~(\ref{eq:ContinuityConstraint})
and (\ref{eq:VelocityConstraint}), respectively. For the momentum
equations, it is noted that (\ref{eq:WeakFormMomentum}) is fufilled
for $-\mathrm{div}_{\Gamma}\,\vek\sigma\left(\vek u,p\right)+\lambda\cdot\vek n_{\Gamma}=\vek f$.
Restricting this to the tangential space by multiplication with the
projector $\mat P$ yields the strong form of the momentum equations
(\ref{eq:MomEqtStatStokes}) because $\mat P\cdot\vek n_{\Gamma}=\vek0$.
It is thus also seen that the Lagrange multiplier field $\lambda$
may be physically interpreted as a force in normal direction.

\subsubsection{Stationary Navier-Stokes flow}

The weak form of the stationary Navier-Stokes equations is similar
to the Stokes problem from above, however, Eq.~(\ref{eq:WeakFormMomentum})
is replaced by
\begin{eqnarray*}
\varrho\cdot\int_{\Gamma}\vek w_{\vek u}\cdot\left(\vek u\cdot\nabla_{\Gamma}^{\mathrm{cov}}\right)\vek u\,\mathrm{d}A+\int_{\Gamma}\nabla_{\Gamma}^{\mathrm{dir}}\vek w_{\vek u}:\vek\sigma\left(\vek u,p\right)\mathrm{d}A+\int_{\Gamma}\lambda\cdot\left(\vek w_{\vek u}\cdot\vek n_{\Gamma}\right)\mathrm{d}A\\
=\int_{\Gamma}\vek w_{\vek u}\cdot\vek f\mathrm{d}A+\int_{\partial\Gamma_{\mathrm{N}}}\!\!\!\vek w_{\vek u}\cdot\hat{\vek t}\,\mathrm{d}s
\end{eqnarray*}
where the added advection term is readily identified.

\subsubsection{Instationary Navier-Stokes flow}

The weak form of the instationary Navier-Stokes problem is: Given
density $\varrho\in\mathbb{R}^{+}$, viscosity $\mu\in\mathbb{R}^{+}$,
body force $\varrho\cdot\vek g\left(\vek x,t\right)$ in $\Gamma\times\tau$,
traction $\hat{\vek t}\left(\vek x,t\right)$ on $\partial\Gamma_{\mathrm{N}}\times\tau$,
and initial condition $u_{0}\left(\vek x\right)$ on $\Gamma$ at
$t=0$ according to (\ref{eq:InitialCond}), find the velocity field
$\vek u\left(\vek x,t\right)\in L_{2}\left(\tau;\mathcal{S}_{\vek u}\right)$,
pressure field $p\left(\vek x,t\right)\in L_{2}\left(\tau;\mathcal{S}_{p}\right)$,
and Lagrange multiplier field $\lambda\left(\vek x,t\right)\in L_{2}\left(\tau;\mathcal{S}_{\lambda}\right)$
such that for all test functions $\left(\vek w_{\vek u},w_{p},w_{\lambda}\right)\in\mathcal{V}_{\vek u}\times\mathcal{V}_{p}\times\mathcal{V}_{\lambda}$,
there holds in $\Gamma\times\tau$ 
\begin{eqnarray}
\varrho\cdot\int_{\Gamma}\vek w_{\vek u}\cdot\left(\partial_{t}\vek u+\left(\vek u\cdot\nabla_{\Gamma}^{\mathrm{cov}}\right)\vek u-\vek g\right)\mathrm{d}A\label{eq:WeakFormMomentum-1}\\
+\int_{\Gamma}\nabla_{\Gamma}\vek w_{\vek u}^{\mathrm{dir}}:\vek\sigma\left(\vek u,p\right)\mathrm{d}A+\int_{\Gamma}\lambda\cdot\left(\vek w_{\vek u}\cdot\vek n_{\Gamma}\right)\mathrm{d}A & = & \int_{\partial\Gamma_{\mathrm{N}}}\!\!\!\vek w_{\vek u}\cdot\hat{\vek t}\,\mathrm{d}s,\nonumber \\
\int_{\Gamma}w_{p}\cdot\mathrm{div}_{\Gamma}\,\vek u\:\mathrm{d}A & = & 0,\label{eq:WeakFormContinuity-1}\\
\int_{\Gamma}w_{\lambda}\cdot\left(\vek u\cdot\vek n_{\Gamma}\right)\mathrm{d}A & = & 0.\label{eq:WeakFormTangVelConstraint-1}
\end{eqnarray}

\subsection{Surface FEM for flows on manifolds\label{XX_SurfaceFEM}}

\subsubsection{Surface meshes\label{XX_SurfaceMeshes}}

Assume that a suitable surface mesh composed by higher-order triangular
or quadrilateral Lagrange elements of order $q$ may be generated
with desired element sizes and all nodes on $\Gamma$. Well-known,
necessary requirements of meshes such as the shape regularity of the
elements and bounds on inner angles, are fulfilled. The shape of each
(physical) element in the mesh results from a map of the corresponding
reference element with $n_{q}$ nodes,
\begin{equation}
\vek x\left(\vek r\right)=\left[\begin{array}{c}
x\left(r,s\right)\\
y\left(r,s\right)\\
z\left(r,s\right)
\end{array}\right]=\sum_{i=1}^{n_{q}}N_{i}^{q}\left(\vek r\right)\vek x_{i}.\label{eq:IsoparamMapping}
\end{equation}
$N_{i}^{q}\left(\vek r\right)$ are classical Lagrangean shape functions
of order $q$ in reference coordinates $\vek r\in\mathbb{R}^{2}$
and $\vek x_{i}\in\Gamma$ are the nodal coordinates. The resulting
mesh is an approximation $\Gamma_{q}^{h}\in C^{0}$ of the exact surface
$\Gamma$. Clearly, $\Gamma_{q}^{h}$ is defined \emph{parametrically}
through the map (\ref{eq:IsoparamMapping}) even if the original $\Gamma$
was implicitly given, e.g., by the zero-isosurface of a level-set
function. See \cite{Fries_2015a,Fries_2016b,Fries_2017a} for the
automatic generation of higher-order meshes on zero-isosurfaces. The
discrete unit normal vector is
\[
\vek n_{\Gamma}^{h}=\frac{\partial_{r}\vek x\times\partial_{s}\vek x}{\left\Vert \partial_{r}\vek x\times\partial_{s}\vek x\right\Vert }
\]
and is not smooth across element edges due to the $C^{0}$-continuity
of the surface mesh. The discrete tangent and co-normal vectors $\vek t_{\partial\Gamma}^{h}$
and $\vek n_{\partial\Gamma}^{h}$ are easily obtained along the element
edges on the boundary of $\Gamma_{q}^{h}$. The definitions of the
surface operators from Section \ref{XX_SurfaceOperators} readily
extend to the case of a discrete manifold $\Gamma_{q}^{h}$ and are
not repeated here.

\subsubsection{Surface FEM}

We use higher-order surface FEM as detailed, e.g., in \cite{Demlow_2009a,Dziuk_2013a}
for the discretization of the weak forms from above. Finite element
spaces of different orders are involved. As mentioned before, suitable
surface meshes of order $q$ may be generated defining approximations
$\Gamma_{q}^{h}\in C^{0}$ of $\Gamma.$ Let there be a ``geometry
mesh'' of order $q=k_{\mathrm{geom}}$ with the sole purpose to approximate
the geometry of the manifold $\Gamma^{h}=\Gamma_{k_{\mathrm{geom}}}^{h}$
and define the element maps (\ref{eq:IsoparamMapping}). In particular,
this mesh is not used to imply a finite element space for the approximation
of the weak forms.

Next, a finite element space of order $k$ is generated on $\Gamma^{h}$
for which it is assumed that there is a second mesh of order $k$.
The two meshes feature the same element types and number of elements
with identical coordinates at the corners, however, the total number
of nodes differs due to the individual orders. It is emphasized that
the coordinates of the nodes in the $k$-th order mesh are, in fact,
never needed and it is only the connectivity which is required to
set up the finite element space.

Associated to triangular or quadrilateral elements in the $k$-th
order mesh, there is a fixed set of \emph{local} basis functions $\left\{ N_{i}^{k}\left(\vek r\right)\right\} $
defined in a reference element with $i=1,\dots,n_{k}$ and $n_{k}$
being the number of nodes per element. Classical Lagrange basis functions
with $N_{i}^{k}\left(\vek r_{j}\right)=\delta_{ij}$ are used herein.
Based on the map (\ref{eq:IsoparamMapping}) which is completely determined
by the geometry mesh, one may generate $\left\{ N_{i}^{k}\left(\vek x\left(\vek r\right)\right)\right\} $
for all $\vek x\in\Gamma^{h}$ and tangential derivatives $\nabla_{\Gamma}N_{i}^{k}\left(\vek x\left(\vek r\right)\right)$
are determined based on Eq.~(\ref{eq:TangGradExplicit}). This is
only an \emph{iso}-parametric map when $k=k_{\mathrm{geom}}$. Summing
up the element contributions for nodes belonging to several elements,
this generates a set of \emph{global}, $C^{0}$-continuous basis functions
$\left\{ M_{i}^{k}\left(\vek x\left(\vek r\right)\right)\right\} $
in $\Gamma^{h}$ with $i=1,\dots,n_{\mathrm{nodes}}^{k}$ and $n_{\mathrm{nodes}}^{k}$
being the number of nodes of the $k$-th order surface mesh. Note
that to generate the nodal basis $\left\{ M_{i}^{k}\left(\vek x\left(\vek r\right)\right)\right\} $,
only the coordinates of the geometry mesh are needed, however, not
from the $k$-th order mesh. A general finite element space of order
$k$ is now defined by
\[
\mathcal{Q}_{k}^{h}=\Big\{ u^{h}\in C_{0}\left(\Gamma_{k_{\mathrm{geom}}}^{h}\right),\:u^{h}=\sum_{i=1}^{n_{\mathrm{nodes}}^{k}}M_{i}^{k}\left(\vek x\left(\vek r\right)\right)\cdot u_{i},\;u_{i}\in\mathbb{R}\Big\}\subset\mathcal{H}^{1}\left(\Gamma_{k_{\mathrm{geom}}}^{h}\right).
\]

Based on this, the following discrete trial and test function spaces
are defined, 
\begin{eqnarray}
\mathcal{S}_{\vek u}^{h} & = & \left\{ \vek u^{h}\in\left[\mathcal{Q}_{k_{\vek u}}^{h}\right]^{3},\:\vek u^{h}=\hat{\vek u}^{h}\:\textrm{on}\:\partial\Gamma_{\mathrm{D}}^{h}\right\} ,\label{eq:TrialU-1}\\
\mathcal{V}_{\vek u}^{h} & = & \left\{ \vek w_{\vek u}^{h}\in\left[\mathcal{Q}_{k_{\vek u}}^{h}\right]^{3},\:\vek w_{\vek u}^{h}=\vek0\:\textrm{on}\:\partial\Gamma_{\mathrm{D}}^{h}\right\} ,\label{eq:TestU-1}\\
\mathcal{S}_{p}^{h}=\mathcal{V}_{p}^{h} & = & \mathcal{Q}_{k_{p}}^{h},\label{eq:TrialAndTestP-1}\\
\mathcal{S}_{\lambda}^{h}=\mathcal{V}_{\lambda}^{h} & = & \mathcal{Q}_{k_{\lambda}}^{h}.\label{eq:TrialAndTestLM-1}
\end{eqnarray}
Although shape functions for the pressure and the Lagrange multiplier
for enforcing the tangential velocity constraint may be discontinuous,
we restrict ourselves to classical $C^{0}$-continuous approximations.
Note that individual orders $k_{\vek u}$, $k_{p}$, and $k_{\lambda}$
are associated to the approximations of velocities $\vek u^{h}$,
pressure $p^{h}$, and Lagrange multiplier field $\lambda^{h}$, respectively.
Analogous to the continuous case, one may impose that the functions
in $\mathcal{S}_{p}^{h}$ have to fulfill $\int_{\Gamma}p^{h}\,\mathrm{d}A=0$
if no Neumann boundary is present.

\subsubsection{Stationary Stokes flow}

The discrete weak form of the Stokes problem reads: Given viscosity
$\mu\in\mathbb{R}^{+}$, body force $\vek f^{h}\left(\vek x\right)$
in $\Gamma^{h}$, and traction $\hat{\vek t}^{h}\left(\vek x\right)$
on $\partial\Gamma_{\mathrm{N}}^{h}$, find the velocity field $\vek u^{h}\left(\vek x\right)\in\mathcal{S}_{\vek u}^{h}$,
pressure field $p^{h}\left(\vek x\right)\in\mathcal{S}_{p}^{h}$,
and Lagrange multiplier field $\lambda^{h}\left(\vek x\right)\in\mathcal{S}_{\lambda}^{h}$
such that for all test functions $\left(\vek w_{\vek u}^{h},w_{p}^{h},w_{\lambda}^{h}\right)\in\mathcal{V}_{\vek u}^{h}\times\mathcal{V}_{p}^{h}\times\mathcal{V}_{\lambda}^{h}$,
there holds in $\Gamma^{h}$ 
\begin{eqnarray}
\int_{\Gamma}\nabla_{\Gamma}^{\mathrm{dir}}\vek w_{\vek u}^{h}:\vek\sigma\left(\vek u^{h},p^{h}\right)\mathrm{d}A+\int_{\Gamma}\lambda^{h}\cdot\left(\vek w_{\vek u}^{h}\cdot\vek n_{\Gamma}^{h}\right)\mathrm{d}A & = & \int_{\Gamma}\vek w_{\vek u}^{h}\cdot\vek f^{h}\mathrm{d}A+\int_{\partial\Gamma_{\mathrm{N}}}\!\!\!\vek w_{\vek u}^{h}\cdot\hat{\vek t^{h}}\,\mathrm{d}s,\label{eq:DiscWeakFormMomentum}\\
\int_{\Gamma}w_{p}^{h}\cdot\mathrm{div}_{\Gamma}\,\vek u^{h}\;\mathrm{d}A & = & 0,\label{eq:DiscWeakFormContinuity}\\
\int_{\Gamma}w_{\lambda}^{h}\cdot\left(\vek u^{h}\cdot\vek n_{\Gamma}^{h}\right)\mathrm{d}A & = & 0.\label{eq:DiscWeakFormTangVelConstraint}
\end{eqnarray}
The usual element assembly yields a linear system of equations in
the form
\begin{equation}
\left[\begin{array}{ccc}
\mat K & \mat G & \mat L\\
\mat G^{\mathrm{T}} & \mat0 & \mat0\\
\mat L^{\mathrm{T}} & \mat0 & \mat0
\end{array}\right]\cdot\left[\begin{array}{c}
\underline{\vek u}\\
\vek p\\
\vek\lambda
\end{array}\right]=\left[\begin{array}{c}
\vek f\\
\vek0\\
\vek0
\end{array}\right],\label{eq:StokesSystemOfEquations}
\end{equation}
with $\left[\underline{\vek u},\vek p,\vek\lambda\right]^{\mathrm{T}}=\left[\vek u,\vek v,\vek w,\vek p,\vek\lambda\right]^{\mathrm{T}}$
being the sought nodal values of the velocity components, pressure,
and Lagrange multiplier. For the implementation, it is interesting
to compare the system (\ref{eq:StokesSystemOfEquations}) with the
system obtained for a classical three-dimensional Stokes problem,
\begin{equation}
\left[\begin{array}{cc}
\mat K_{\mathrm{3D}} & \mat G_{\mathrm{3D}}\\
\mat G_{\mathrm{3D}}^{\mathrm{T}} & \mat0
\end{array}\right]\cdot\left[\begin{array}{c}
\underline{\vek u}\\
\vek p
\end{array}\right]=\left[\begin{array}{c}
\vek f\\
\vek0
\end{array}\right].\label{eq:StokesSystemOfEquations3D}
\end{equation}
Assume a function which generates $\mat K_{\mathrm{3D}}$ and $\mat G_{\mathrm{3D}}$
based on three-dimensional FE shape functions (including classical
partial derivatives with respect to $\vek x$) evaluated at given
integration points in 3D. The same function may be used for generating
$\mat K$ and $\mat G$ provided that (i) the integration points are
restricted to $\Gamma^{h}$ with proper weights, (ii) the classical
partial derivatives in $\nabla$ are replaced by the tangential derivatives
as in $\nabla_{\Gamma}^{\mathrm{dir}}$, and (iii) the contribution
to $\mat K$ at the current integration point, $\mat K\left(\vek x_{i}\right)$,
is projected as $\mat K\left(\vek x_{i}\right)=\mat P\left(\vek x_{i}\right)\cdot\mat K_{\mathrm{3D}}\left(\vek x_{i}\right)\cdot\mat P\left(\vek x_{i}\right)$
which is due to Eq.~(\ref{eq:DiffContractionVariantC}). The same
shall later hold for the advection matrix $\mat C\left(\underline{\vek u}\right)$
in the Navier-Stokes equations.

As expected in the context of the Lagrange multiplier method, the
matrix in Eq.~(\ref{eq:StokesSystemOfEquations}) has a saddle-point
structure and is typical for a mixed FEM. The well-known Babu\v ska-Brezzi
condition \cite{Babuska_1971a,Brezzi_1974,Franca_1988a} must be fulfilled
to obtain useful solutions for all involved fields. This may be achieved
by adjusting the orders of the approximation spaces for the different
fields and is further detailed in the numerical results. It is noted
that stabilization may be employed to circumvent the Babu\v ska-Brezzi
condition rather than to fulfill it, see, e.g., \cite{Franca_1988a,Hughes_1987b,Hughes_1986e}
which is, however, beyond the scope of this work.

\subsubsection{Stationary Navier-Stokes flow}

The discrete weak form of the stationary Navier-Stokes problem reads:
Given density $\varrho\in\mathbb{R}^{+}$, viscosity $\mu\in\mathbb{R}^{+}$,
body force $\varrho\cdot\vek g^{h}\left(\vek x\right)$ in $\Gamma^{h}$,
and traction $\hat{\vek t}^{h}\left(\vek x\right)$ on $\partial\Gamma_{\mathrm{N}}^{h}$,
find the velocity field $\vek u^{h}\left(\vek x\right)\in\mathcal{S}_{\vek u}^{h}$,
pressure field $p^{h}\left(\vek x\right)\in\mathcal{S}_{p}^{h}$,
and Lagrange multiplier field $\lambda^{h}\left(\vek x\right)\in\mathcal{S}_{\lambda}^{h}$
such that for all test functions $\left(\vek w_{\vek u}^{h},w_{p}^{h},w_{\lambda}^{h}\right)\in\mathcal{V}_{\vek u}^{h}\times\mathcal{V}_{p}^{h}\times\mathcal{V}_{\lambda}^{h}$,
there holds in $\Gamma^{h}$
\begin{eqnarray*}
 &  & \varrho\cdot\int_{\Gamma}\vek w_{\vek u}^{h}\cdot\left(\left(\vek u^{h}\cdot\nabla_{\Gamma}^{\mathrm{cov}}\right)\vek u^{h}-\vek g^{h}\right)\mathrm{d}A+\int_{\Gamma}\nabla_{\Gamma}^{\mathrm{dir}}\vek w_{\vek u}^{h}:\vek\sigma\left(\vek u^{h},p^{h}\right)\mathrm{d}A+\int_{\Gamma}\lambda^{h}\!\cdot\!\left(\vek w_{\vek u}^{h}\cdot\vek n_{\Gamma}^{h}\right)\mathrm{d}A\\
 &  & -\int_{\partial\Gamma_{\mathrm{N}}}\!\!\!\vek w_{\vek u}^{h}\cdot\hat{\vek t}^{h}\,\mathrm{d}s+\int_{\Gamma}w_{p}^{h}\cdot\mathrm{div}_{\Gamma}\,\vek u^{h}\;\mathrm{d}A+\int_{\Gamma}w_{\lambda}^{h}\cdot\left(\vek u^{h}\cdot\vek n_{\Gamma}^{h}\right)\mathrm{d}A\\
 &  & +\sum_{\mathrm{e}=1}^{n_{\mathrm{el}}}\int_{\Gamma_{\mathrm{e}}}\tau_{\mathrm{SUPG}}\left(\left(\vek u^{h}\cdot\nabla_{\Gamma}^{\mathrm{cov}}\right)\vek w_{\vek u}^{h}\right)\cdot\left[\varrho\cdot\left(\left(\vek u^{h}\cdot\nabla_{\Gamma}^{\mathrm{cov}}\right)\vek u^{h}-\vek g^{h}\right)-\mathrm{div}_{\Gamma}\,\vek\sigma\left(\vek u^{h},p^{h}\right)\right]=0.
\end{eqnarray*}
The equations related to the different field equations were added
up for brevity. The last row adds a stabilization term which is needed
to obtain stable solutions for flows at high Reynolds numbers \cite{Donea_2003a,Gresho_2000a}.
In particular, the streamline upwind Petrov-Galerkin (SUPG) method
is used for the stabilization. Different definitions of the stabilization
parameter $\tau_{\mathrm{SUPG}}$ are found \cite{Shakib_1991a,Tezduyar_2000a,Tezduyar_2003a}
and 
\[
\tau_{\mathrm{SUPG}}=\left[\left(\frac{2}{\Delta t}\right)^{2}+\left(\frac{2\left\Vert \vek u_{\mathrm{e}}\right\Vert }{h_{\mathrm{e}}}\right)^{2}+\left(\frac{4\mu}{h_{\mathrm{e}}{}^{2}}\right)^{2}\right]^{-1/2}
\]
is used herein with element-averaged velocity $\vek u_{\mathrm{e}}$,
element length $h_{\mathrm{e}}$ and $\Delta t\rightarrow\infty$
for the stationary case. When stabilization is not necessary because
no oscillations occur, $\tau_{\mathrm{SUPG}}=0.$ Note that in the
stabilization term, second-order derivatives appear (only in the element
interiors). The definition of \emph{tangential} second-order derivatives
is given, e.g., in \cite{Delfour_2011a}.

Element assembly results in a \emph{non}-linear system of equations
of the form
\begin{equation}
\left[\begin{array}{ccc}
\mat K^{\star}+\mat C\left(\underline{\vek u}\right) & \mat G^{\star} & \mat L\\
\mat G^{\mathrm{T}} & \mat0 & \mat0\\
\mat L^{\mathrm{T}} & \mat0 & \mat0
\end{array}\right]\cdot\left[\begin{array}{c}
\underline{\vek u}\\
\vek p\\
\vek\lambda
\end{array}\right]=\left[\begin{array}{c}
\vek f\\
\vek0\\
\vek0
\end{array}\right],\label{eq:NavStokesSystemOfEquations}
\end{equation}
which is no longer symmetric (partly) due to the advection matrix
$\mat C\left(\underline{\vek u}\right)$. The distinguishing feature
of $\mat K^{\star}$ and $\mat G^{\star}$ (compared to $\mat K$
and $\mat G$ of the Stokes problem) are the added SUPG-stabilization
terms. The issues related to mixed FEMs and the Babu\v ska-Brezzi
condition remain relevant.

\subsubsection{Instationary Navier-Stokes flow}

The discrete weak form of the instationary Navier-Stokes problem is:
Given density $\varrho\in\mathbb{R}^{+}$, viscosity $\mu\in\mathbb{R}^{+}$,
body force $\varrho\cdot\vek g^{h}\left(\vek x,t\right)$ in $\Gamma^{h}\times\tau$,
traction $\hat{\vek t}^{h}\left(\vek x,t\right)$ on $\partial\Gamma_{\mathrm{N}}^{h}\times\tau$,
and initial condition $u_{0}^{h}\left(\vek x\right)$ on $\Gamma^{h}$
at $t=0$ according to (\ref{eq:InitialCond}), find the velocity
field $\vek u^{h}\left(\vek x,t\right)\in L_{2}\left(\tau;\mathcal{S}_{\vek u}^{h}\right)$,
pressure field $p^{h}\left(\vek x,t\right)\in L_{2}\left(\tau;\mathcal{S}_{p}^{h}\right)$,
and Lagrange multiplier field $\lambda^{h}\left(\vek x,t\right)\in L_{2}\left(\tau;\mathcal{S}_{\lambda}^{h}\right)$
such that for all test functions $\left(\vek w_{\vek u}^{h},w_{p}^{h},w_{\lambda}^{h}\right)\in\mathcal{V}_{\vek u}^{h}\times\mathcal{V}_{p}^{h}\times\mathcal{V}_{\lambda}^{h}$,
there holds in $\Gamma^{h}\times\tau$
\begin{eqnarray*}
 &  & \varrho\cdot\int_{\Gamma}\vek w_{\vek u}^{h}\cdot\left(\partial_{t}\vek u^{h}+\left(\vek u^{h}\cdot\nabla_{\Gamma}^{\mathrm{cov}}\right)\vek u^{h}-\vek g^{h}\right)\mathrm{d}A+\int_{\Gamma}\nabla_{\Gamma}^{\mathrm{dir}}\vek w_{\vek u}^{h}:\vek\sigma\left(\vek u^{h},p^{h}\right)\mathrm{d}A+\int_{\Gamma}\lambda^{h}\!\cdot\!\left(\vek w_{\vek u}^{h}\cdot\vek n_{\Gamma}^{h}\right)\mathrm{d}A\\
 &  & -\int_{\partial\Gamma_{\mathrm{N}}}\!\!\!\vek w_{\vek u}^{h}\cdot\hat{\vek t}^{h}\,\mathrm{d}s+\int_{\Gamma}w_{p}^{h}\cdot\mathrm{div}_{\Gamma}\,\vek u^{h}\;\mathrm{d}A+\int_{\Gamma}w_{\lambda}^{h}\cdot\left(\vek u^{h}\cdot\vek n_{\Gamma}^{h}\right)\mathrm{d}A\\
 &  & +\sum_{\mathrm{e}=1}^{n_{\mathrm{el}}}\int_{\Gamma_{\mathrm{e}}}\tau_{\mathrm{SUPG}}\left(\left(\vek u^{h}\cdot\nabla_{\Gamma}^{\mathrm{cov}}\right)\vek w_{\vek u}^{h}\right)\cdot\left[\varrho\cdot\left(\partial_{t}\vek u^{h}+\left(\vek u^{h}\cdot\nabla_{\Gamma}^{\mathrm{cov}}\right)\vek u^{h}-\vek g^{h}\right)-\mathrm{div}_{\Gamma}\,\vek\sigma\left(\vek u^{h},p^{h}\right)\right]=0.
\end{eqnarray*}
This yields a system of non-linear semi-discrete equations for $t\in\tau$
\begin{eqnarray*}
\mat M\cdot\dot{\underline{\vek u}}\left(t\right)+\left(\mat K^{\star}+\mat C\left(\underline{\vek u}\right)\right)\cdot\underline{\vek u}\left(t\right)+\mat G^{\star}\cdot\vek p\left(t\right)+\mat L\cdot\vek\lambda\left(t\right) & = & \vek f\left(t\right),\\
\mat G^{\mathrm{T}}\cdot\underline{\vek u}\left(t\right) & = & \vek0,\\
\mat L^{\mathrm{T}}\cdot\underline{\vek u}\left(t\right) & = & \vek0,
\end{eqnarray*}
with initial condition $\underline{\vek u}\left(0\right)$. This system
may be advanced in time by using finite difference schemes and the
Crank-Nicolson method is employed herein.

\section{Numerical results\label{X_NumericalResults}}

The following error measures are computed in the convergence studies.
When analytic (exact) velocity and pressure fields, $\vek u_{\mathrm{ex}}$
and $p_{\mathrm{ex}}$, are known, the velocity error is determined
by 
\begin{equation}
\varepsilon_{\vek u}=\sum_{i=1}^{3}\sqrt{\int_{\Gamma}\left(u_{i}^{h}\left(\vek x\right)-u_{i,\mathrm{ex}}\left(\vek x\right)\right)^{2}\mathrm{d}A}\label{eq:ErrorVelocity}
\end{equation}
and the pressure error calculated as
\begin{equation}
\varepsilon_{p}=\sqrt{\int_{\Gamma}\left(p^{h}\left(\vek x\right)-p_{\mathrm{ex}}\left(\vek x\right)\right)^{2}\mathrm{d}A}.\label{eq:ErrorPressure}
\end{equation}
When analytic solutions are not available, it is useful to evaluate
the error of the FE approximations in the strong form of the momentum
or continuity equations, integrated over the domain. For the example
of stationary Stokes flow, the corresponding residual errors are defined
as 
\begin{equation}
\varepsilon_{\mathrm{mom}}=\sqrt{\int_{\Gamma}\left(\mat P\cdot\mathrm{div}_{\Gamma}\,\vek\sigma\left(\vek u^{h},p^{h}\right)+\vek f^{h}\right)^{2}\mathrm{d}A}\label{eq:ErrorMomentum}
\end{equation}
and 
\begin{equation}
\varepsilon_{\mathrm{cont}}=\sqrt{\int_{\Gamma}\left(\mathrm{div}_{\Gamma}\,\vek u^{h}\right)^{2}\mathrm{d}A}.\label{eq:ErrorContinuity}
\end{equation}
This can be easily extended to the case of Navier-Stokes flows where
the advection term is added to the integrand in (\ref{eq:ErrorMomentum}).
Also the error in the tangential velocity constraint from Eq.~(\ref{eq:VelocityConstraint})
may be computed in a similar manner. The evaluation of the error $\varepsilon_{\mathrm{mom}}$
involves second-order derivatives and convergence can only be expected
for higher-order elements and sufficiently smooth solutions.

\subsection{Stokes flow on an axisymmetric surface}

\begin{figure}
\centering

\subfigure[domain]{\includegraphics[height=5cm]{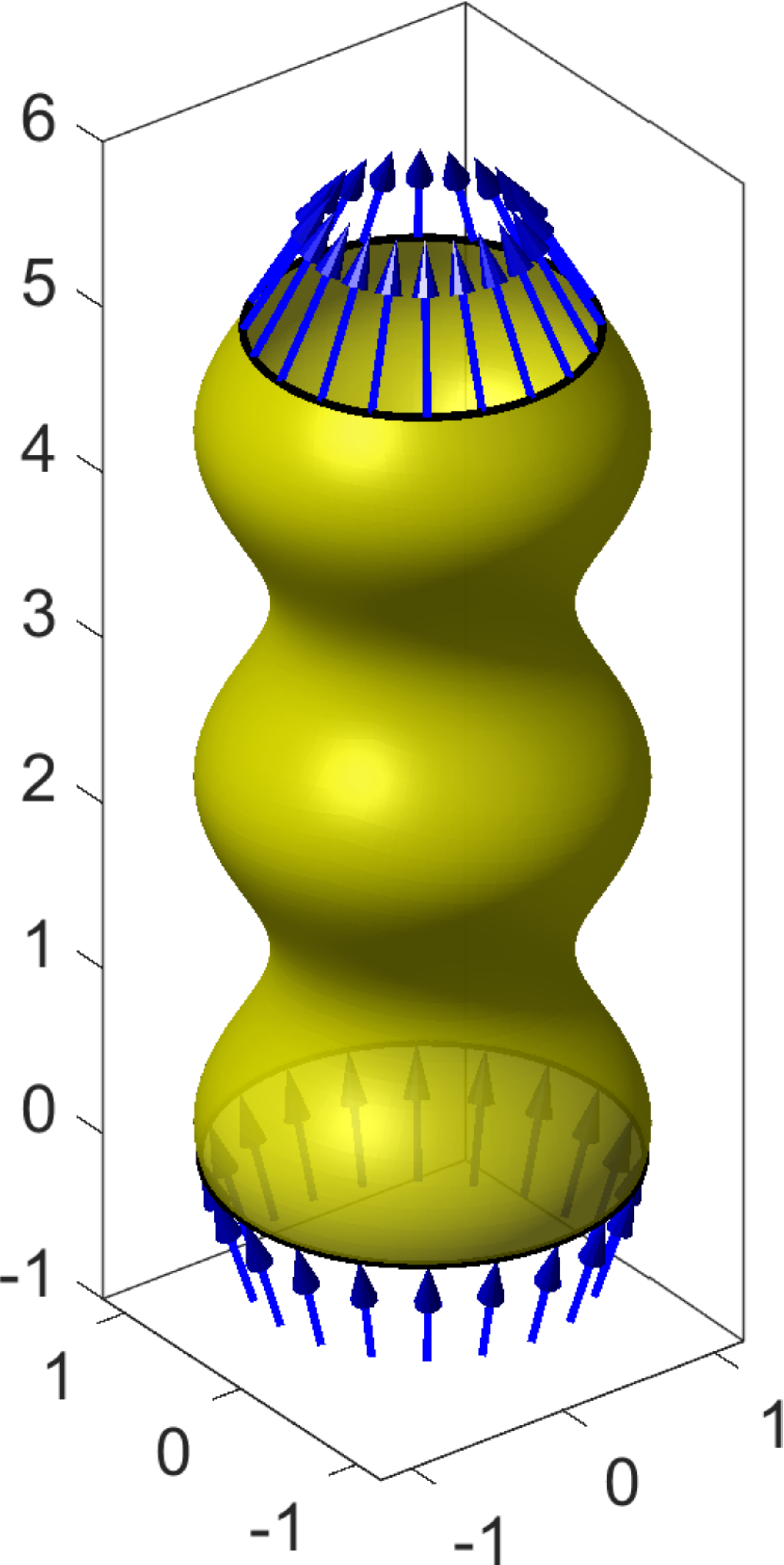}}\qquad\subfigure[quad-mesh, $n_z=6$]{\includegraphics[height=4cm]{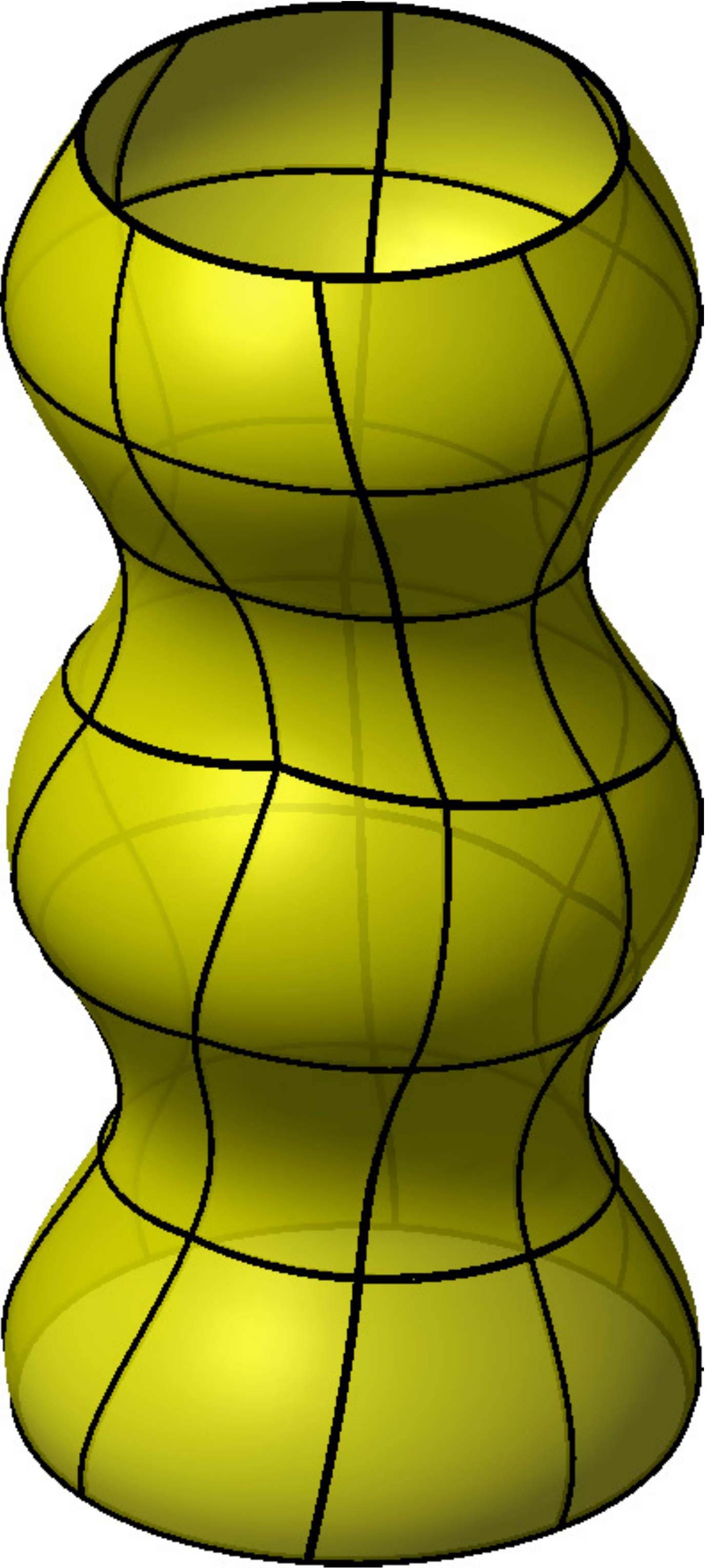}}\qquad\subfigure[quad-mesh, $n_z=10$]{\includegraphics[height=4cm]{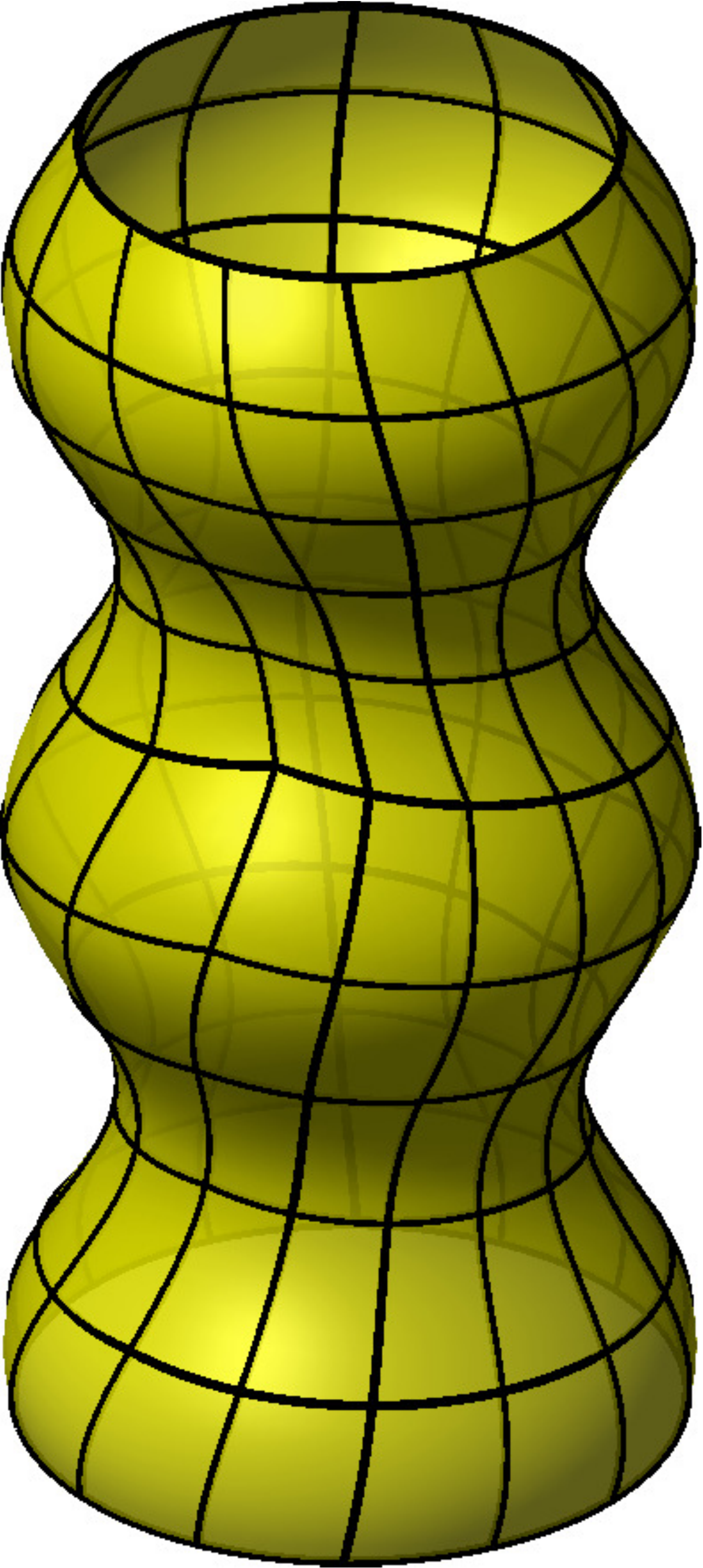}}\qquad\subfigure[quad-mesh, $n_z=20$]{\includegraphics[height=4cm]{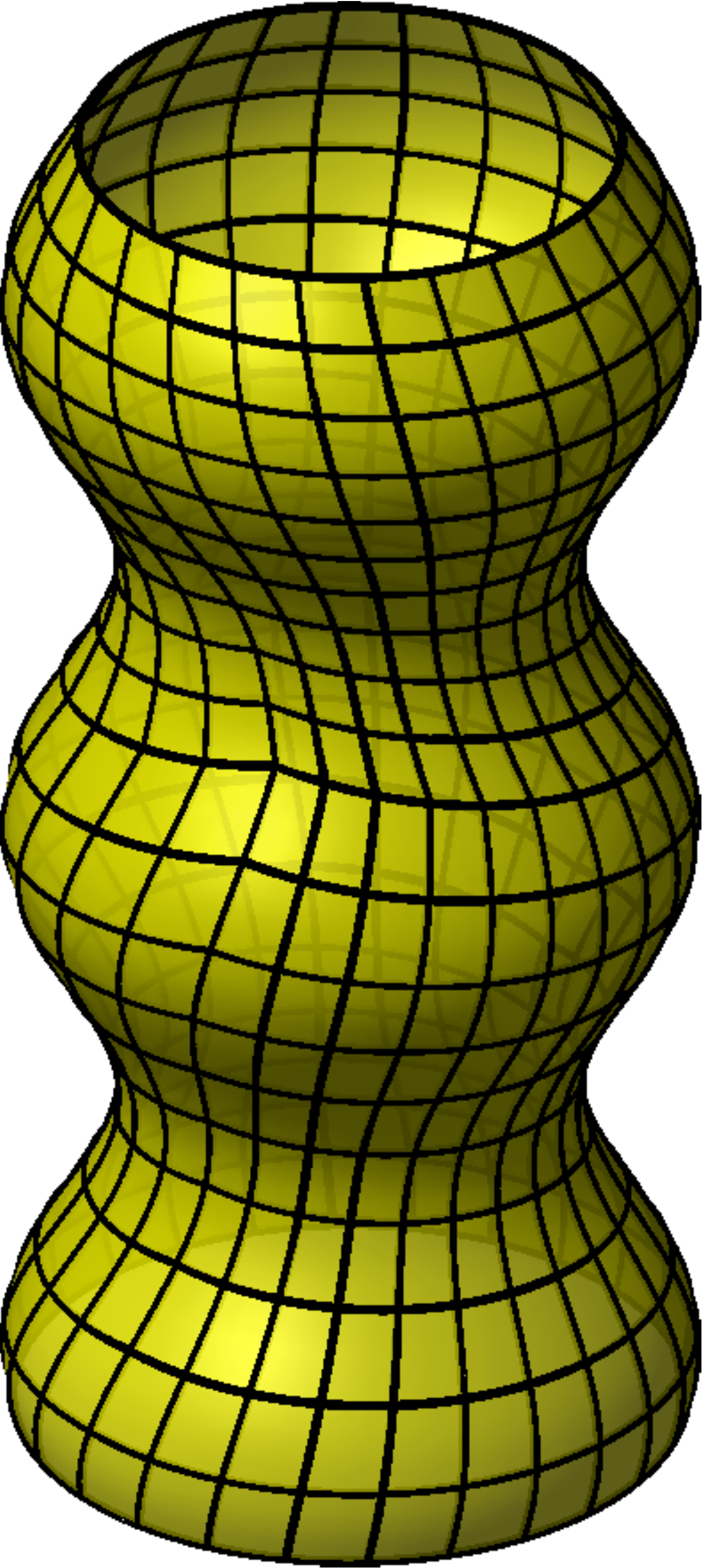}}\qquad\subfigure[tri-mesh, $n_z=6$]{\includegraphics[height=4cm]{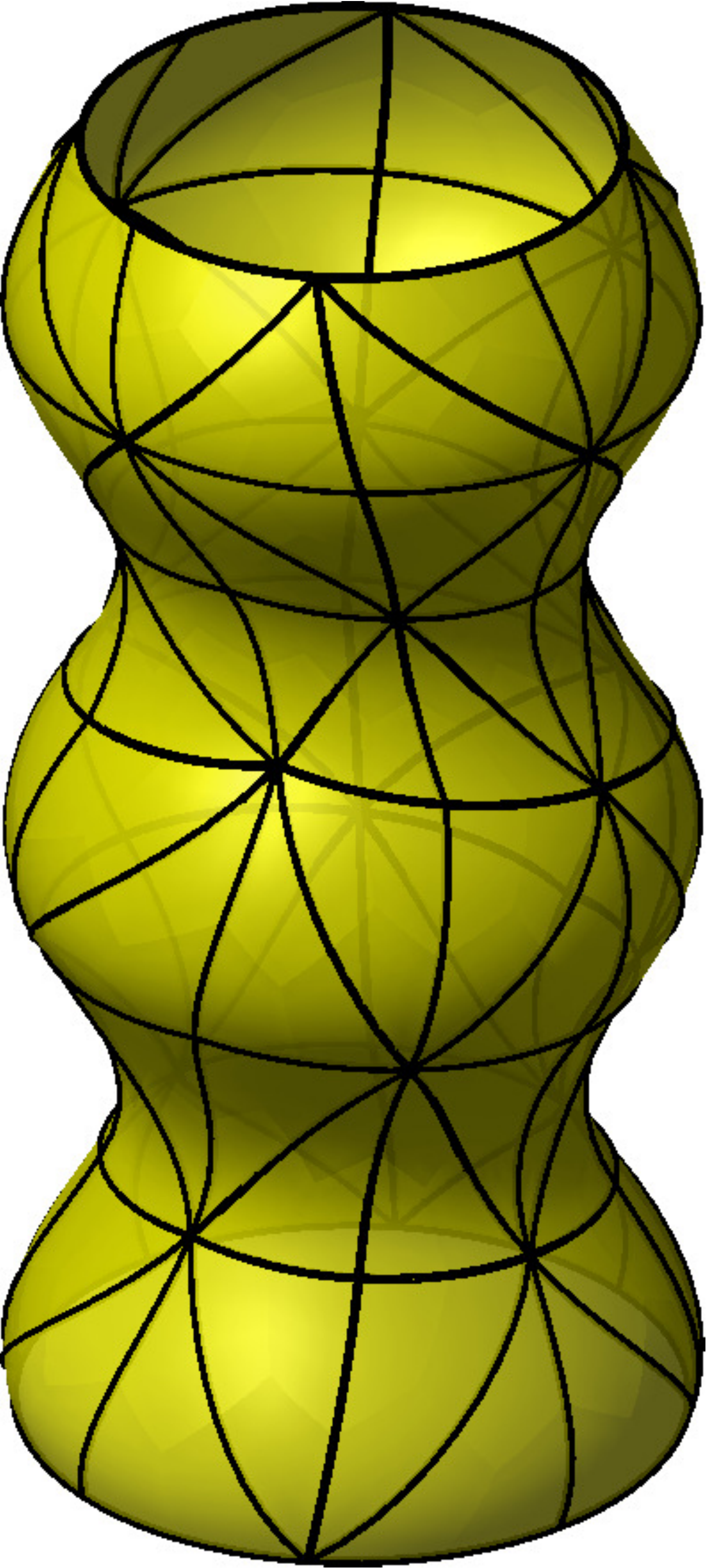}}

\caption{\label{fig:RotSymmDomainAndMeshes}Manifold for the axisymmetric test
case and meshes with different resolutions. In (a), the blue arrows
indicate the direction of the flow at the inflow (bottom) and outflow
(top) boundaries.}
\end{figure}

A test case is developed for which analytic solutions are available.
An axisymmetric surface with height $L=5$ and radius
\[
r\left(z\right)=1+\nicefrac{1}{5}\cdot\sin\left(1+3\cdot z\right),\qquad z\in\left[0,L\right],
\]
is generated as illustrated in Fig.~\ref{fig:RotSymmDomainAndMeshes}(a).
Let $r_{0}=r(0)$ and $r_{0}^{\prime}=\frac{\mathrm{d}r(0)}{\mathrm{d}z}$.
In parametrized form, one may also define $\Gamma$ based on the map
$\vek x\left(\vek a\right):\mathbb{R}^{2}\rightarrow\mathbb{R}^{3}$,
\[
\vek x\left(\vek a\right)=\left[\begin{array}{l}
\cos a\cdot r(b)\\
\sin a\cdot r(b)\\
b
\end{array}\right]\;\text{with}\;a\in\left[0,2\pi\right],\;b\in\left[0,5\right].
\]

The lower boundary at $z=0$ is the Dirichlet boundary $\partial\Gamma_{\mathrm{D}}$,
where the inflow in co-normal direction of the manifold is prescribed
as
\[
\hat{\vek u}\left(\vek x\right)=\frac{\vek u^{\star}}{\left\Vert \vek u^{\star}\right\Vert }\quad\text{and}\quad\vek u^{\star}=\left[\begin{array}{l}
r_{0}^{\prime}\cdot\cos\theta\\
r_{0}^{\prime}\cdot\sin\theta\\
1
\end{array}\right],
\]
with angle $\theta$ given by $\tan\theta=\nicefrac{y}{x}$. The upper
boundary at $z=L$ is the outflow boundary where zero-tractions are
applied as Neumann boundary conditions. The density and viscosity
are set to $\varrho=1$ and $\mu=0.01$, respectively.

\begin{figure}
\centering

\subfigure[$u\left(\vek x\right)$]{\includegraphics[height=4cm]{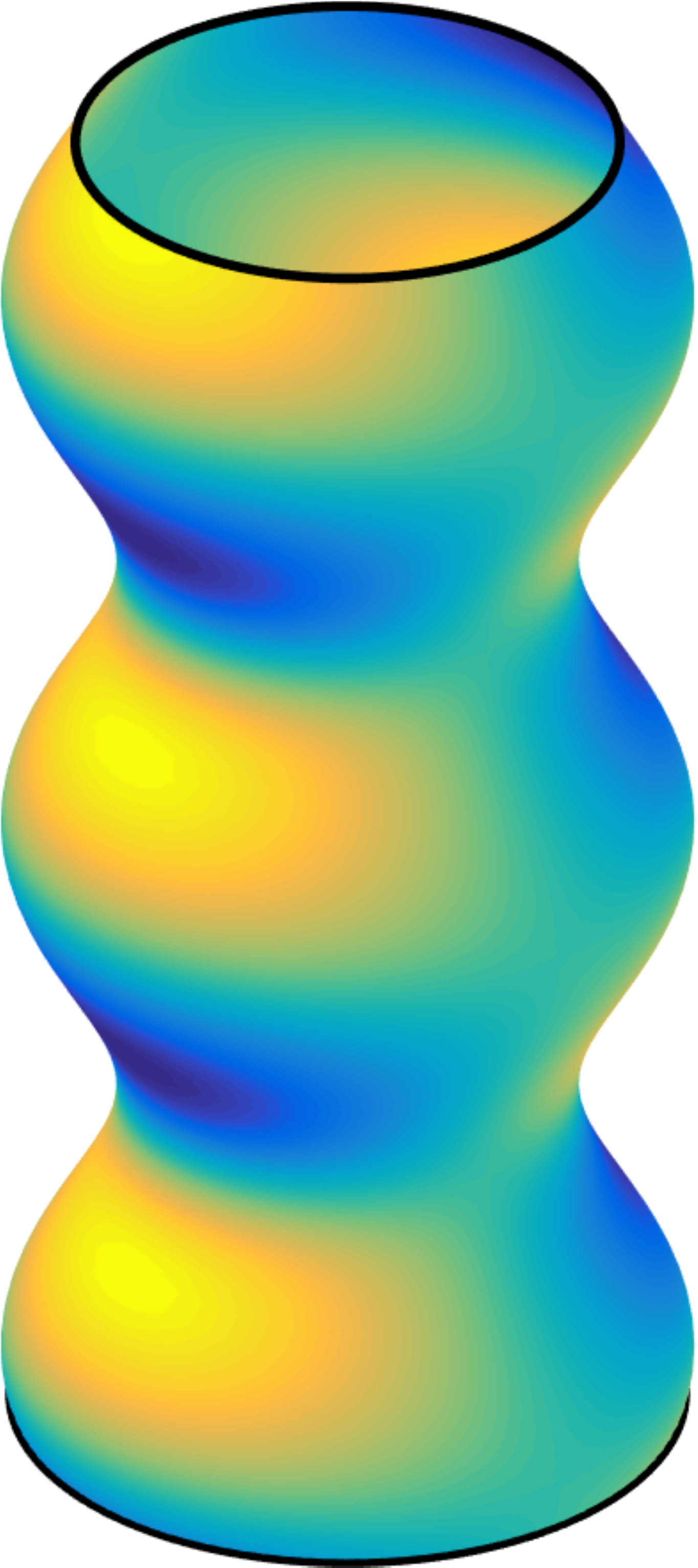}}\qquad\subfigure[$v\left(\vek x\right)$]{\includegraphics[height=4cm]{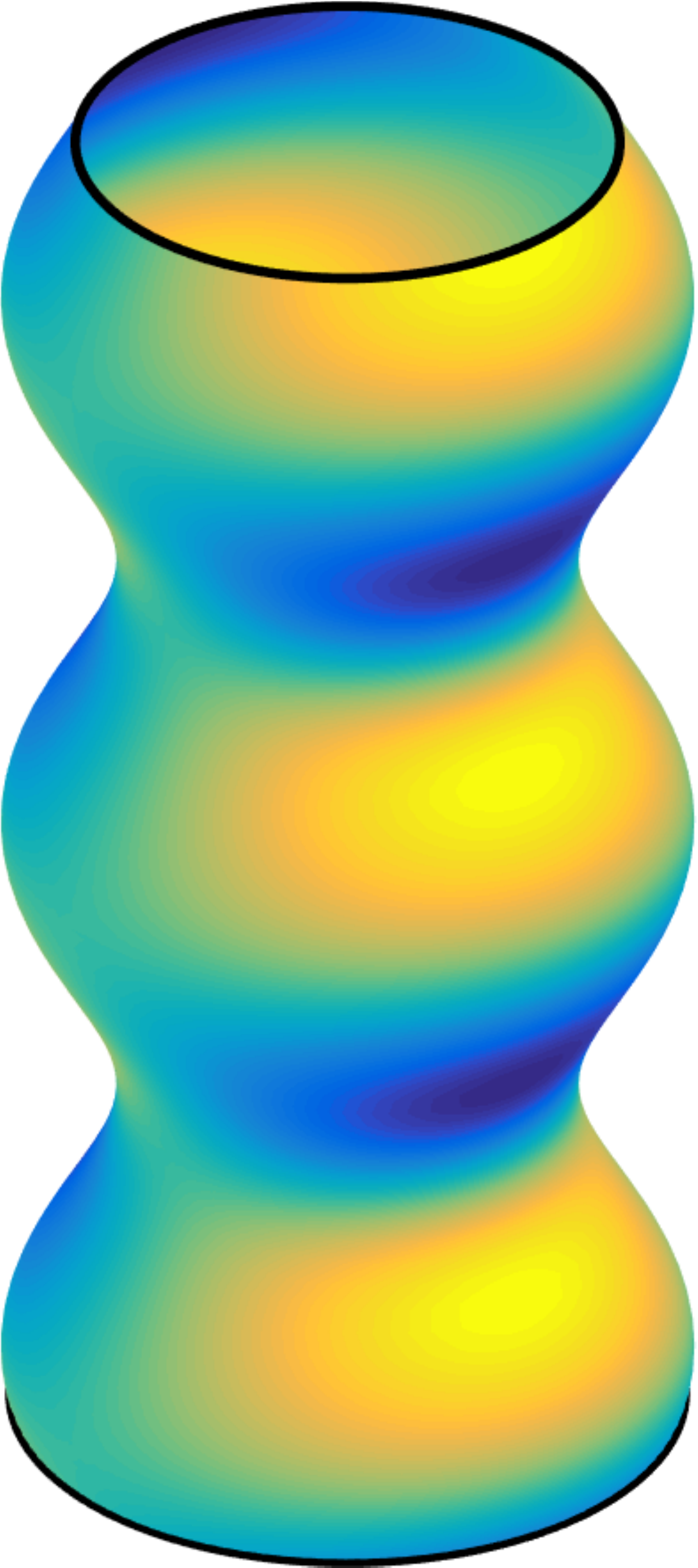}}\qquad\subfigure[$w\left(\vek x\right)$]{\includegraphics[height=4cm]{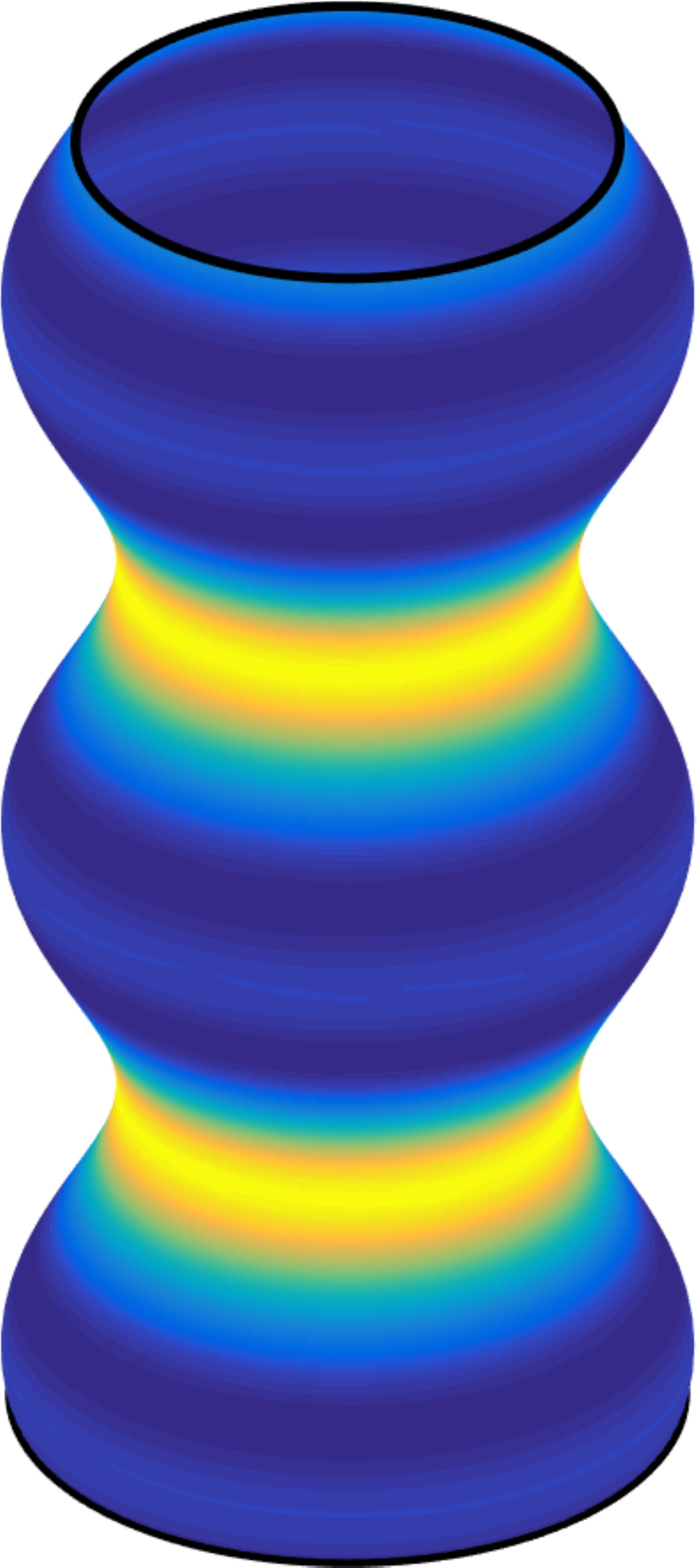}}\qquad\subfigure[$\left\Vert \vek u\left(\vek x\right)\right\Vert$]{\includegraphics[height=4cm]{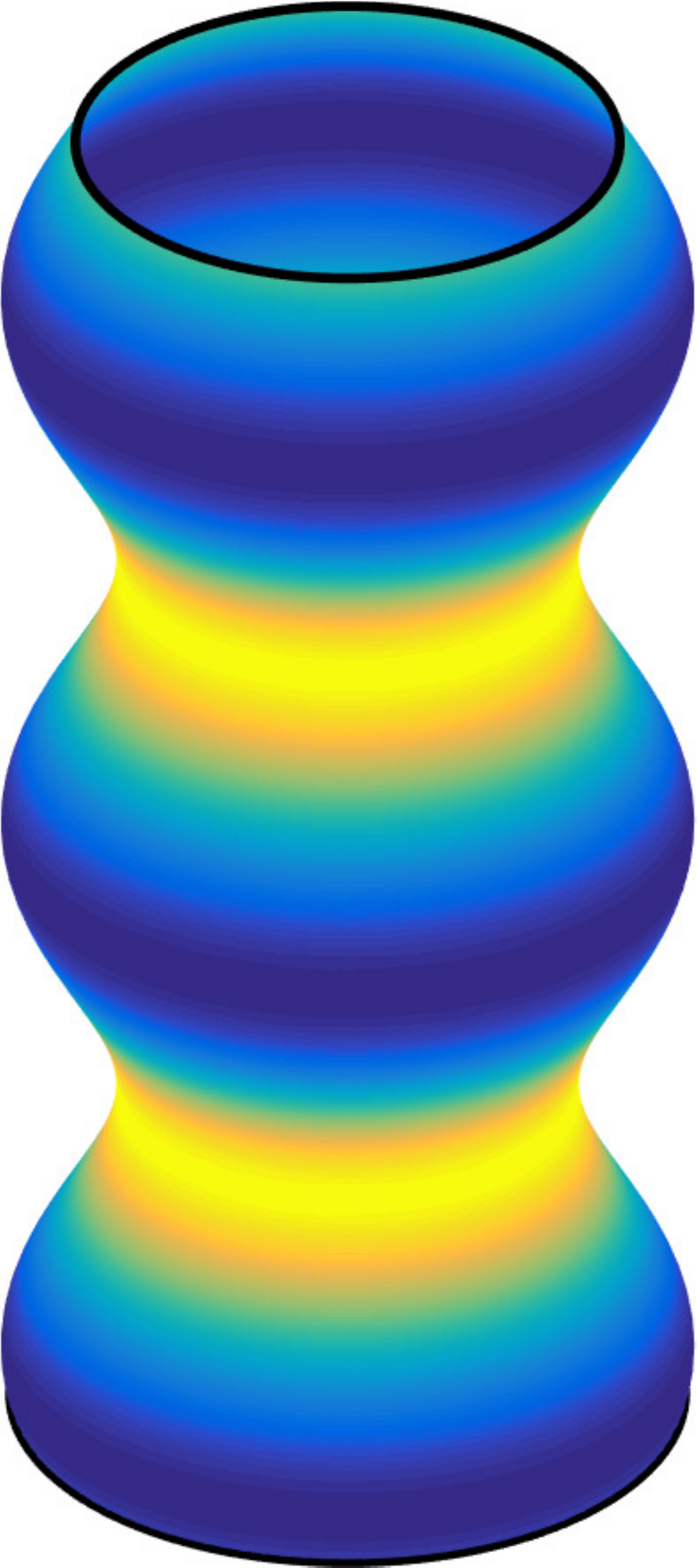}}\qquad\subfigure[$w\left(z\right)$, $\left\Vert \vek u\left(z\right)\right\Vert$]{\includegraphics[height=5cm]{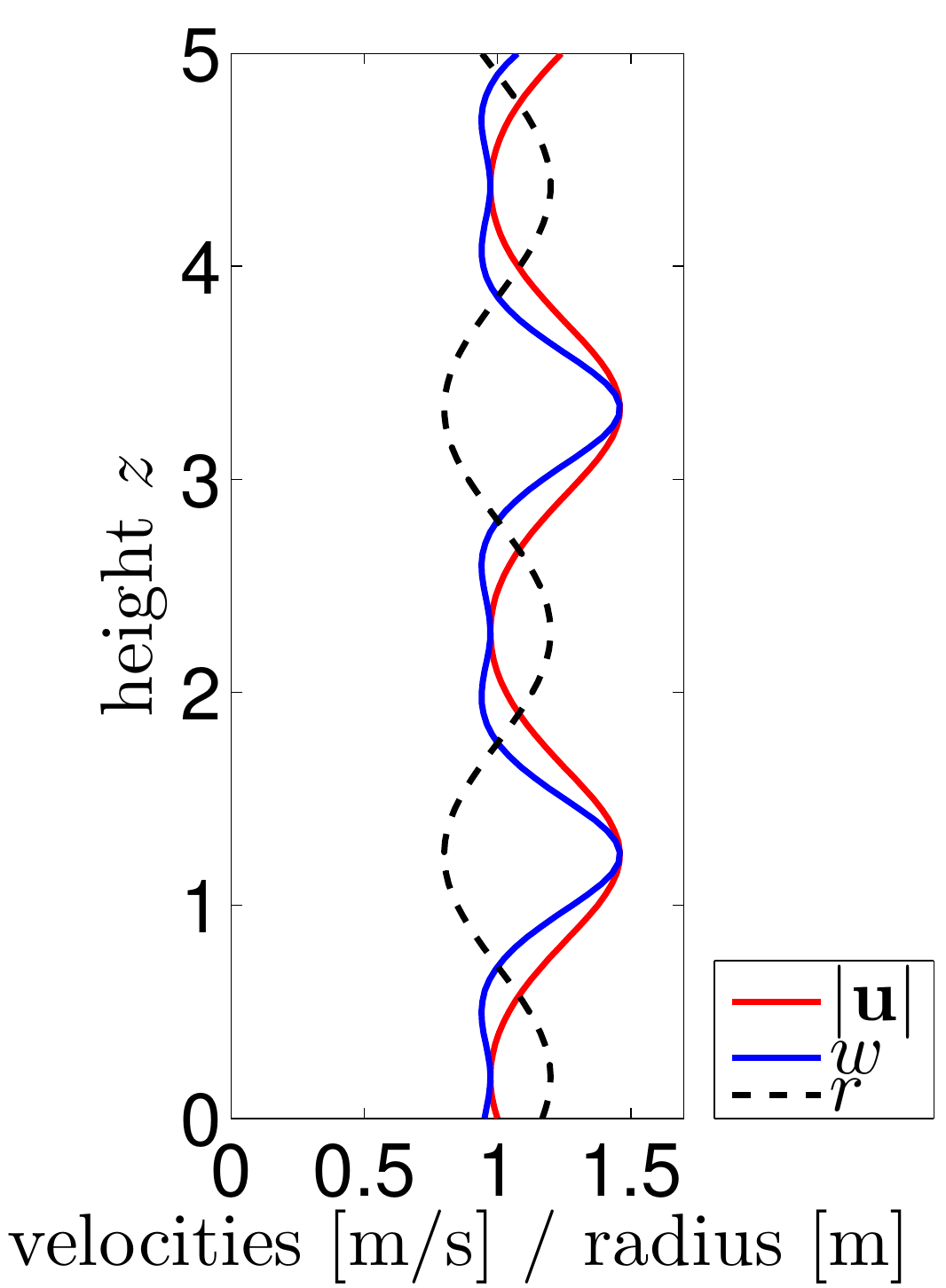}}

\caption{\label{fig:RotSymmSolution}Exact solution for the axisymmetric test
case.}
\end{figure}

The mass flow on the lower boundary is
\[
Q_{0}=\int_{\partial\Gamma_{\mathrm{D}}}\hat{\vek u}\left(\vek x\right)\cdot\vek n_{\partial\Gamma}\,\mathrm{d}s=2\pi\cdot r_{0}
\]
and due to mass conservation, the mass flow along the height follows
as $Q\left(z\right)=2\pi\cdot\nicefrac{r_{0}}{r(z)}$. As the flow
field is expected to be axisymmetric for this test case, and the tangential
velocity constraint applies, one may compute the velocity components
as
\[
\left[\begin{array}{l}
u_{\mathrm{ex}}(\vek x)\\
v_{\mathrm{ex}}(\vek x)\\
w_{\mathrm{ex}}(\vek x)
\end{array}\right]=\frac{r_{0}}{r\cdot\sqrt{1+\left(\frac{\mathrm{d}r}{\mathrm{d}z}\right)^{2}}}\cdot\left[\begin{array}{l}
\frac{\mathrm{d}r}{\mathrm{d}z}\cdot\nicefrac{x}{r}\\
\frac{\mathrm{d}r}{\mathrm{d}z}\cdot\nicefrac{y}{r}\\
1
\end{array}\right].
\]
See Fig.~\ref{fig:RotSymmSolution} for a graphical representation.
It is noted that the mass flow $Q\left(z\right)$, velocity magnitude
$\left\Vert \vek u\right\Vert $, and the vertical velocity component
$w$ are only functions of $z$, that is, they do not vary in $x$-
and $y$-directions.

Finite element approximations are carried out on various meshes composed
by triangular or quadrilateral Lagrange elements of different orders.
For the convergence studies, meshes with $n_{z}=\left\{ 4,6,10,14,20,30,40,60\right\} $
elements over the height are chosen; the number of elements in circumferential
direction is $n_{\theta}=\mathrm{round}\left(\nicefrac{2\pi r_{0}}{L}\cdot n_{z}\right)$.
The meshes are perturbed, as illustrated in Figs.~\ref{fig:RotSymmDomainAndMeshes}(b)
to (d), to avoid perfectly axisymmetric meshes which, otherwise, could
have improved the convergence rates for this special case.

The individual element orders used for the convergence studies are
indicated by a $4$-tuple $\left\{ k_{\mathrm{geom}},k_{\vek u},k_{p},k_{\lambda}\right\} $.
To be precise, this tuple summarizes the employed orders for the geometry,
$k_{\mathrm{geom}}$, the velocities, $k_{\vek u}$, the pressure,
$k_{p}$, and the Lagrange multiplier for enforcing the tangential
velocity constraint, $k_{\lambda}$. For each tuple, meshes with different
resolutions (given by $n_{z}$ and $n_{\theta}$) are considered and
errors calculated, each time resulting in one curve in the convergence
plots as indicated in the legends.

Systematic studies of different combinations of element orders showed
that equal-order approximations for the velocity and pressure, i.e.,
$k_{p}=k_{\vek u}$ do not converge satisfactory (or at all), which
is well-known from the standard context of the incompressible Navier-Stokes
equations in 2D and 3D due to the Babu\v ska-Brezzi condition. For
the studies outlined in this paper, we shall choose $k_{p}=k_{\vek u}-1$
which is a popular choice for FEM approximations of classical incompressible
flows and known as Taylor-Hood elements \cite{Taylor_1973a}.

\begin{figure}
\centering

\subfigure[tri, $\varepsilon_{\vek u}$]{\includegraphics[width=0.45\textwidth]{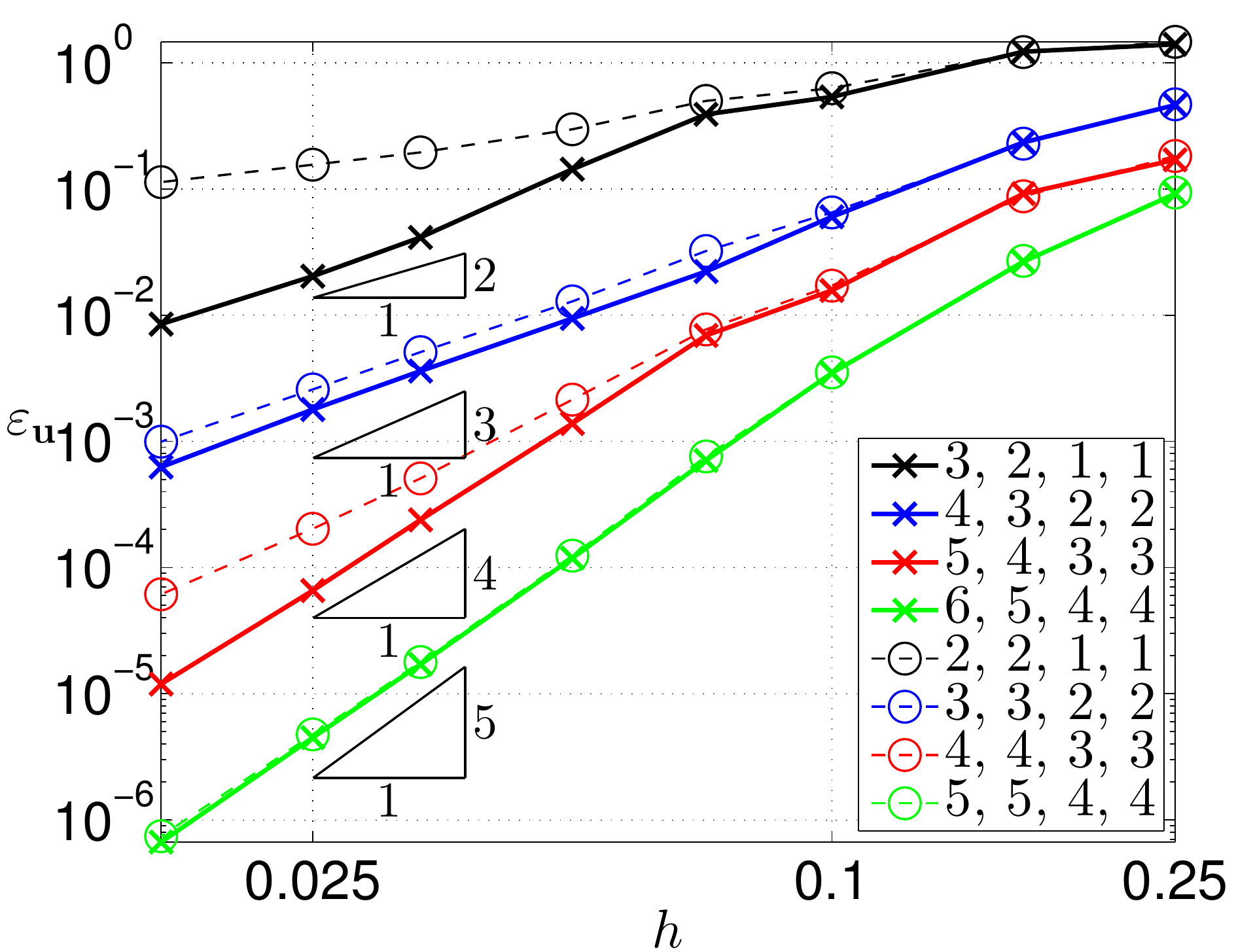}}\qquad\subfigure[tri, $\varepsilon_{p}$]{\includegraphics[width=0.45\textwidth]{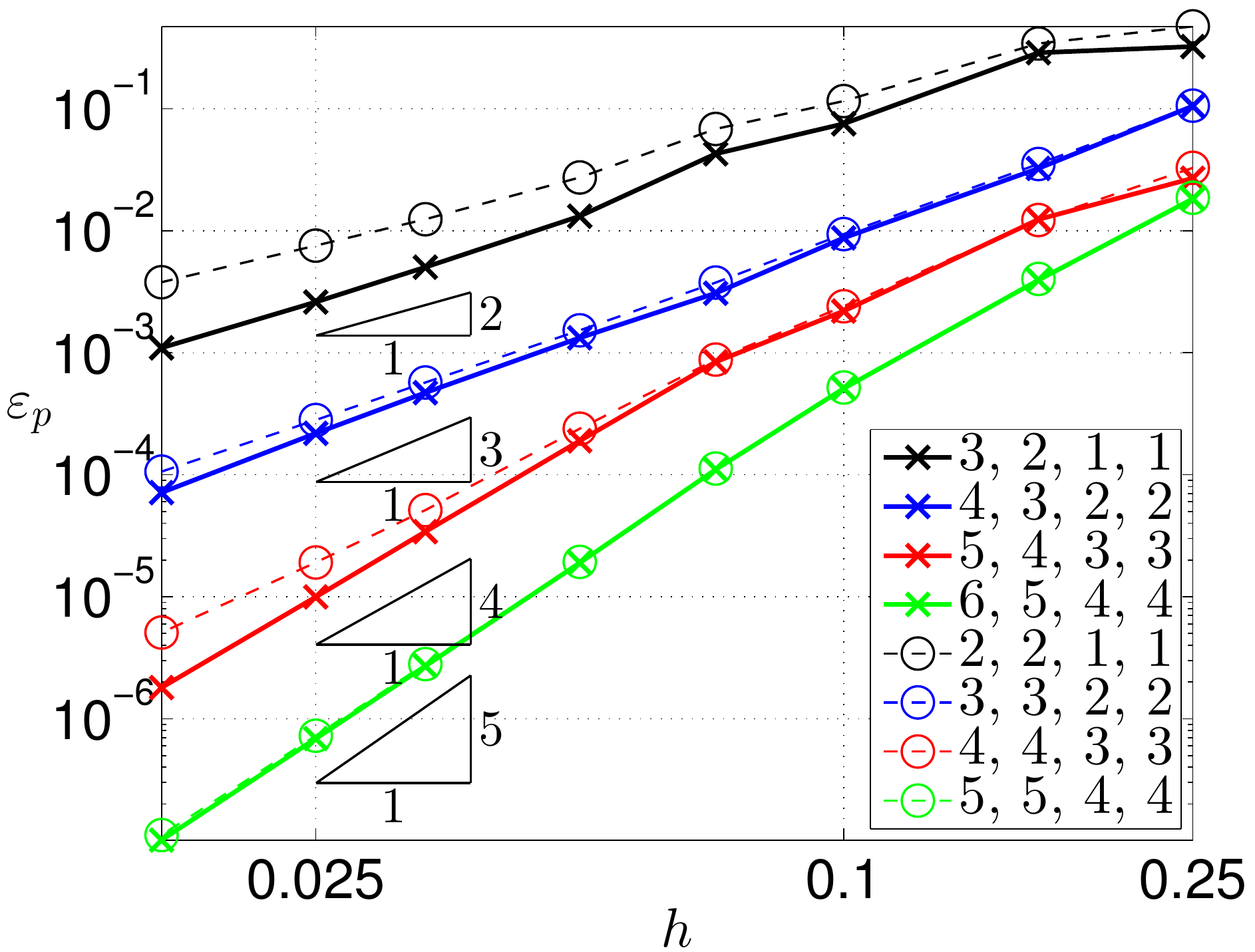}}

\subfigure[quad, $\varepsilon_{\vek u}$]{\includegraphics[width=0.45\textwidth]{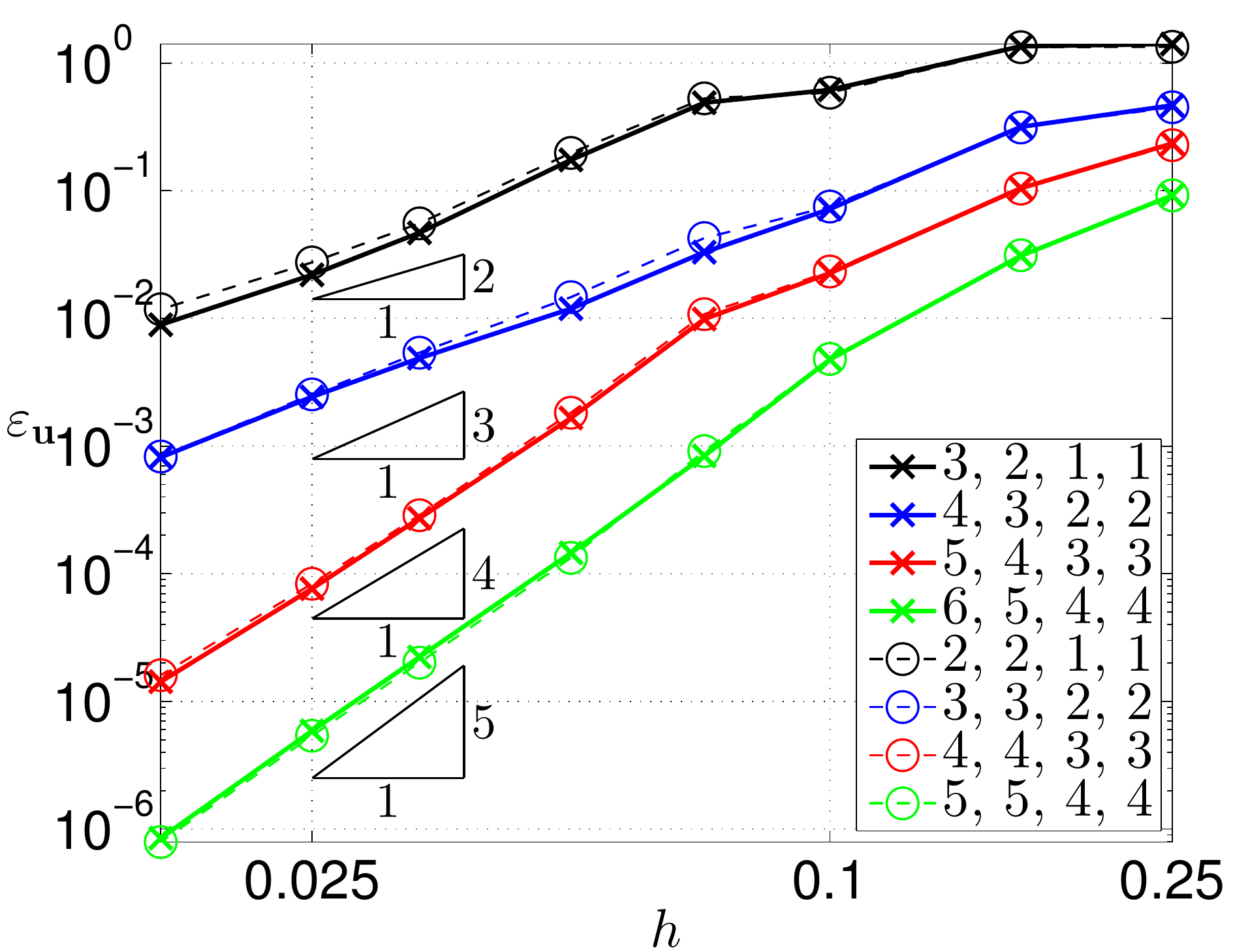}}\qquad\subfigure[quad, $\varepsilon_{p}$]{\includegraphics[width=0.45\textwidth]{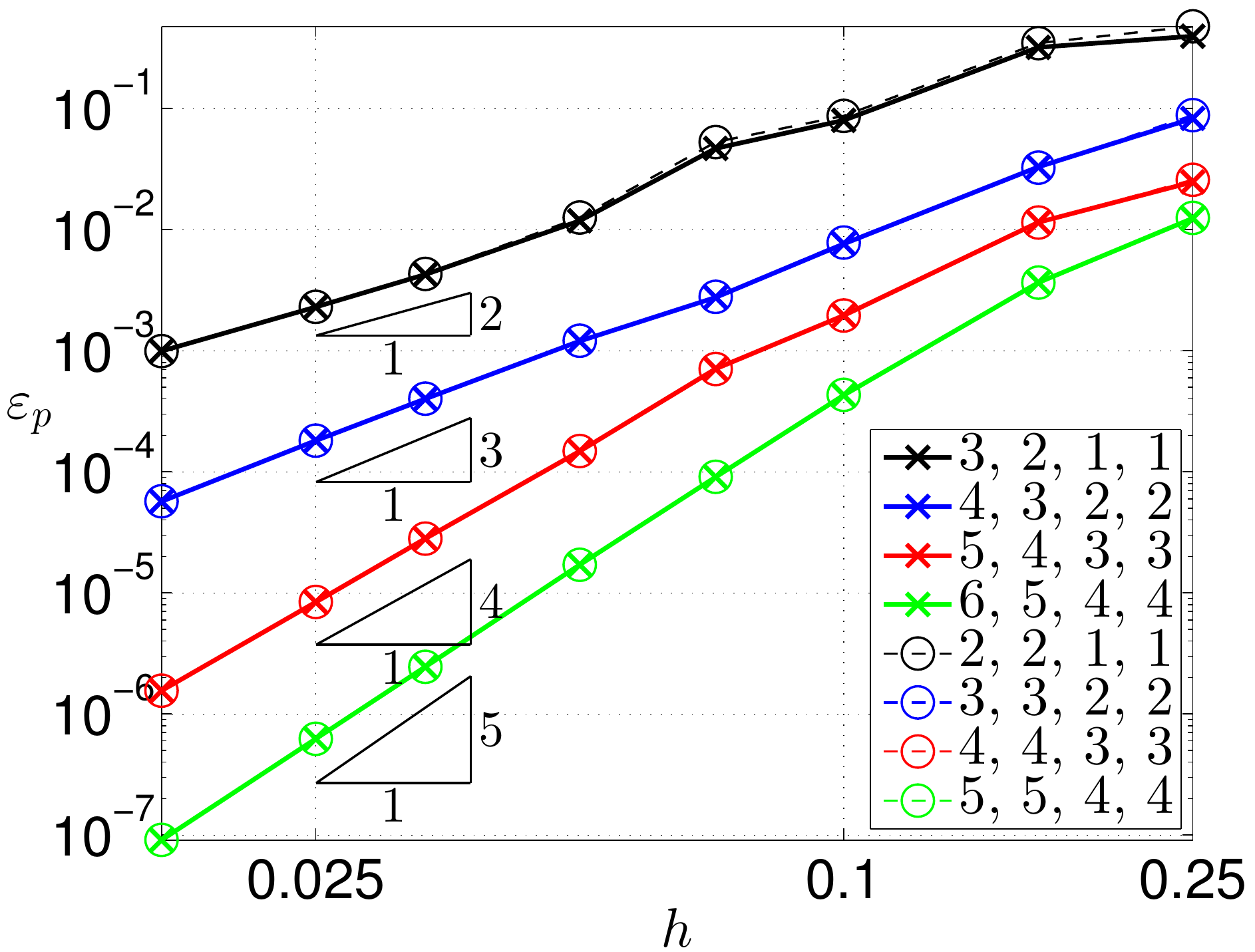}}

\caption{\label{fig:RotSymmResNewA}Convergence results in $\varepsilon_{\vek u}$
and $\varepsilon_{p}$ for the axisymmetric test case, (a) and (b)
for triangular elements, (c) and (d) for quadrilateral elements. The
legends decode the orders $\left\{ k_{\mathrm{geom}},k_{\vek u},k_{p},k_{\lambda}\right\} $
of the meshes.}
\end{figure}

\begin{figure}
\centering

\subfigure[tri, $\varepsilon_{\mathrm{mom}}$]{\includegraphics[width=0.45\textwidth]{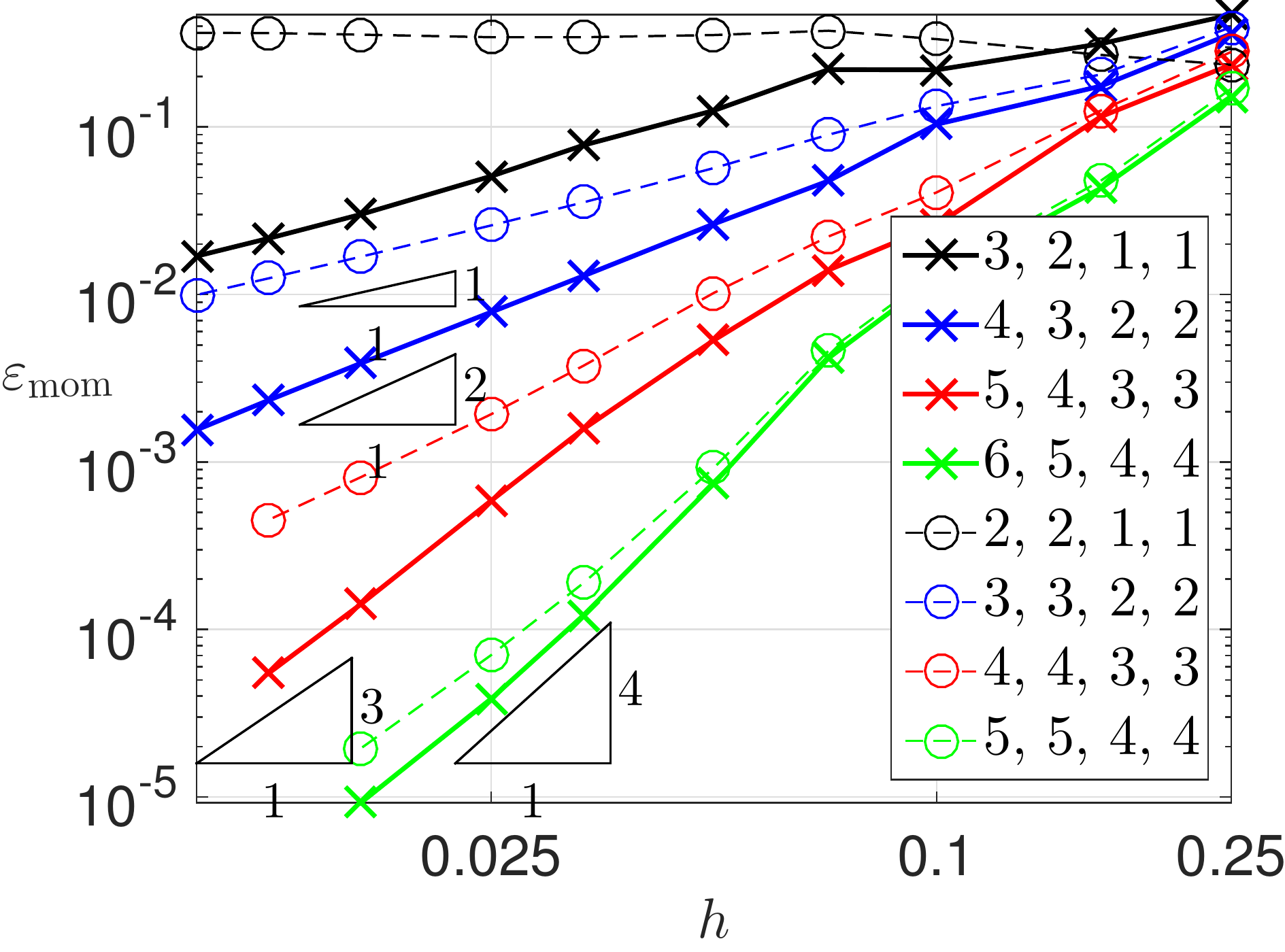}}\qquad\subfigure[tri, $\varepsilon_{\mathrm{cont}}$]{\includegraphics[width=0.45\textwidth]{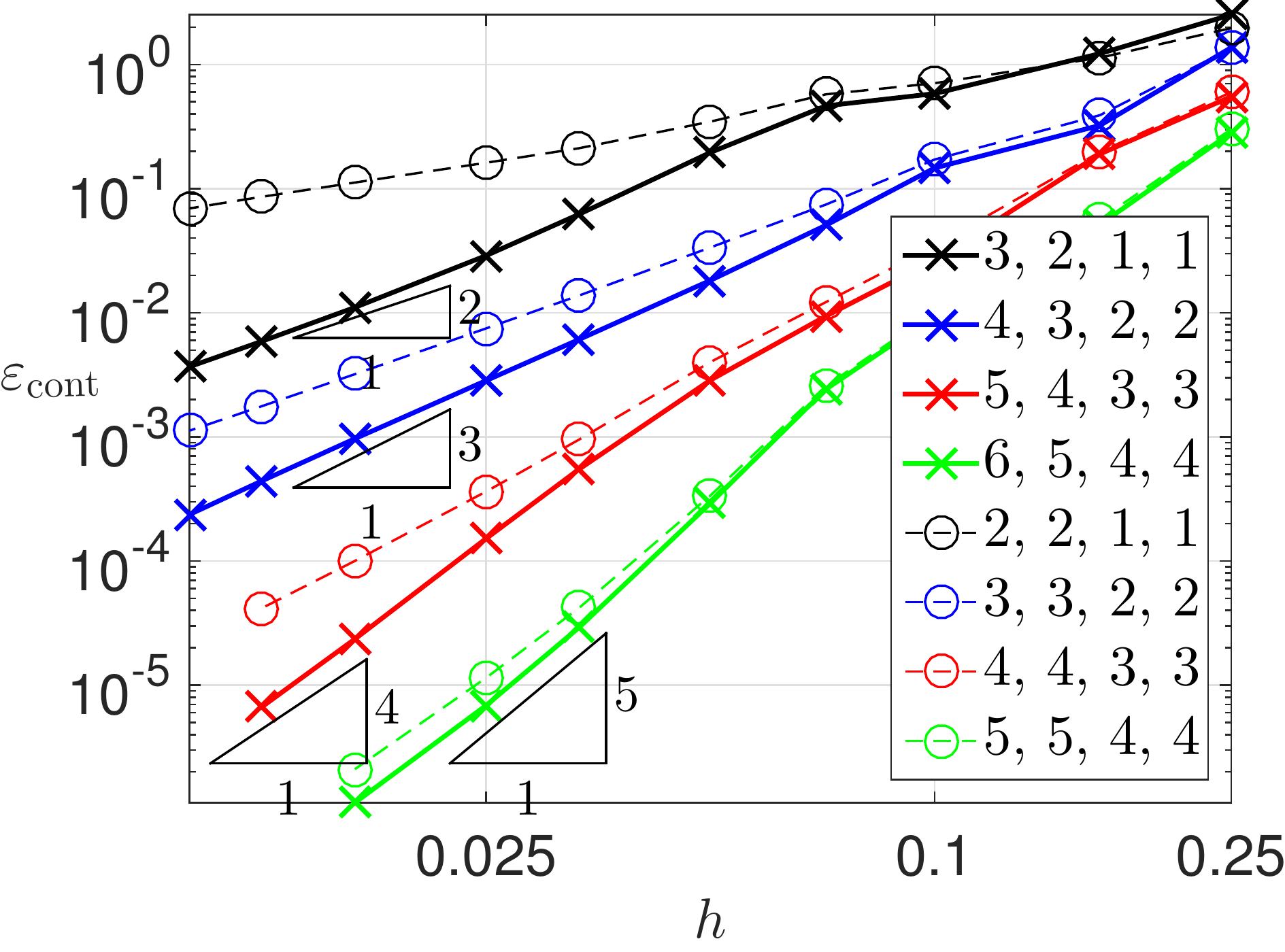}}

\subfigure[quad, $\varepsilon_{\mathrm{mom}}$]{\includegraphics[width=0.45\textwidth]{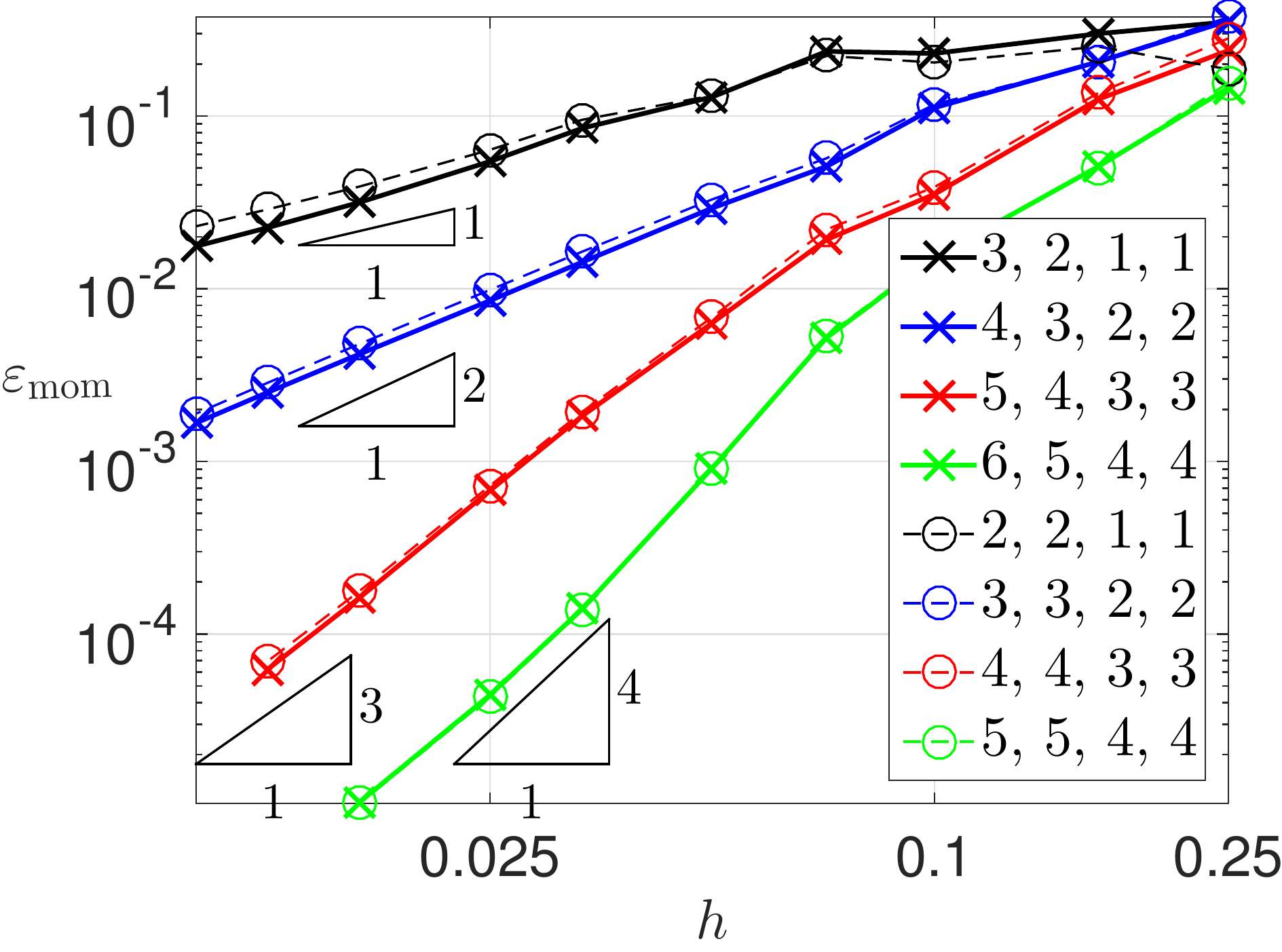}}\qquad\subfigure[quad, $\varepsilon_{\mathrm{cont}}$]{\includegraphics[width=0.45\textwidth]{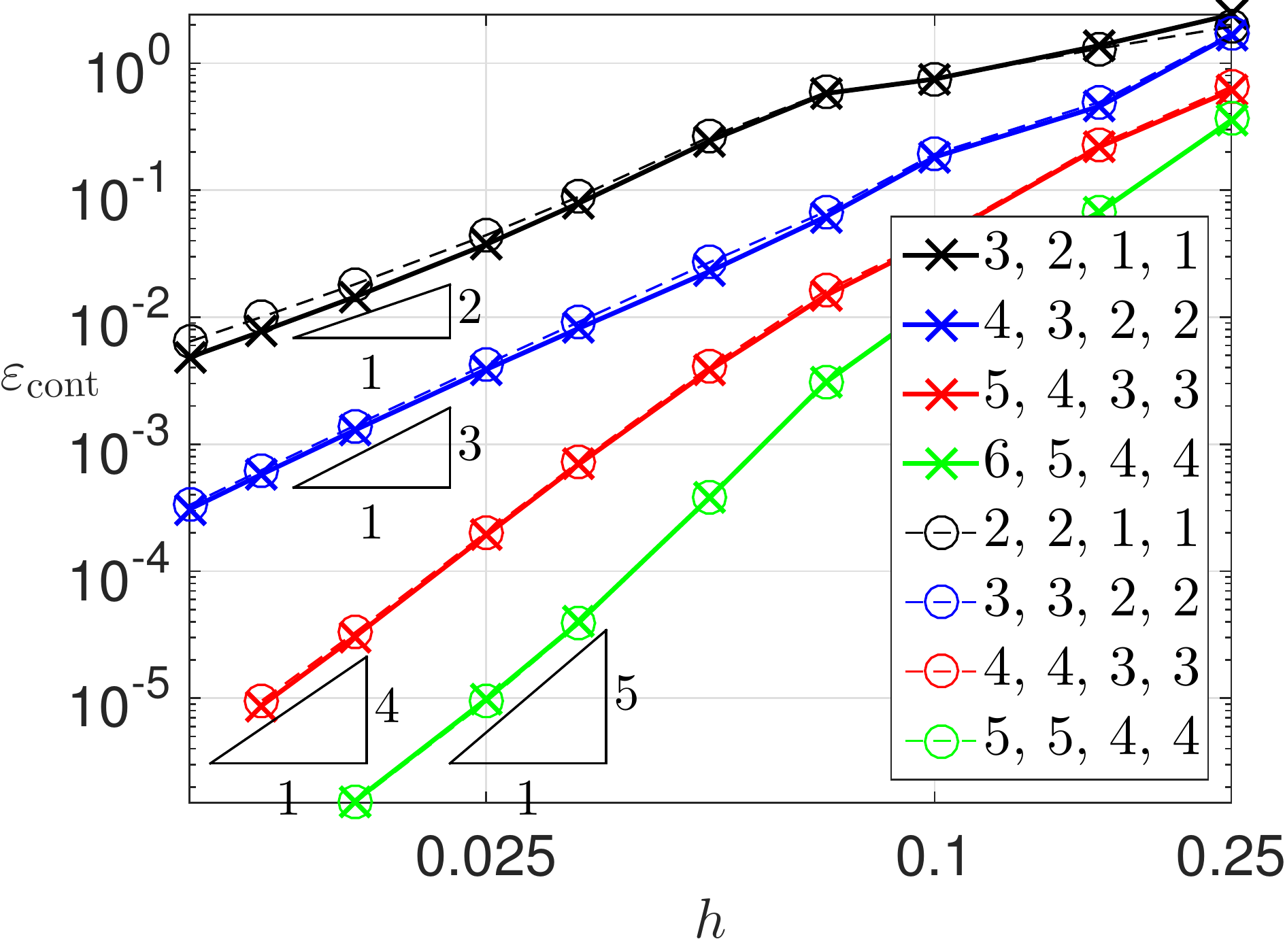}}

\caption{\label{fig:RotSymmResNewB}Convergence results in $\varepsilon_{\mathrm{mom}}$
and $\varepsilon_{\mathrm{cont}}$ for the axisymmetric test case,
(a) and (b) for triangular elements, (c) and (d) for quadrilateral
elements. The legends decode the orders $\left\{ k_{\mathrm{geom}},k_{\vek u},k_{p},k_{\lambda}\right\} $
of the meshes.}
\end{figure}

For the first study, we use $2\leq k_{\vek u}\leq5$, $k_{p}=k_{\vek u}-1$,
and $k_{\lambda}=k_{\vek u}-1$ which, lateron, becomes the recommended
standard setting. Convergence plots for $\varepsilon_{\vek u}$ and
$\varepsilon_{p}$ are given in Fig.~\ref{fig:RotSymmResNewA}. The
thick solid lines are for $k_{\mathrm{geom}}=k_{\vek u}+1$. It is
noteworthy that for quadrilateral elements, setting $k_{\mathrm{geom}}=k_{\vek u}$
leads to almost identical results as seen from the thin dashed lines
in Figs.~\ref{fig:RotSymmResNewA}(c) and (d). This does not necessarily
hold for triangular elements, see Figs.~\ref{fig:RotSymmResNewA}(a)
and (b), where the convergence may drop by one order when setting
$k_{\mathrm{geom}}=k_{\vek u}$ rather than $k_{\mathrm{geom}}=k_{\vek u}+1$.
This is later confirmed for the errors $\varepsilon_{\mathrm{mom}}$
and $\varepsilon_{\mathrm{cont}}$ in Fig.~\ref{fig:RotSymmResNewB}.
Therefore, we recommend to choose the geometry one order higher than
$k_{\vek u}$ which is done in the remainder of this work. Another
reason is that the normal vector $\vek n$ is present in the governing
equations and is computed based on the Jacobi matrix, i.e., first
derivatives of the element mappings of order $k_{\mathrm{geom}}$
are involved.

\begin{figure}
\centering

\subfigure[quad, $\varepsilon_{\mathrm{mom}}$]{\includegraphics[width=0.45\textwidth]{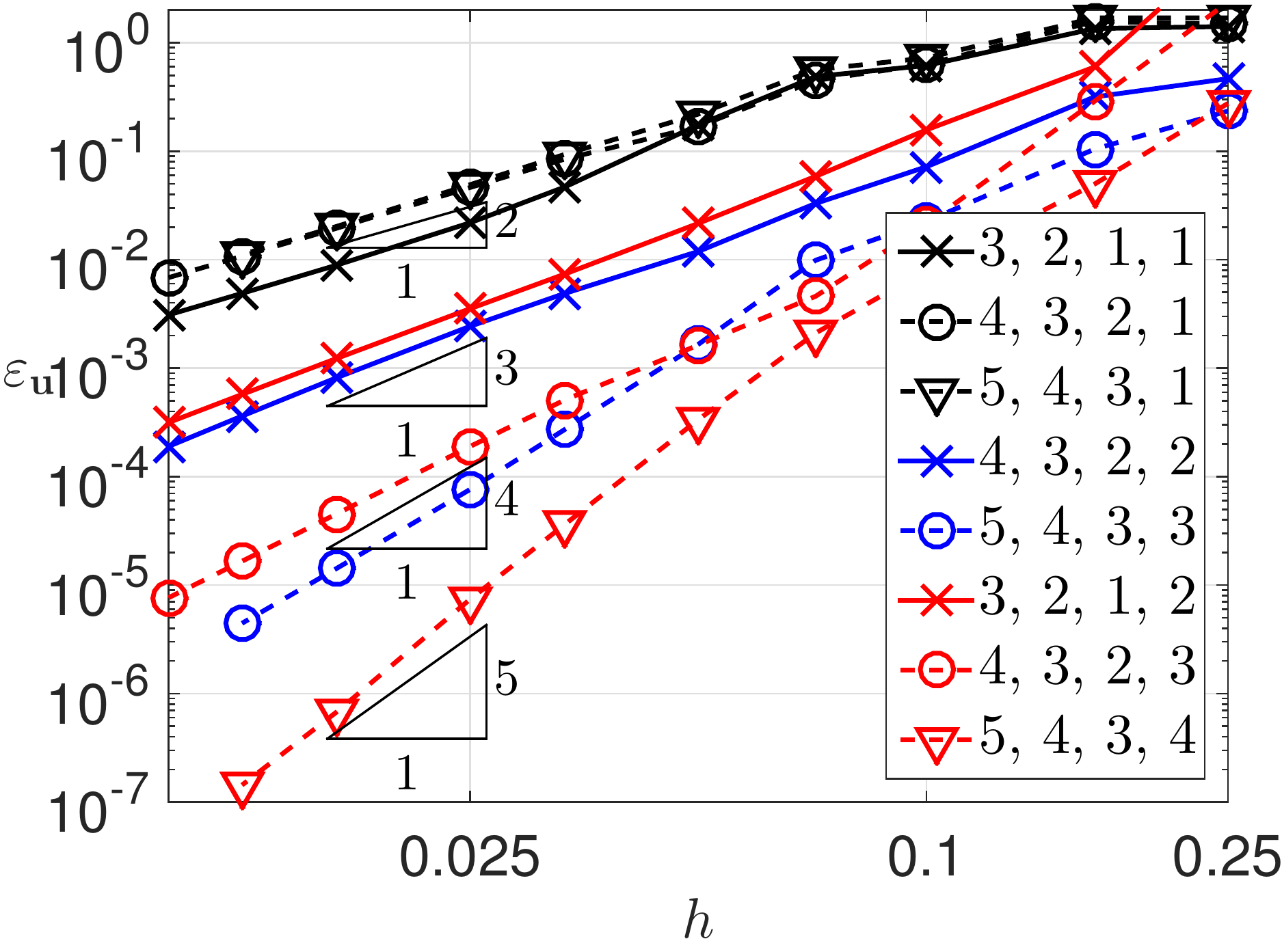}}\qquad\subfigure[quad, $\kappa$]{\includegraphics[width=0.45\textwidth]{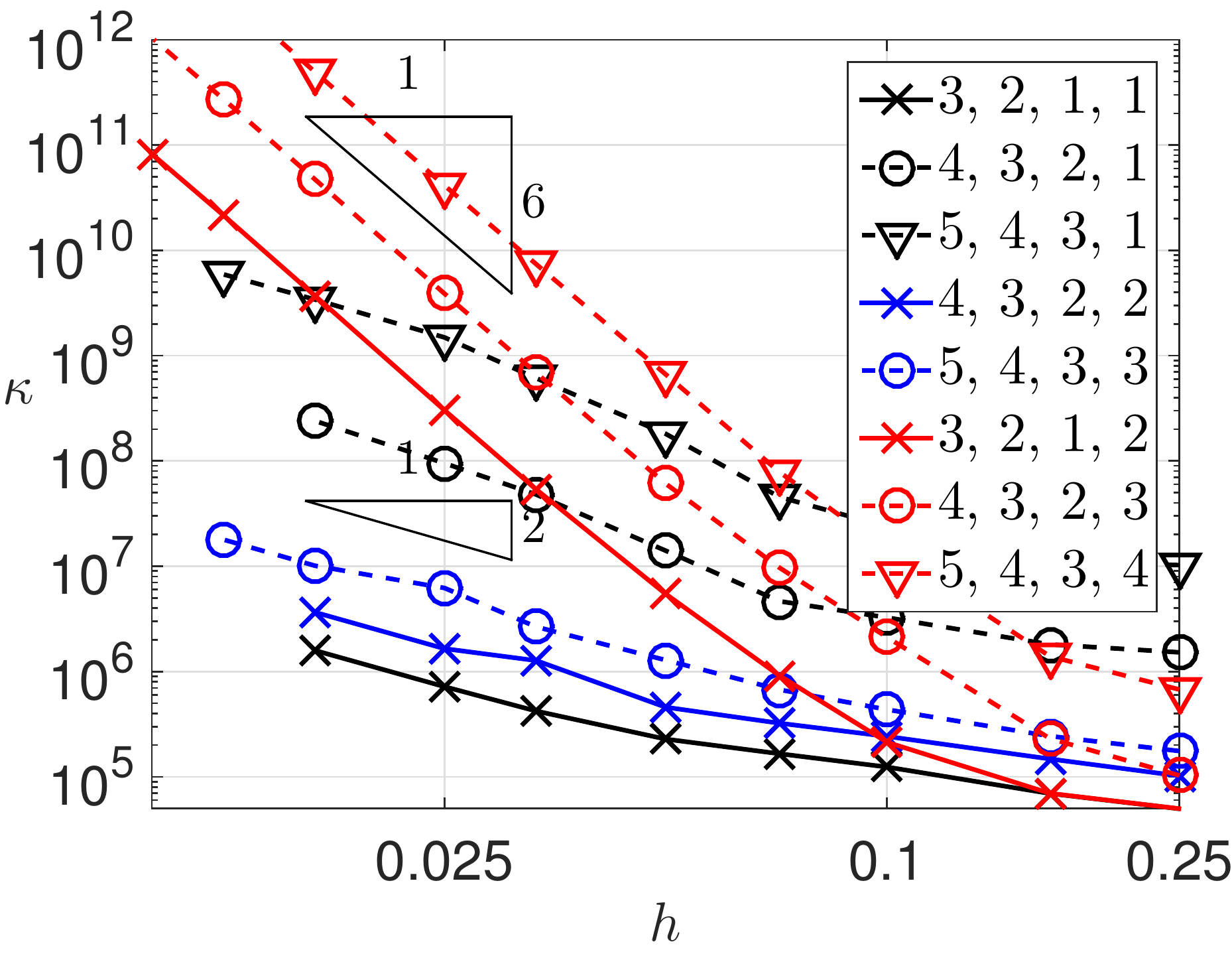}}

\caption{\label{fig:RotSymmResNewC}Influence of the order of the Lagrange
multiplier field for enforcing the tangential velocity constraint:
(a) convergence results in $\varepsilon_{\vek u}$ and (b) conditioning
$\kappa$ for the axisymmetric test case.}
\end{figure}

It is important to note in Fig.~\ref{fig:RotSymmResNewA} that the
convergence rates in the pressure are optimal, $m_{p}=k_{\vek p}+1$,
however, in the velocities one order sub-optimal, $m_{\vek u}=k_{\vek u}$.
We have traced this back to the influence of the order $k_{\lambda}$
of the Lagrange  multiplier field. This is demonstrated in Fig.~\ref{fig:RotSymmResNewC}
where (a) shows the error $\varepsilon_{\vek u}$ and (b) the condition
number $\kappa$ of the corresponding system of equations (obtained
with Matlab's \texttt{condest}-function). As before, $k_{\mathrm{geom}}=k_{\vek u}+1$
and $k_{p}=k_{\vek u}-1$. Fig.~\ref{fig:RotSymmResNewA} shows that
setting $k_{\lambda}=1$ yields convergence rates $m_{\vek u}=2$
independent of the other orders (black lines). Setting $k_{\lambda}=k_{\vek u}$
yields \emph{optimal} convergence rates $m_{\vek u}=k_{\vek u}+1$
for the velocities (red lines), however, there is a dramatic influence
on the conditioning which scales with $\kappa\sim O\left(h^{-6}\right)$
in this case rather than with $O\left(h^{-2}\right)$ for all choices
where $k_{\lambda}<k_{\vek u}$. Therefore, we set $k_{\lambda}=k_{\vek u}-1$
in the following and accept the sub-optimal convergence in the velocities.

Next, the error is observed in the strong form of the momentum and
continuity equations, $\varepsilon_{\mathrm{mom}}$ and $\varepsilon_{\mathrm{cont}}$,
see Eqs.~(\ref{eq:ErrorMomentum}) and (\ref{eq:ErrorContinuity}).
Results for $k_{p}=k_{\lambda}=k_{\vek u}-1$ are depicted in Fig.~\ref{fig:RotSymmResNewB}
for triangular and quadrilateral elements. Again, the thick lines
refer to $k_{\mathrm{geom}}=k_{\vek u}+1$ and the thin dashed lines
to $k_{\mathrm{geom}}=k_{\vek u}$. As mentioned above for the $L_{2}$-errors
in the velocities and pressure, this makes a difference (of one order)
for triangular elements, however, not for quadrilateral elements.
When using $k_{\mathrm{geom}}=k_{\vek u}+1$ on the safe side, the
convergence rate in $\varepsilon_{\mathrm{mom}}$ is $m_{\mathrm{mom}}=k_{\vek u}-1$
as expected due to the presence of second-order derivatives of $\vek u$
in the momentum equations. The expected convergence rate in $\varepsilon_{\mathrm{cont}}$
is $m_{\mathrm{cont}}=k_{\vek u}$ due to the presence of first order
derivatives of $\vek u$ in the continuity equation.

\subsection{Driven cavity flows on manifolds}

\begin{figure}
\centering

\subfigure[flat]{\includegraphics[width=0.3\textwidth]{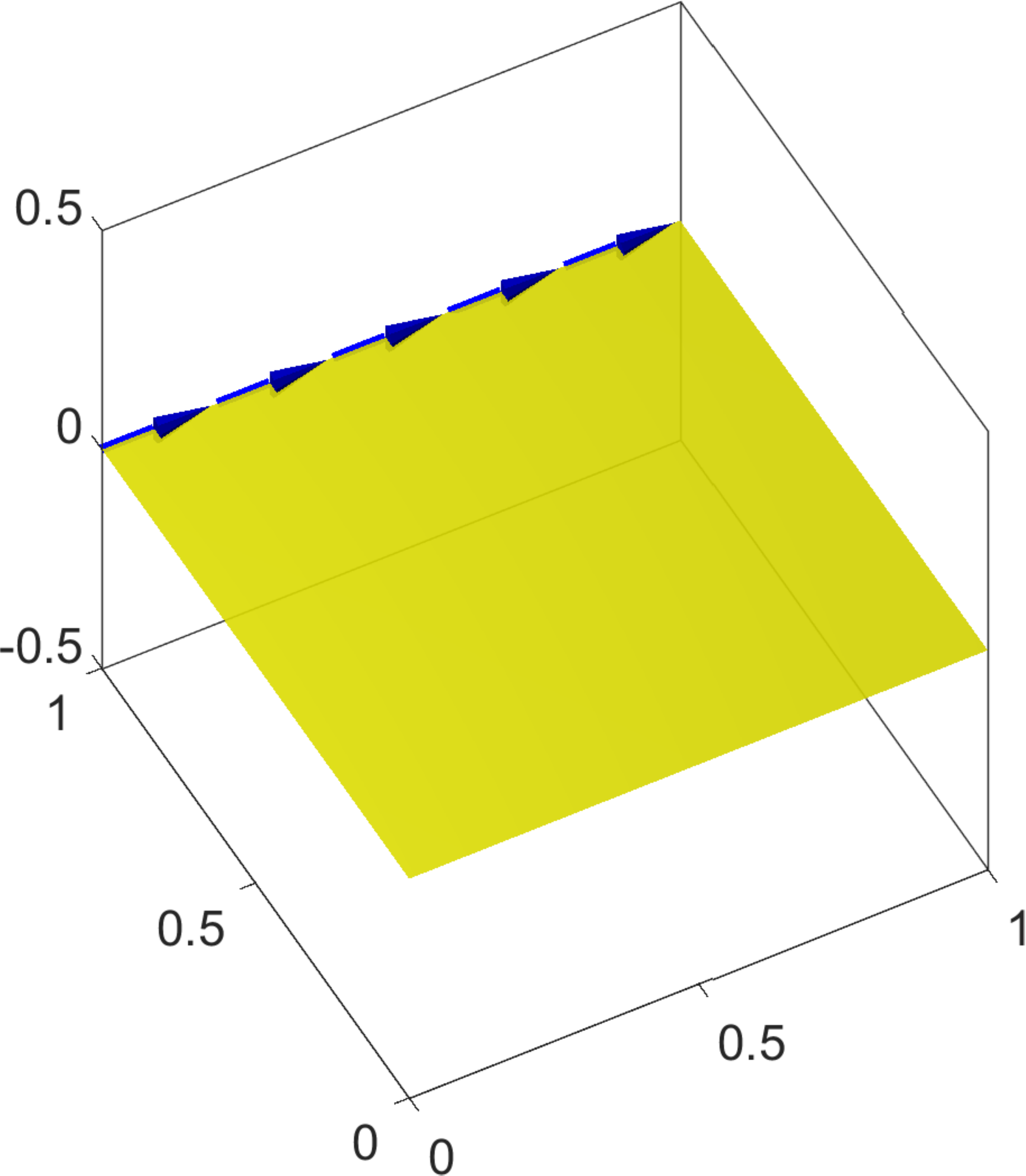}}\hfill\subfigure[$\alpha=0.4$]{\includegraphics[width=0.3\textwidth]{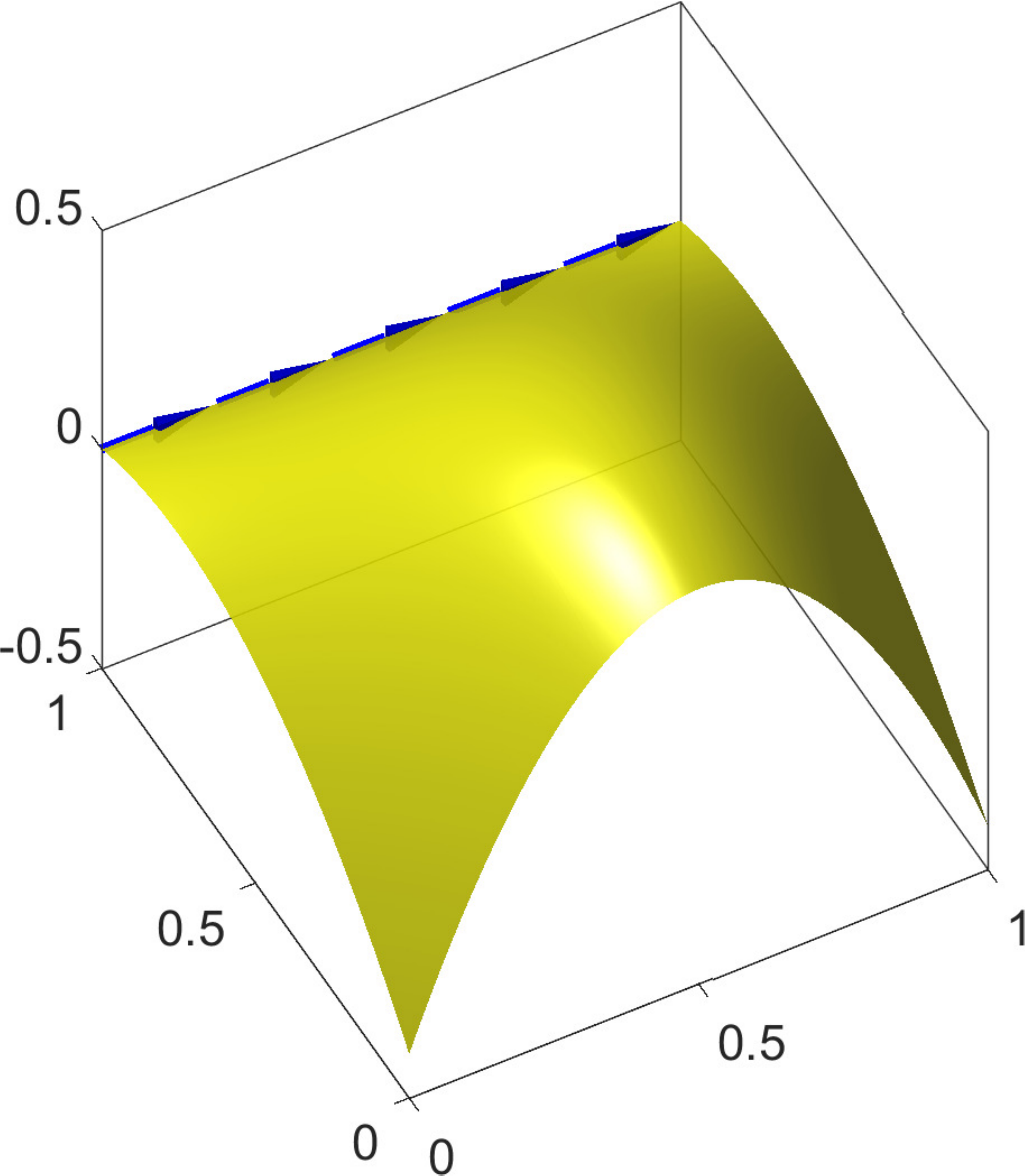}}\hfill\subfigure[$\beta=0.4$]{\includegraphics[width=0.3\textwidth]{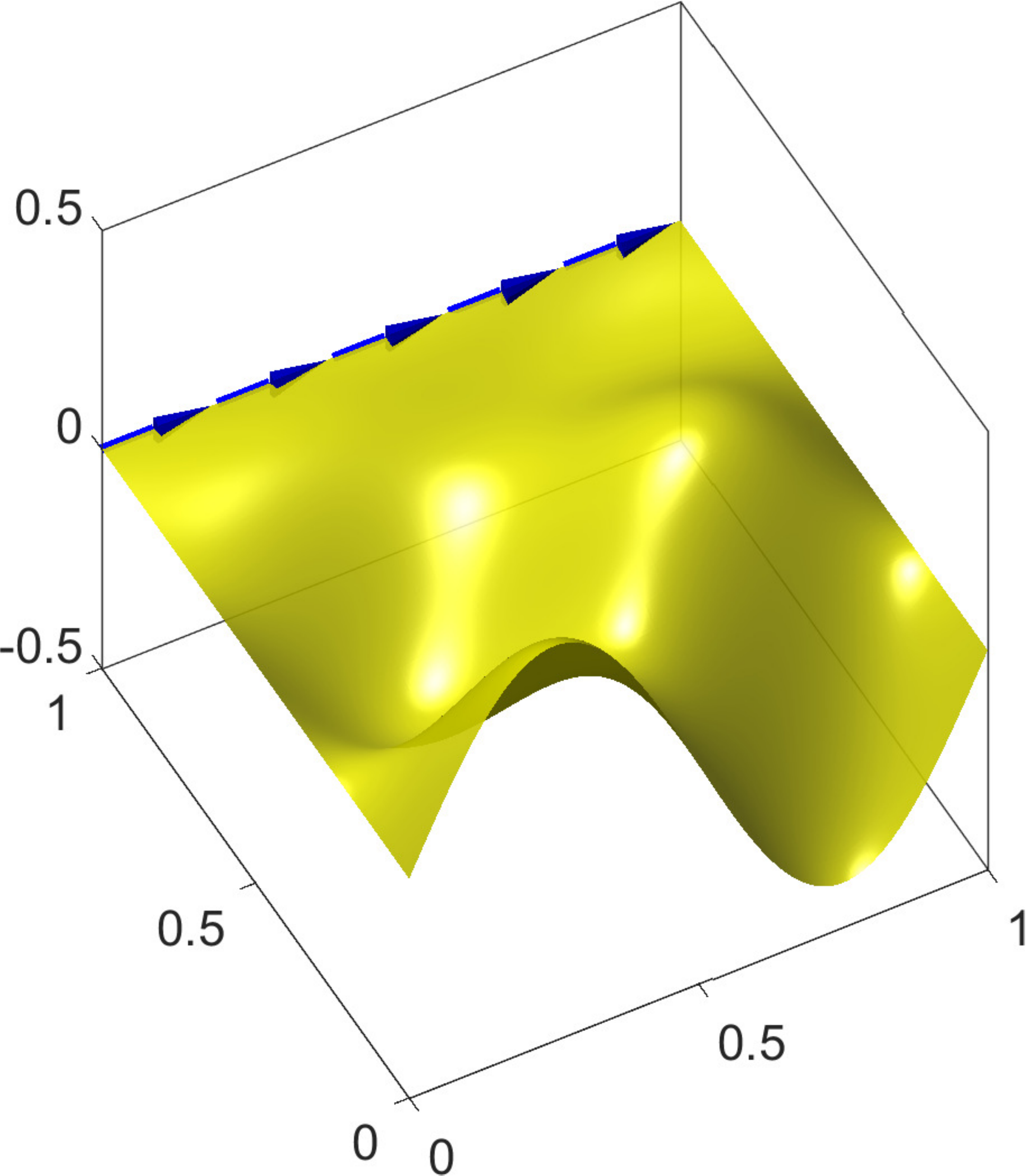}}

\caption{\label{fig:DrivCavDomains}Different manifolds for the driven cavity
test case, (a) flat, (b) map A with $\alpha=0.4$, and (c) map B with
$\beta=0.4$.}
\end{figure}

The stationary Navier-Stokes model is considered in this example.
Starting point is the driven cavity for the case of a flat 2D domain
as depicted in Fig.~\ref{fig:DrivCavDomains}(a). This case has well-documented
reference solutions for a variety of Reynolds numbers \cite{Ghia_1982a}.
There, a flow inside a quadratic domain $\Omega_{\mathrm{2D}}=\left(0,1\right)\times\left(0,1\right)$
with no-slip boundary conditions on the left, right and lower wall
develops under a shear flow of $u=1.0$ and $v=0.0$ applied on the
upper boundary until a stationary solution is reached. The Reynolds
number is computed as $\mathrm{Re}=\varrho\cdot u\cdot L/\mu$.

\begin{figure}
\centering

\subfigure[$\alpha=0.4$, $\left\Vert \vek u(\vek x)\right\Vert$]{\includegraphics[width=0.2\textwidth]{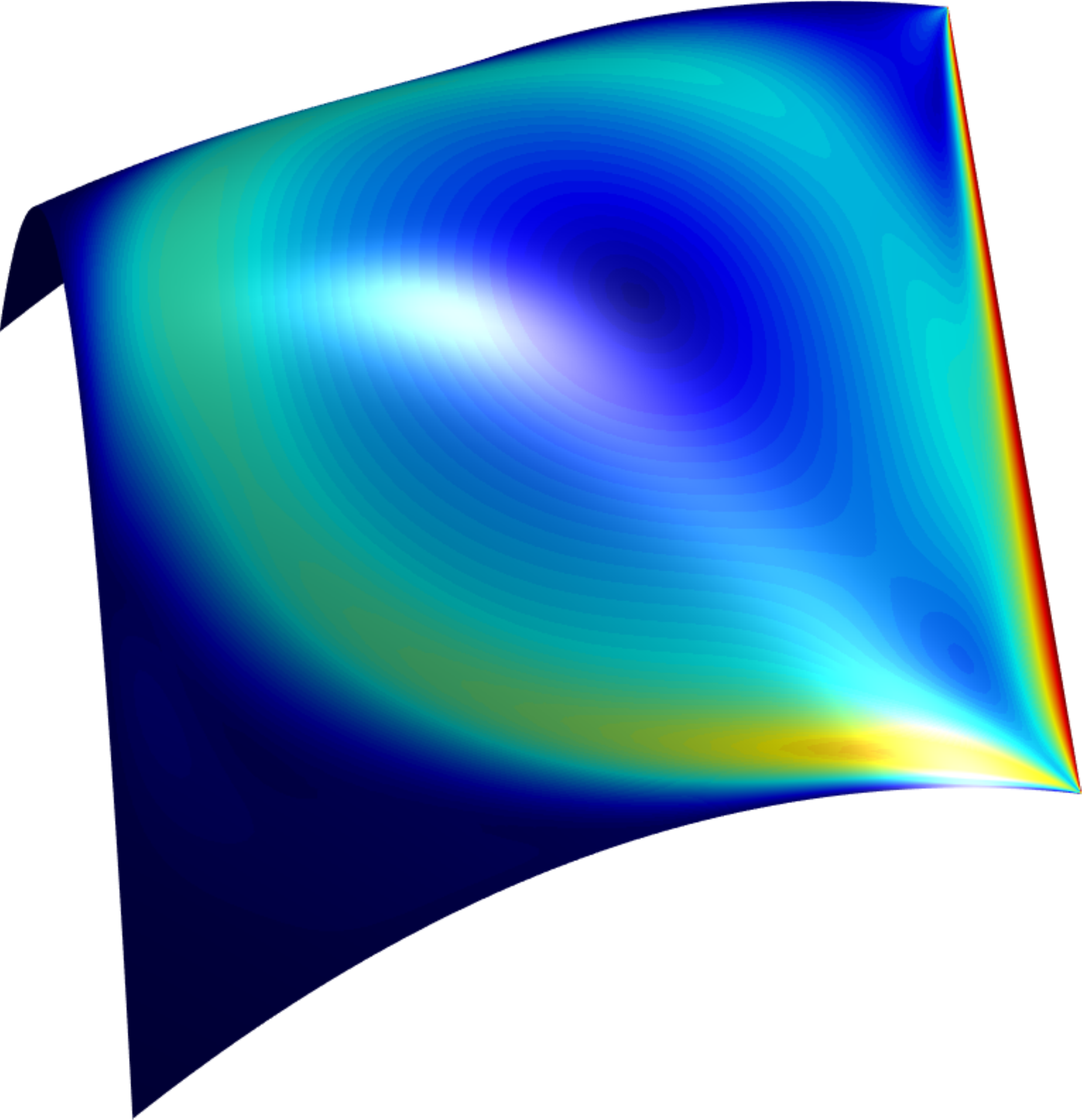}}\hfill\subfigure[$\alpha=0.4$, $p(\vek x)$]{\includegraphics[width=0.2\textwidth]{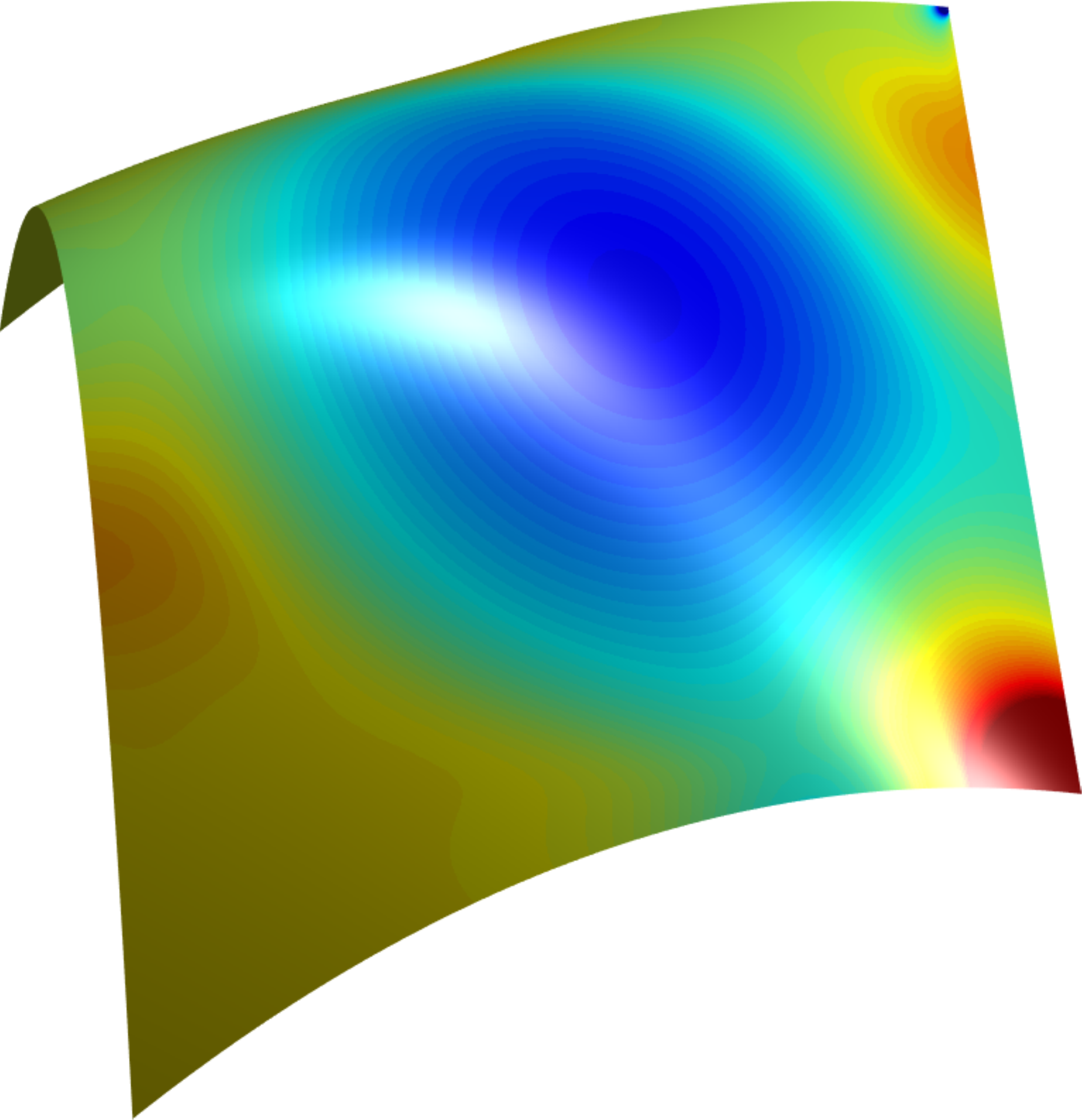}}\hfill\subfigure[$\beta=0.4$, $\left\Vert \vek u(\vek x)\right\Vert$]{\includegraphics[width=0.25\textwidth]{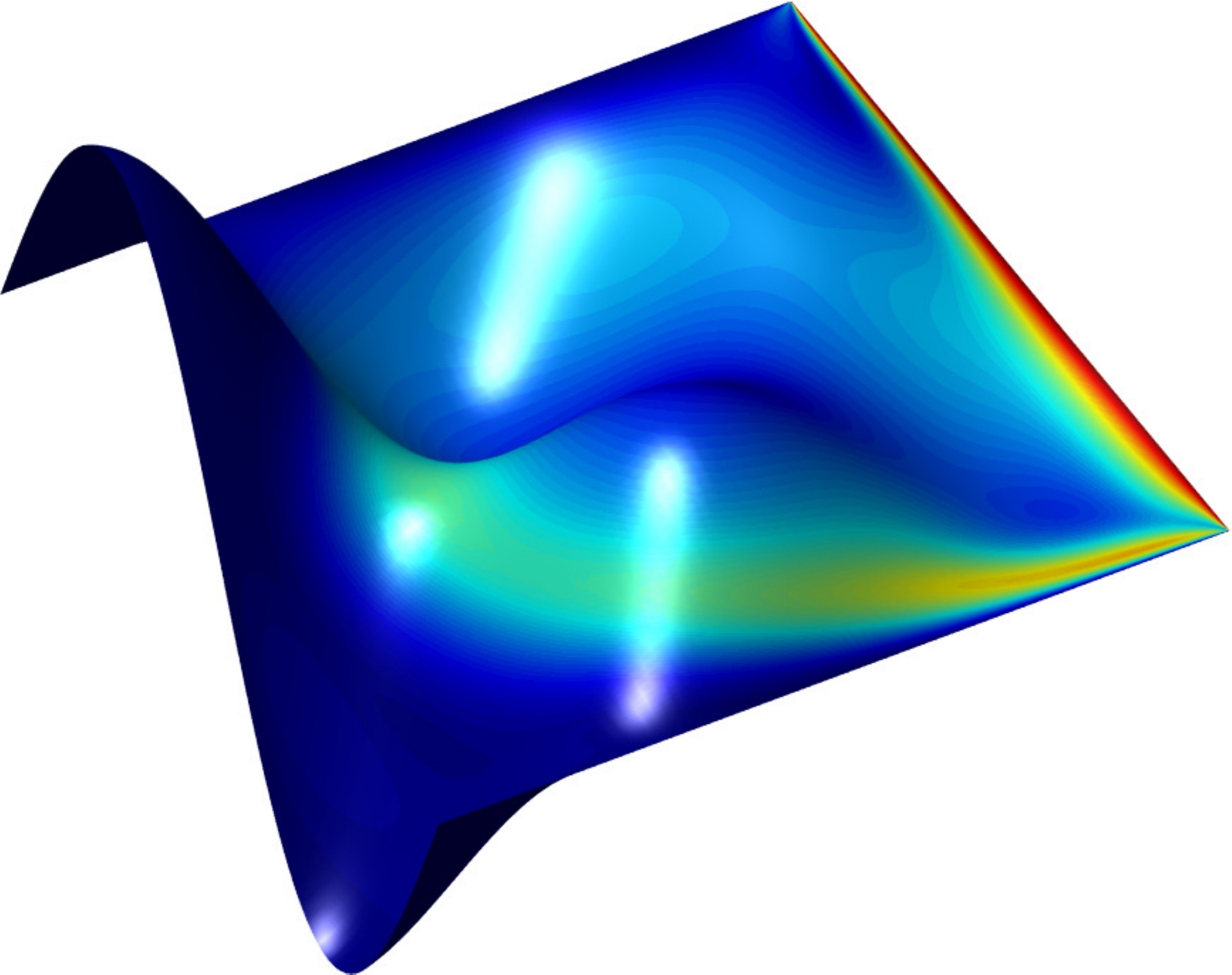}}\hfill\subfigure[$\beta=0.4$, $p(\vek x)$]{\includegraphics[width=0.25\textwidth]{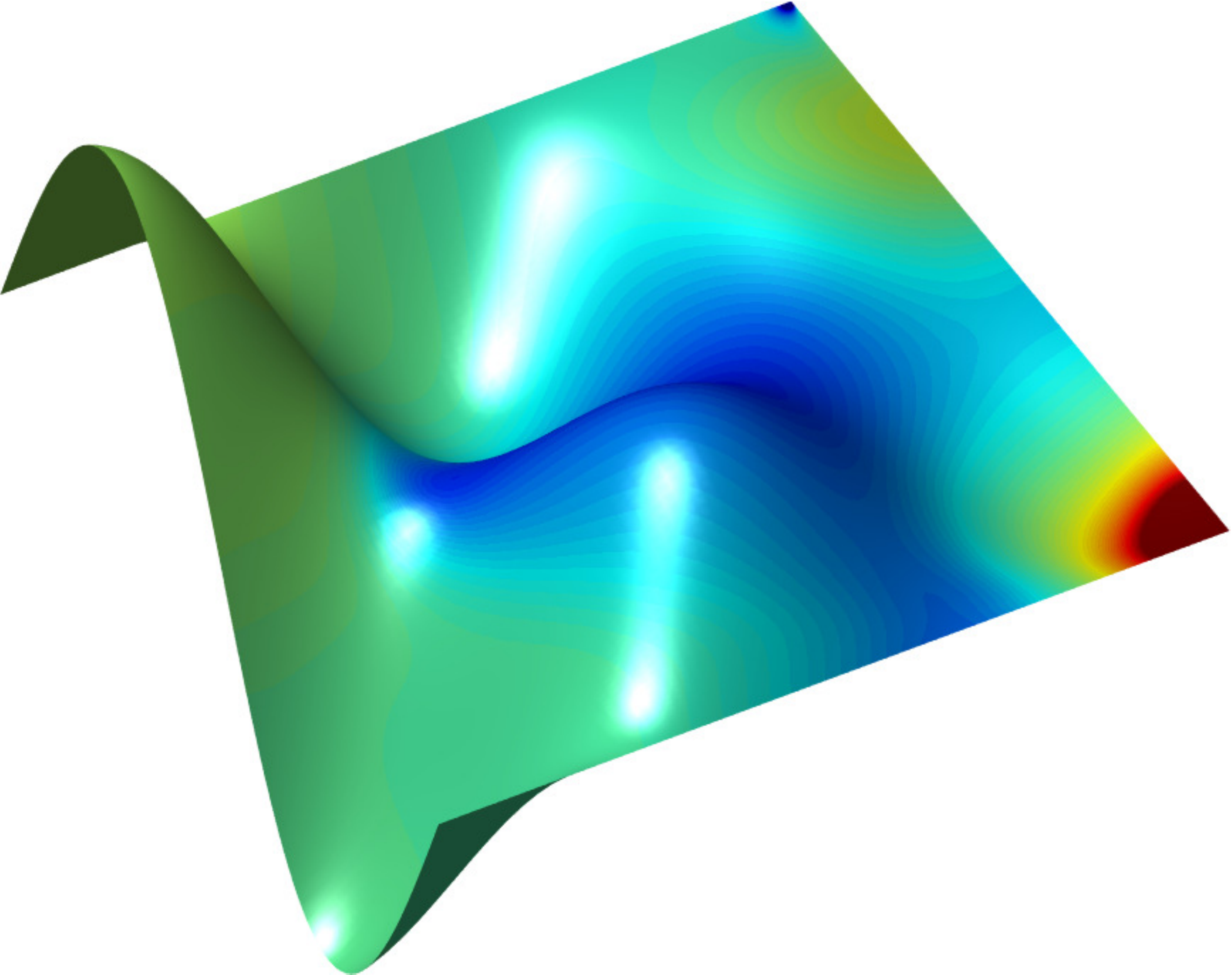}}

\caption{\label{fig:DrivCavFields}Velocity and pressure fields for the driven
cavity test case with $\ensuremath{\mu=0.001}$ for map A with $\alpha=0.4$
and map B with $\beta=0.4$.}
\end{figure}

Herein, the situation is extended to curved surfaces in 3D by deforming
the flat 2D domain $\Omega_{\mathrm{2D}}$ in $z$-direction using
functions $z(x,y)$. In particular, two different maps A and B are
used,
\begin{eqnarray*}
\text{map A}:\qquad z(x,y) & = & \alpha\cdot\left(-1+8x+2y-8x^{2}\right)\cdot\left(1-y\right),\\
\text{map B}:\qquad z(x,y) & = & \beta\cdot\left(1-y\right)\cdot\sin\left(\left(2x-1\right)\pi\right)\cdot\cos\left(\left(2y-1\right)\pi\right),
\end{eqnarray*}
where $\alpha$ and $\beta$ scale the height in $z$-direction, see
Figs.~\ref{fig:DrivCavDomains}(b) and (c) for examples. The advantage
is that for $\alpha=0$ and $\beta=0$, the flat situation is recovered
and the reference solutions in \cite{Ghia_1982a} are relevant. We
have confirmed that these solutions are recovered with great accuracy
also for any rigid body tranformation of $\Omega_{\mathrm{2D}}$ into
three dimensions. The density is chosen as $\varrho=1$ and two different
viscosities of $\mu=0.01$ and $\mu=0.001$ leading to Reynolds numbers
of $\mathrm{Re}=100$ and $\mathrm{Re}=1000$ for the flat case, respectively.
Solutions for the velocity magnitude and pressure field for some example
manifolds are displayed in Fig.~\ref{fig:DrivCavFields}.

\begin{figure}
\centering

\subfigure[$10\times 10$ elements]{\includegraphics[width=0.25\textwidth]{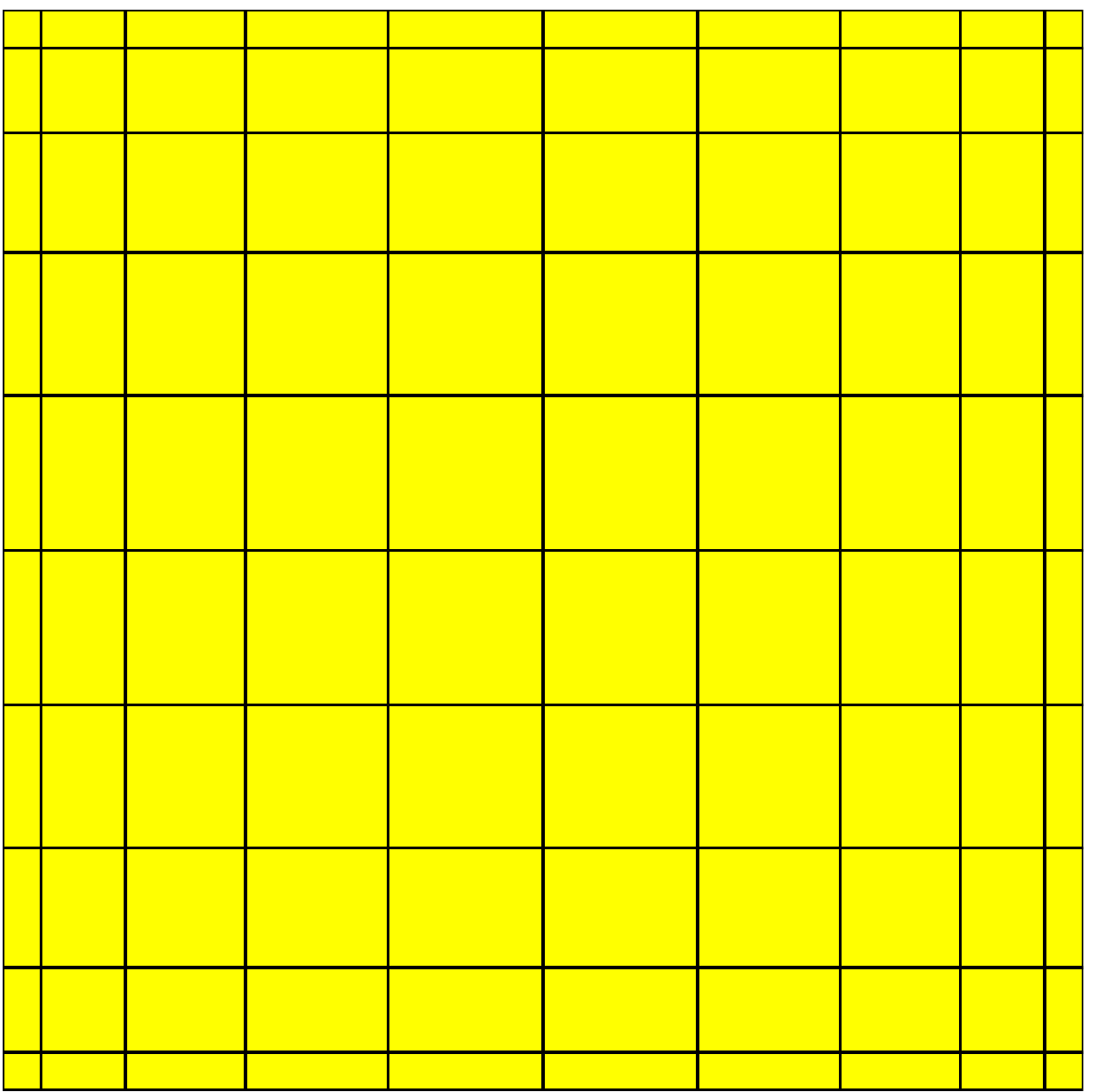}}\hfill\subfigure[$20\times 20$ elements]{\includegraphics[width=0.25\textwidth]{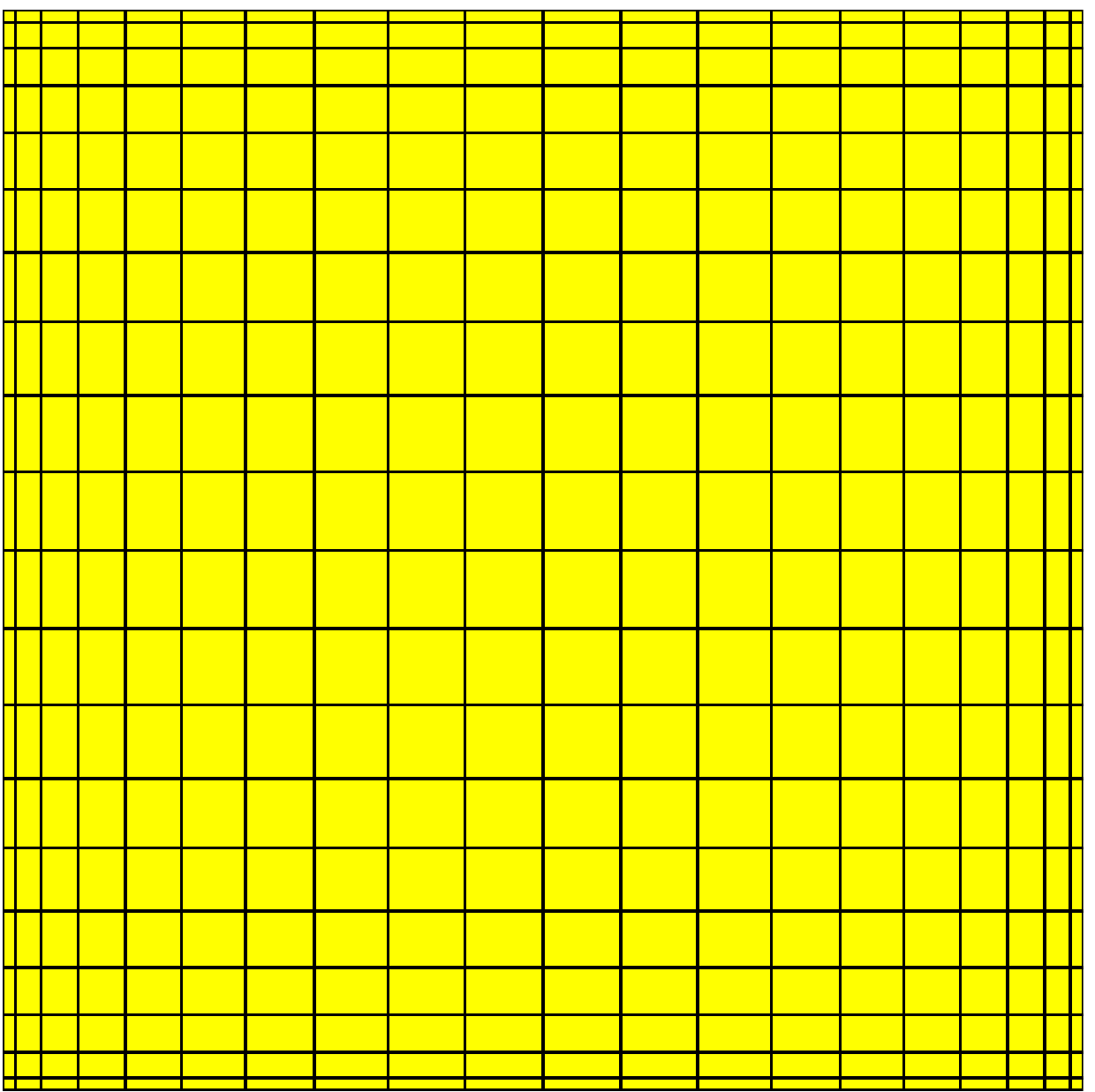}}\hfill\subfigure[$50\times 50$ elements]{\includegraphics[width=0.25\textwidth]{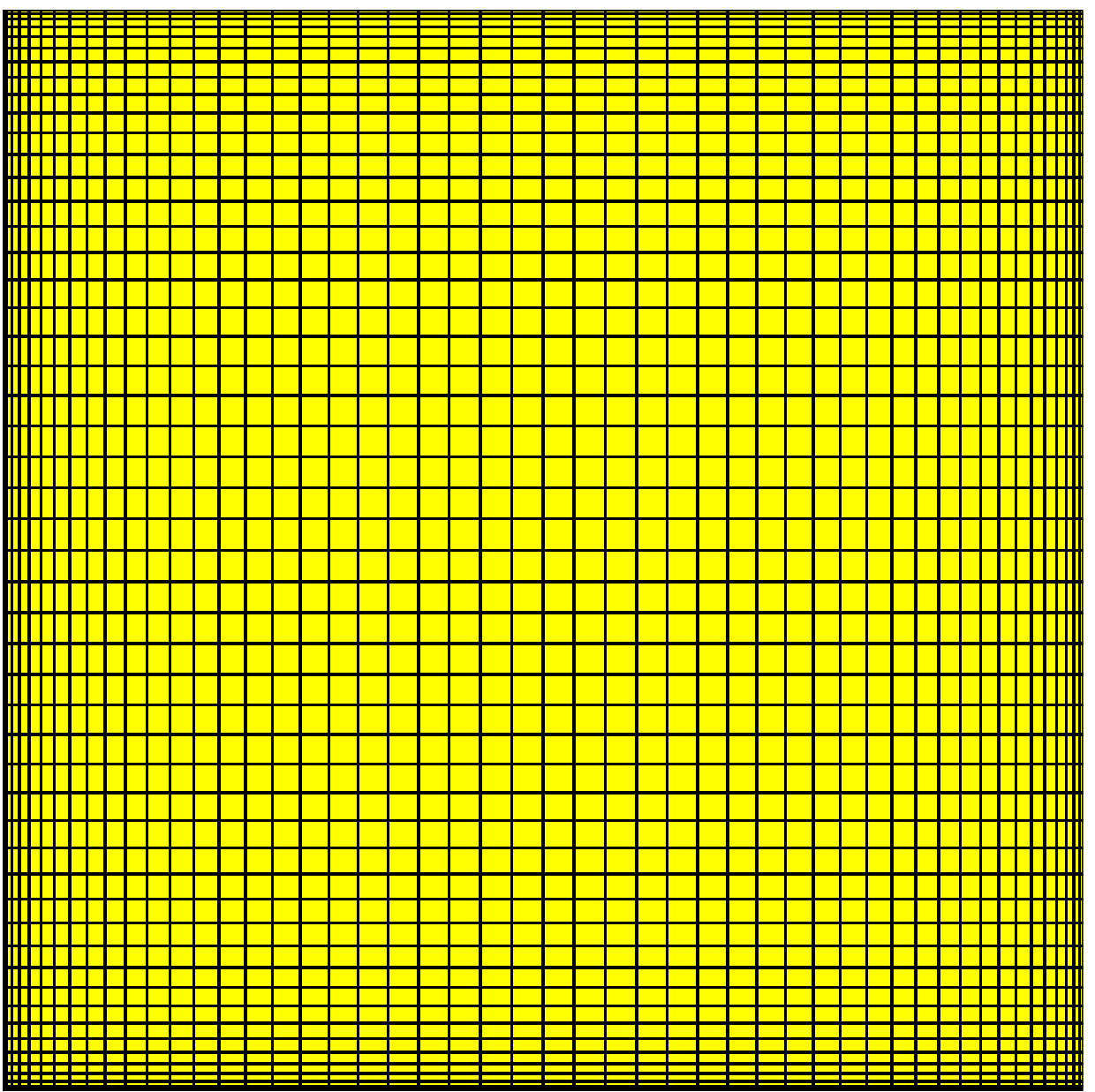}}

\caption{\label{fig:DrivCavMeshes}Different meshes for the driven cavity test
case in top view.}
\end{figure}

The meshes feature quadrilateral elements of different orders and
are refined towards the boundaries to capture the resulting boundary
layers. See Fig.~\ref{fig:DrivCavMeshes} for the meshes in $\Omega_{\mathrm{2D}}$
which are mapped to 3D according to map A and B from above for various
scaling coefficients $\alpha$ and $\beta$. The number of elements
per dimension is $n=\left\{ 10,20,30,50,70,100\right\} $. For the
numerical studies, $k_{p}=k_{\lambda}=k_{\vek u}-1$ and $k_{\mathrm{geom}}=k_{\vek u}+1$
is used as recommended above.

\begin{figure}
\centering

\subfigure[map A, $\mu=0.01$]{\includegraphics[width=0.4\textwidth]{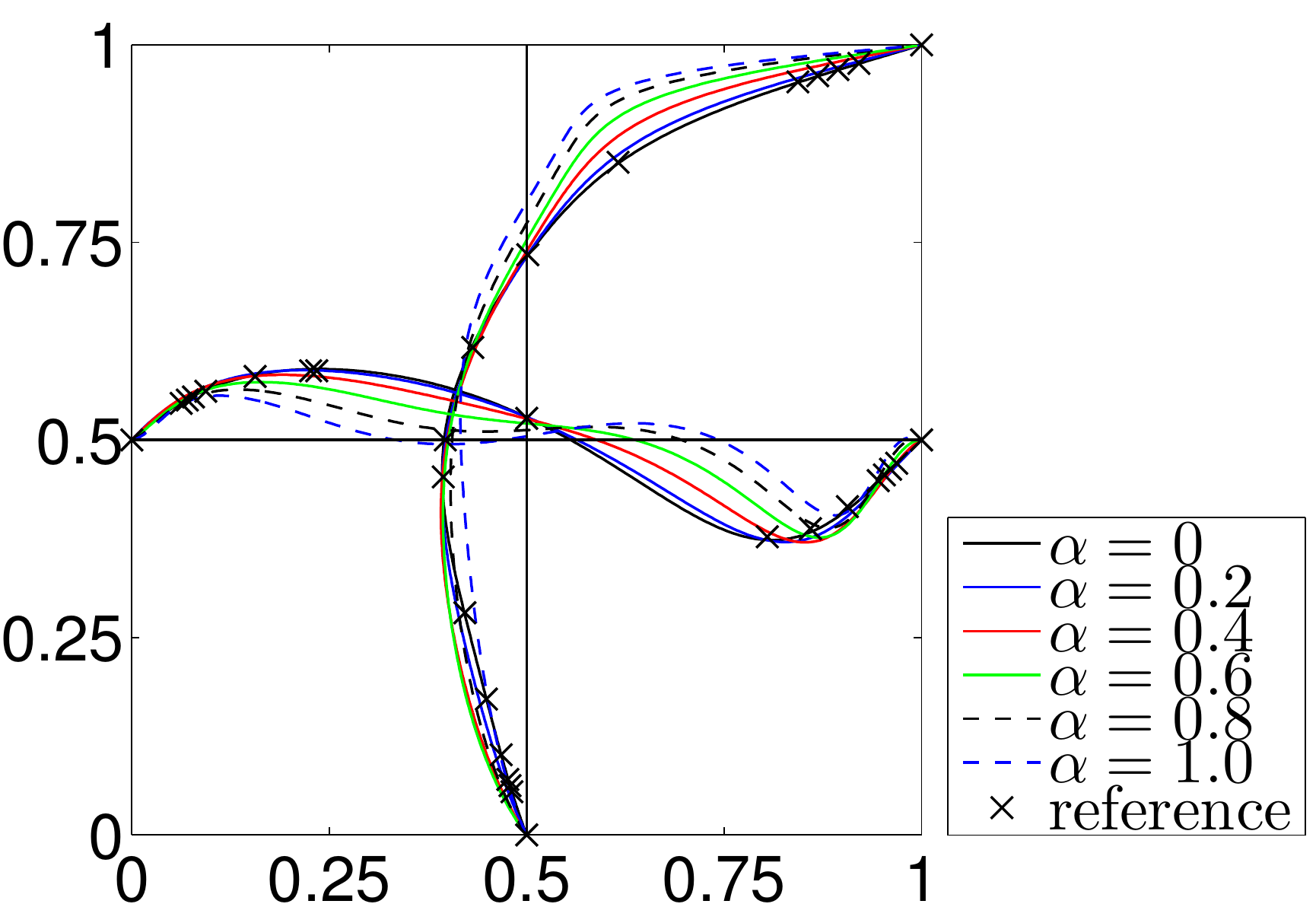}}\quad\subfigure[map B, $\mu=0.01$]{\includegraphics[width=0.4\textwidth]{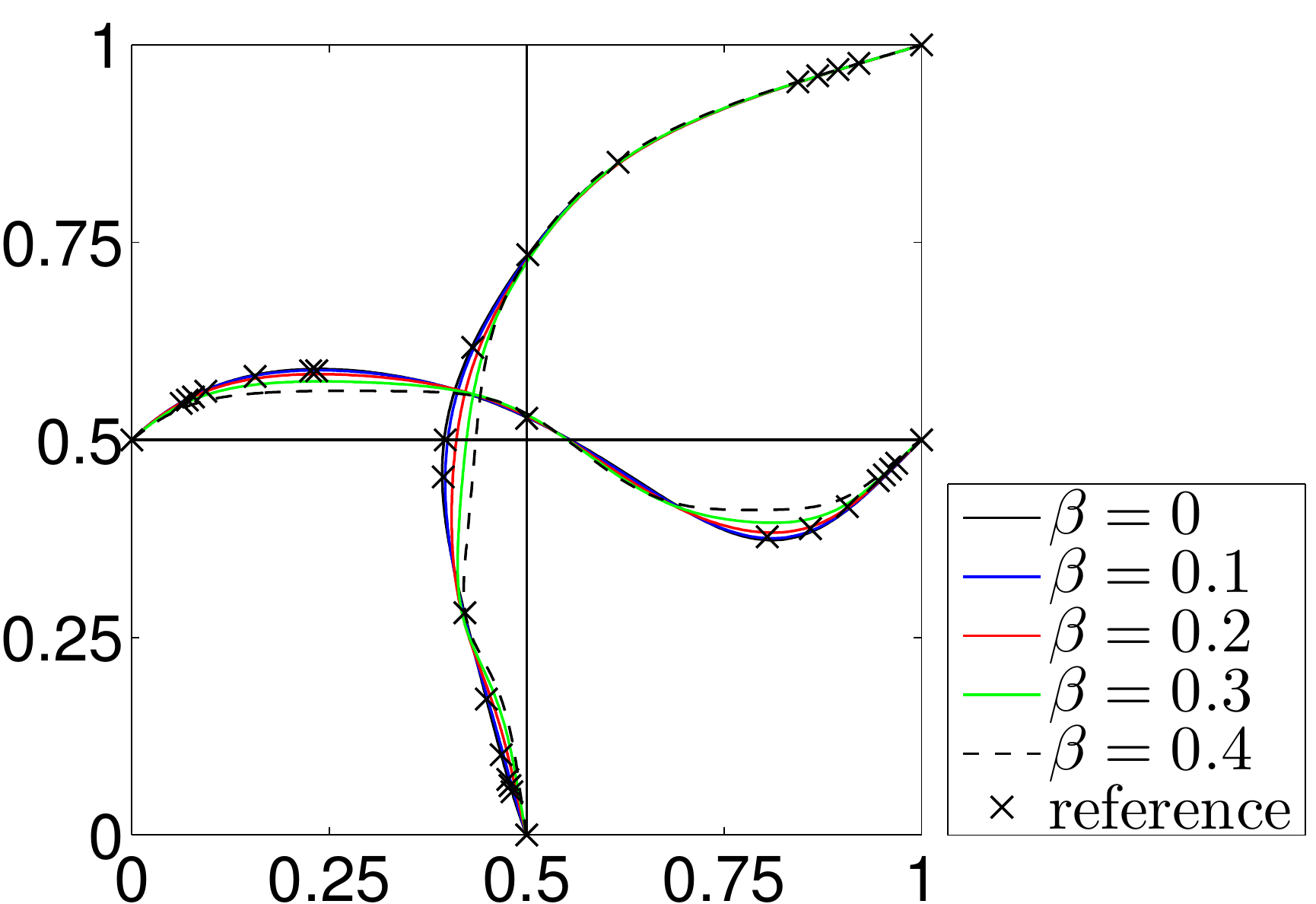}}

\subfigure[map A, $\mu=0.001$]{\includegraphics[width=0.4\textwidth]{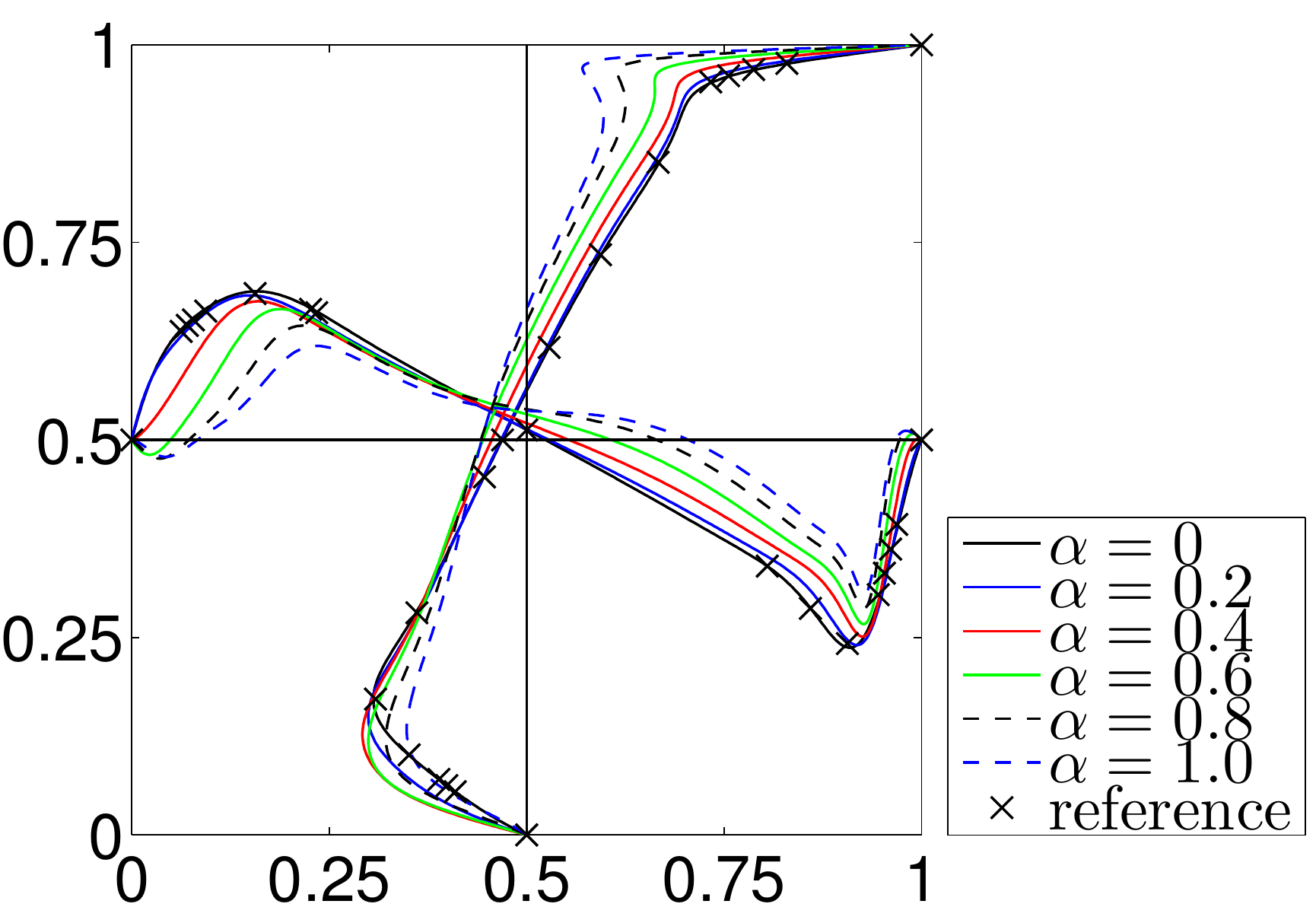}}\quad\subfigure[map B, $\mu=0.001$]{\includegraphics[width=0.4\textwidth]{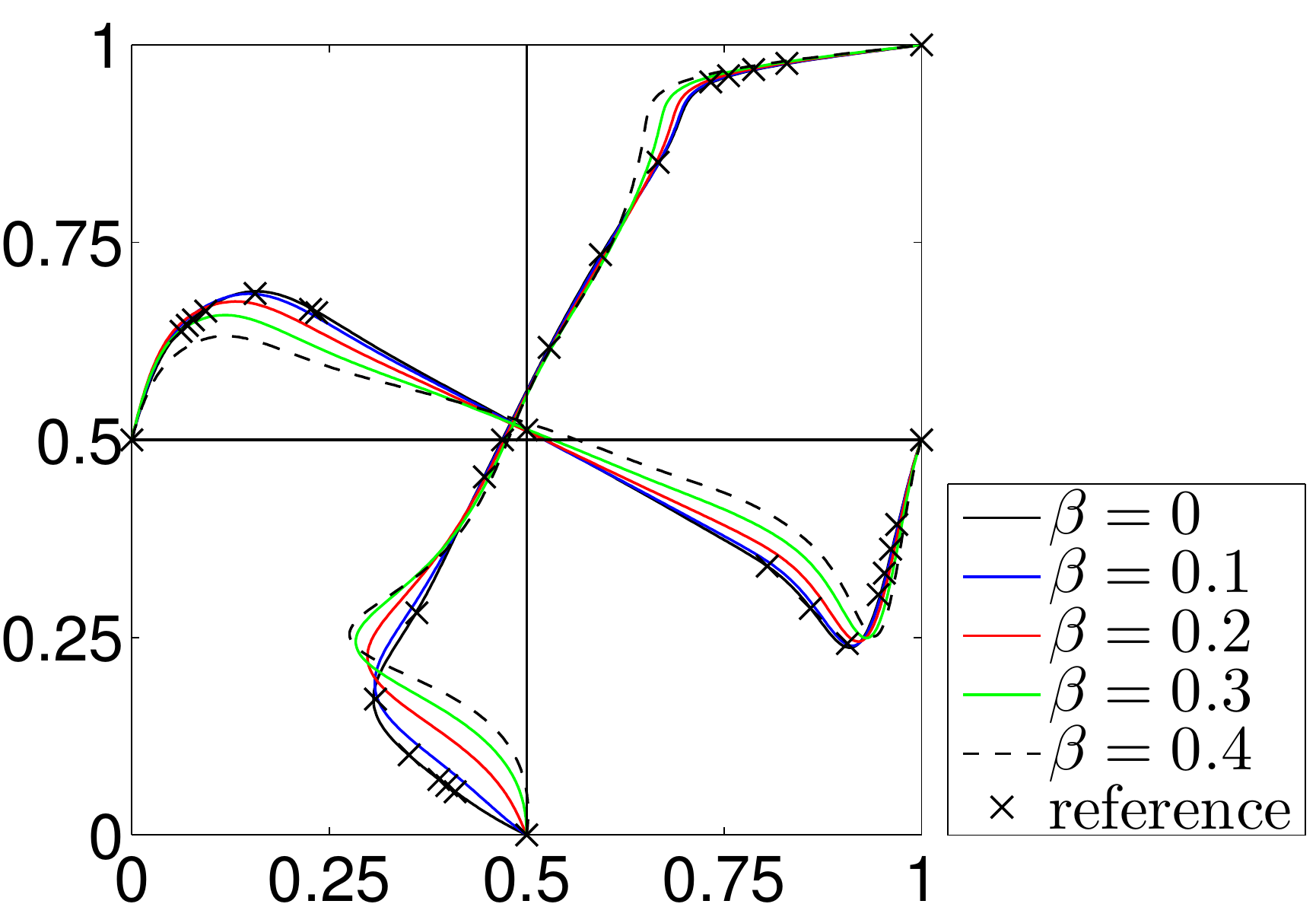}}

\caption{\label{fig:DrivCavVisResA}Velocity profiles for the driven cavity
test case for different $\alpha$ and $\beta$. The vertical profiles
show the velocity component $u\left(\vek x\right)$, the horizontal
profiles $v\left(\vek x\right)$. The scaling factor of the velocities
is $0.5$.}
\end{figure}

\begin{figure}
\centering

\subfigure[map A, $\mu=0.01$]{\includegraphics[width=0.4\textwidth]{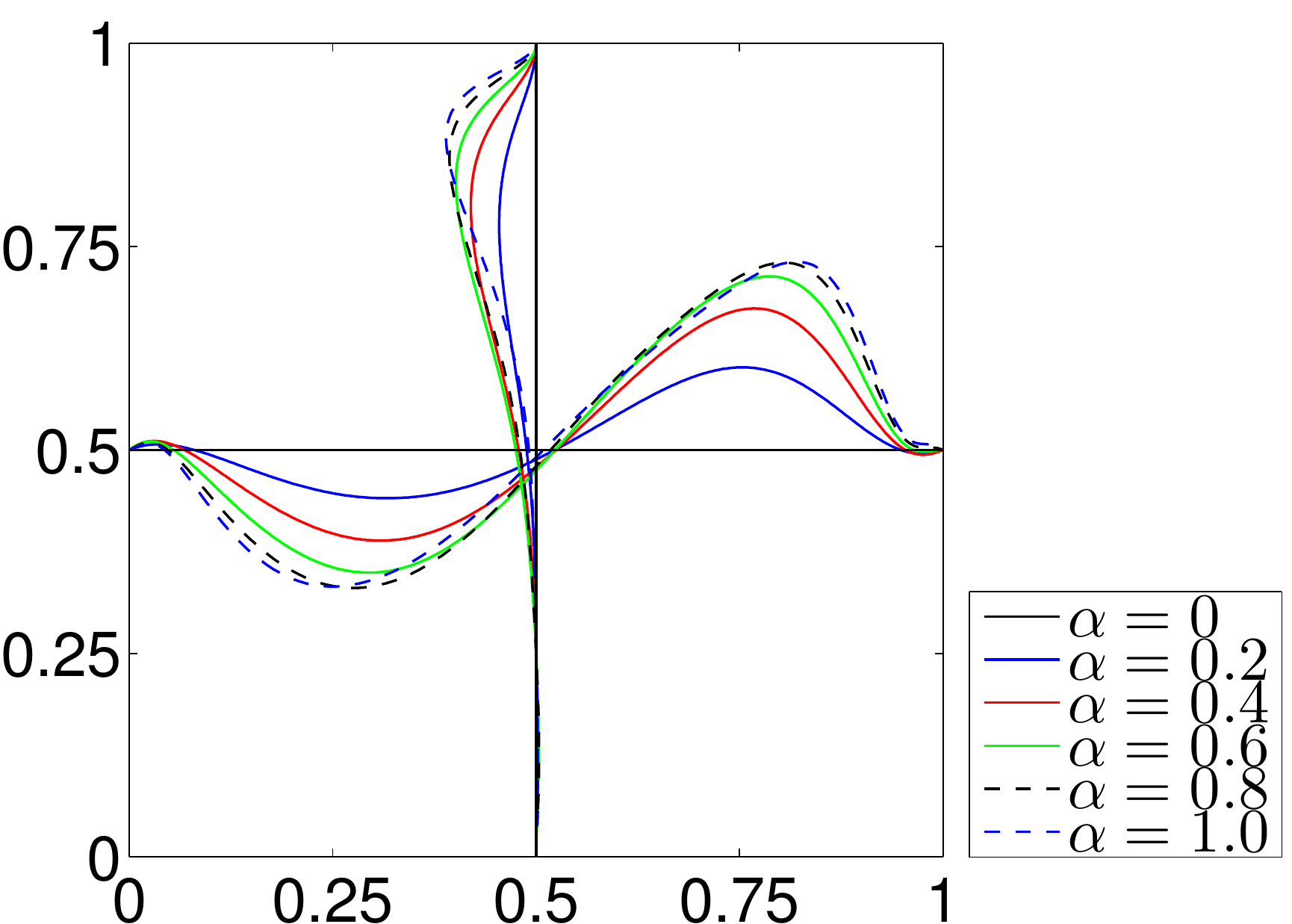}}\quad\subfigure[map B, $\mu=0.01$]{\includegraphics[width=0.4\textwidth]{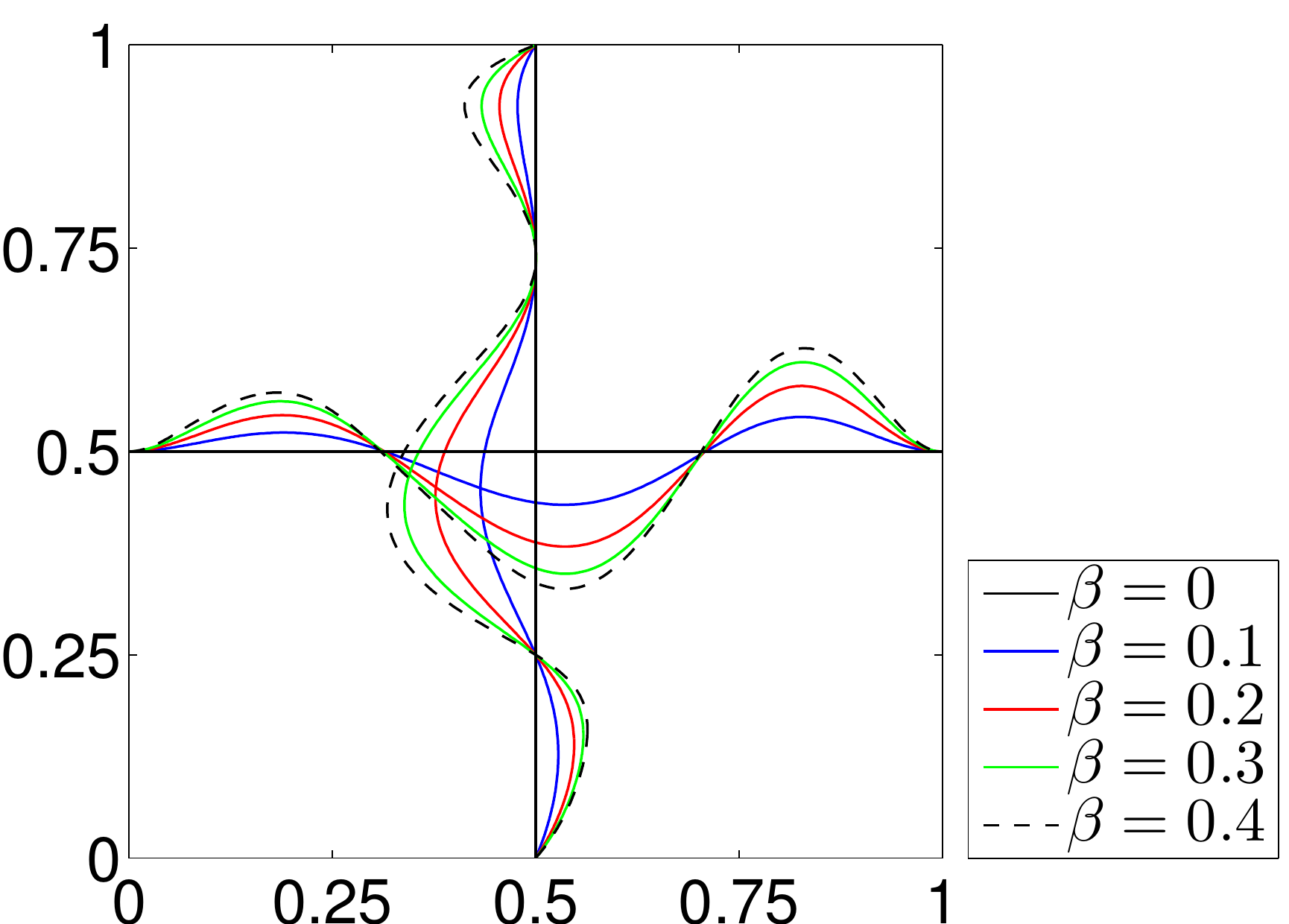}}

\subfigure[map A, $\mu=0.001$]{\includegraphics[width=0.4\textwidth]{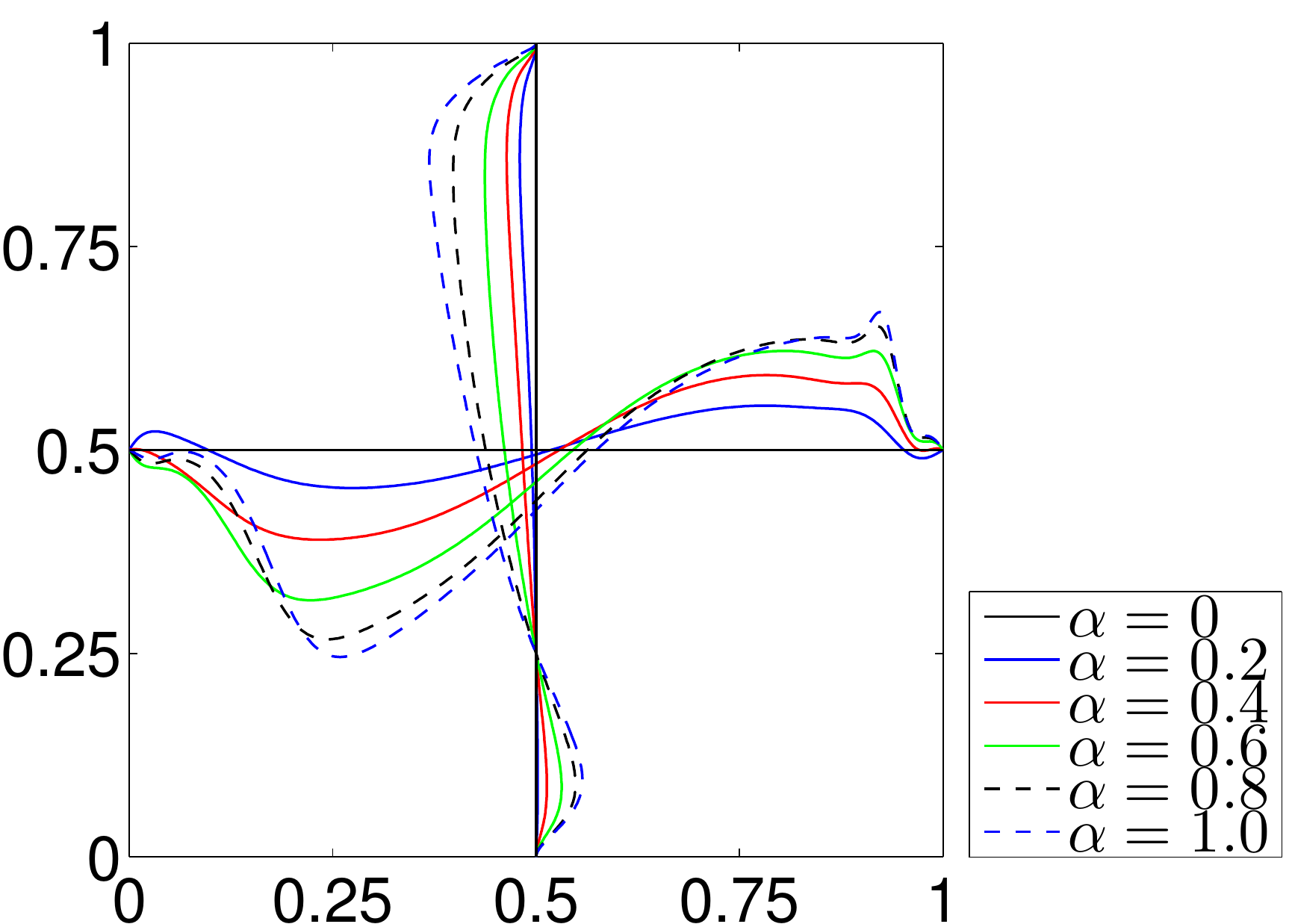}}\quad\subfigure[map B, $\mu=0.001$]{\includegraphics[width=0.4\textwidth]{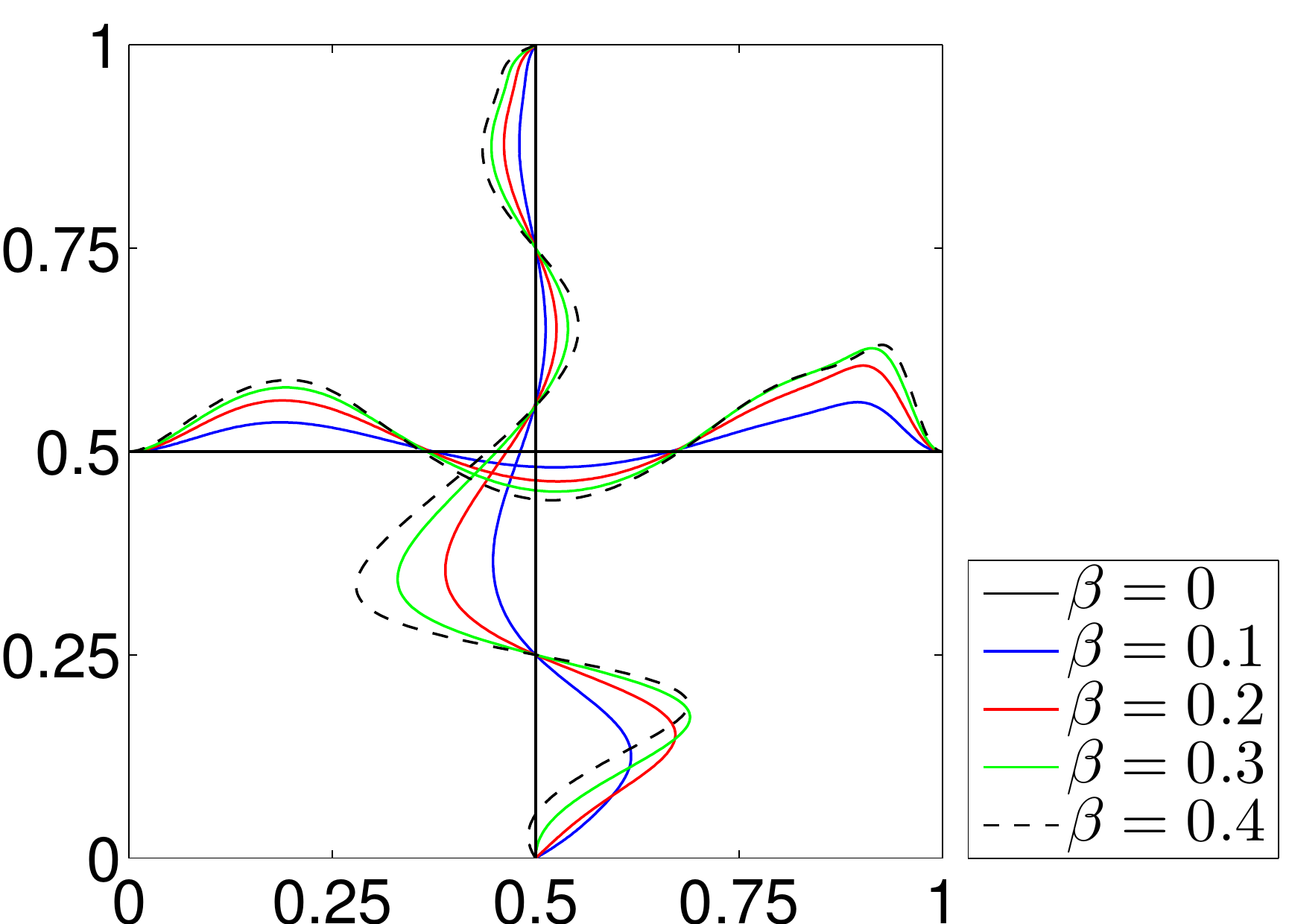}}

\caption{\label{fig:DrivCavVisResB}Velocity profiles for the driven cavity
test case for different $\alpha$ and $\beta$. The horizontal and
vertical profiles show the velocity component $w\left(\vek x\right)$
scaled by the factor $1$.}
\end{figure}

\begin{figure}
\centering

\subfigure[flat]{\includegraphics[width=0.47\textwidth]{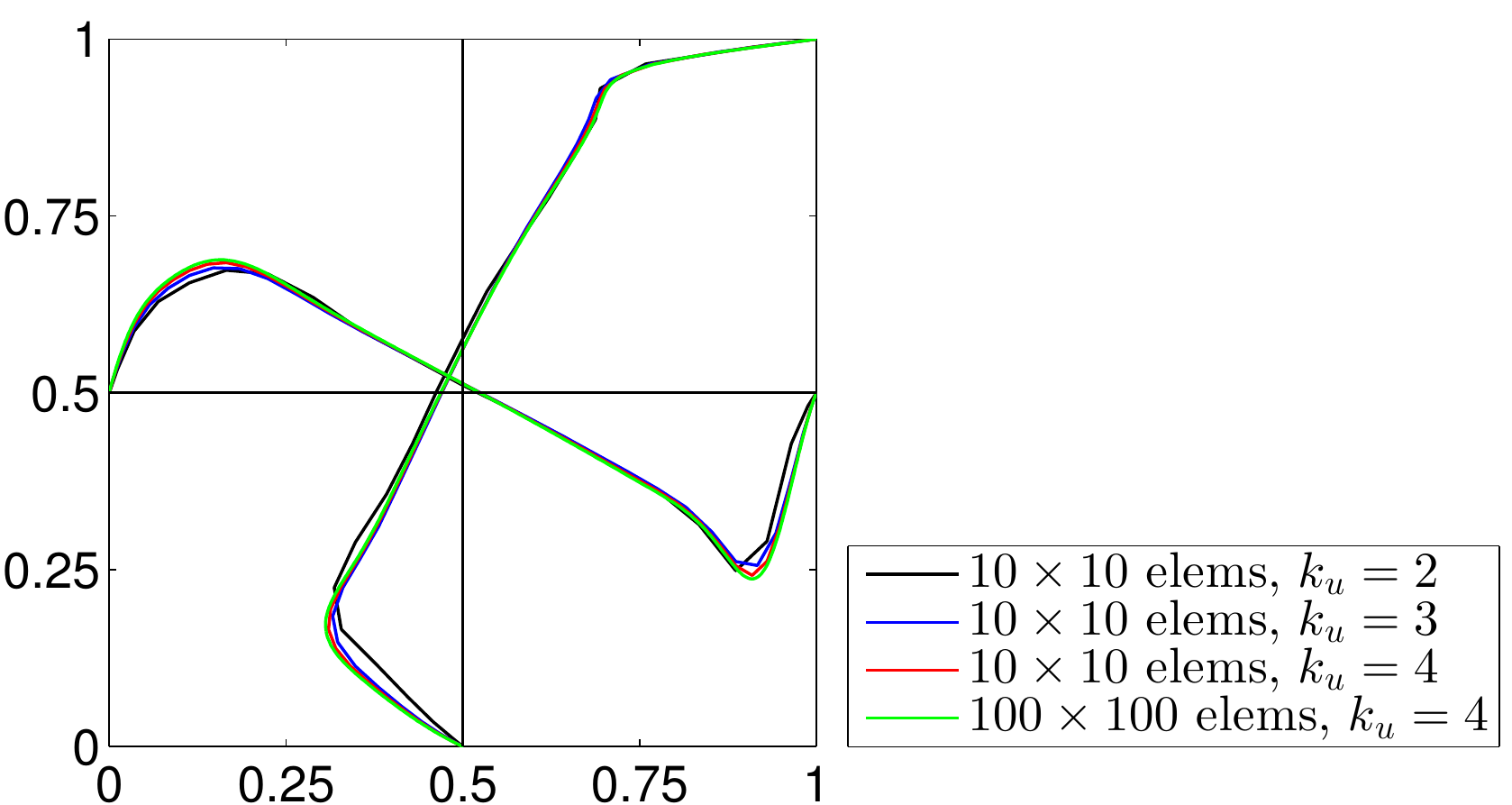}}\quad\subfigure[map A, $\alpha=1.0$]{\includegraphics[width=0.47\textwidth]{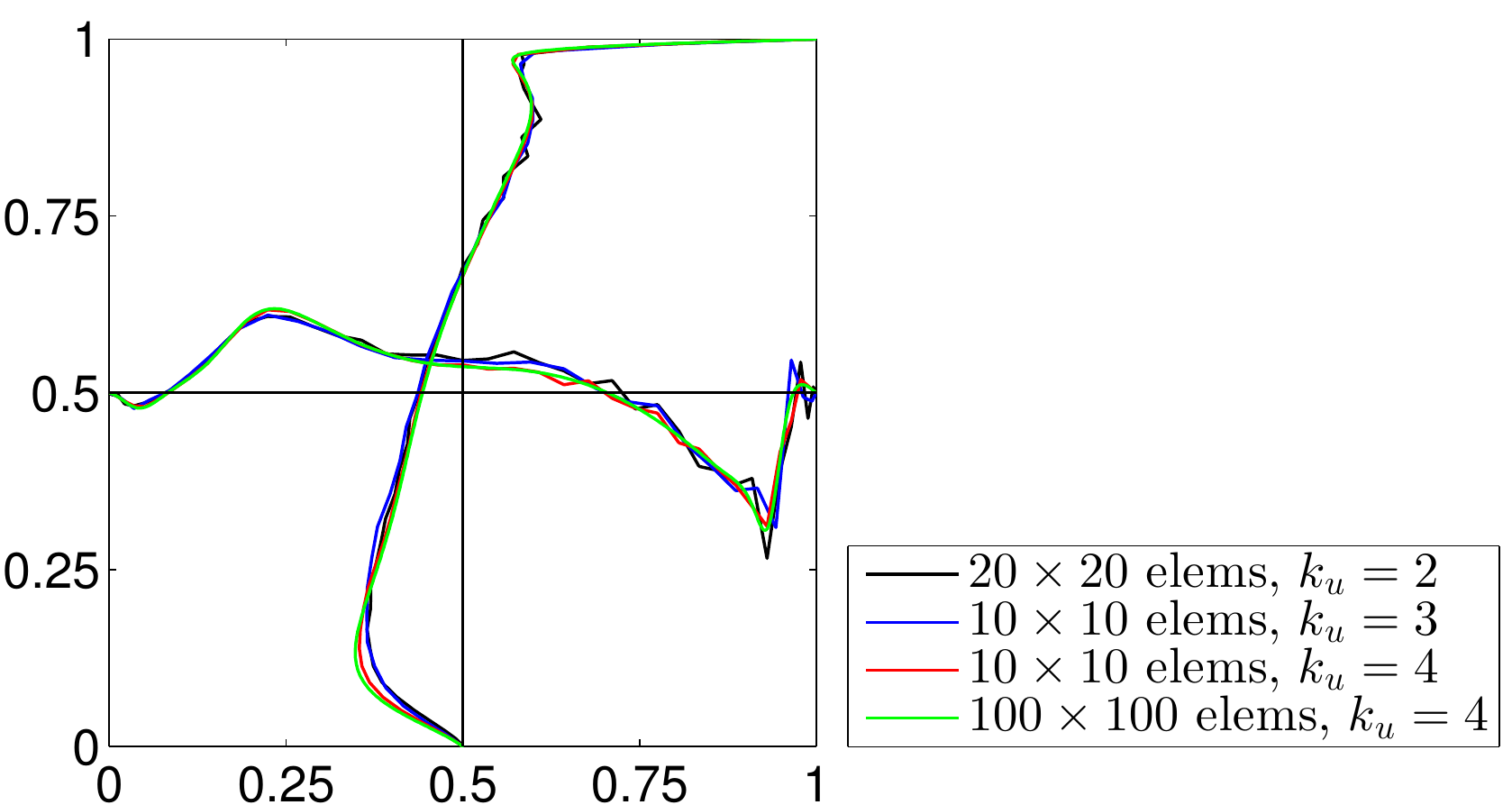}}

\caption{\label{fig:DrivCavVisResC}Velocity profiles for the driven cavity
test case following Fig.~\ref{fig:DrivCavVisResA}. Results of coarse
meshes with $10\times10$ elements are compared to the high-accuracy
results from above (with $100\times100$ elements with $k_{\vek u}=4$),
(a) flat manifold, (b) manifold according to map A with $\alpha=1.0$.}
\end{figure}

Just as for the reference solutions in \cite{Ghia_1982a}, the results
are presented as velocity profiles along the horizontal and vertical
centerlines in $\Omega_{\mathrm{2D}}$. Fig.~\ref{fig:DrivCavVisResA}
shows the profiles for the velocity component $u$ along the vertical
centerline and $v$ along the horizontal centerline for the two maps
with different scaling factors $\alpha$ and $\beta$, respectively.
The crosses indicating the reference solution from \cite{Ghia_1982a}
are only relevant for the flat case where $\alpha=\beta=0$. The results
for the velocity component $w$ along the two centerlines are given
in Fig.~\ref{fig:DrivCavVisResB}. These results have the quality
of benchmark solutions and have been obtained with $k_{\vek u}=4$
and $100$ elements per dimensions. The convergence of other element
orders and mesh resolutions towards these profiles has been confirmed,
and a small selection is shown in Fig.~\ref{fig:DrivCavVisResC}.
Without stabilization, the typical oscillations are seen for this
rather high Reynolds number for coarse meshes with low order. As no
analytical solutions for the velocities and pressure are available,
it is impossible to provide convergence results in $\varepsilon_{\vek u}$
and $\varepsilon_{p}$. Furthermore, the singular pressure in the
upper left and right corners lead to singularities in the derivatives
of other physical fields. Thus, it cannot be expected that (optimal)
convergence in $\varepsilon_{\mathrm{mom}}$ and $\varepsilon_{\mathrm{cont}}$
is achieved.

\subsection{Flows on zero-level sets}

\begin{figure}
\centering

\subfigure[domain]{\includegraphics[width=0.35\textwidth]{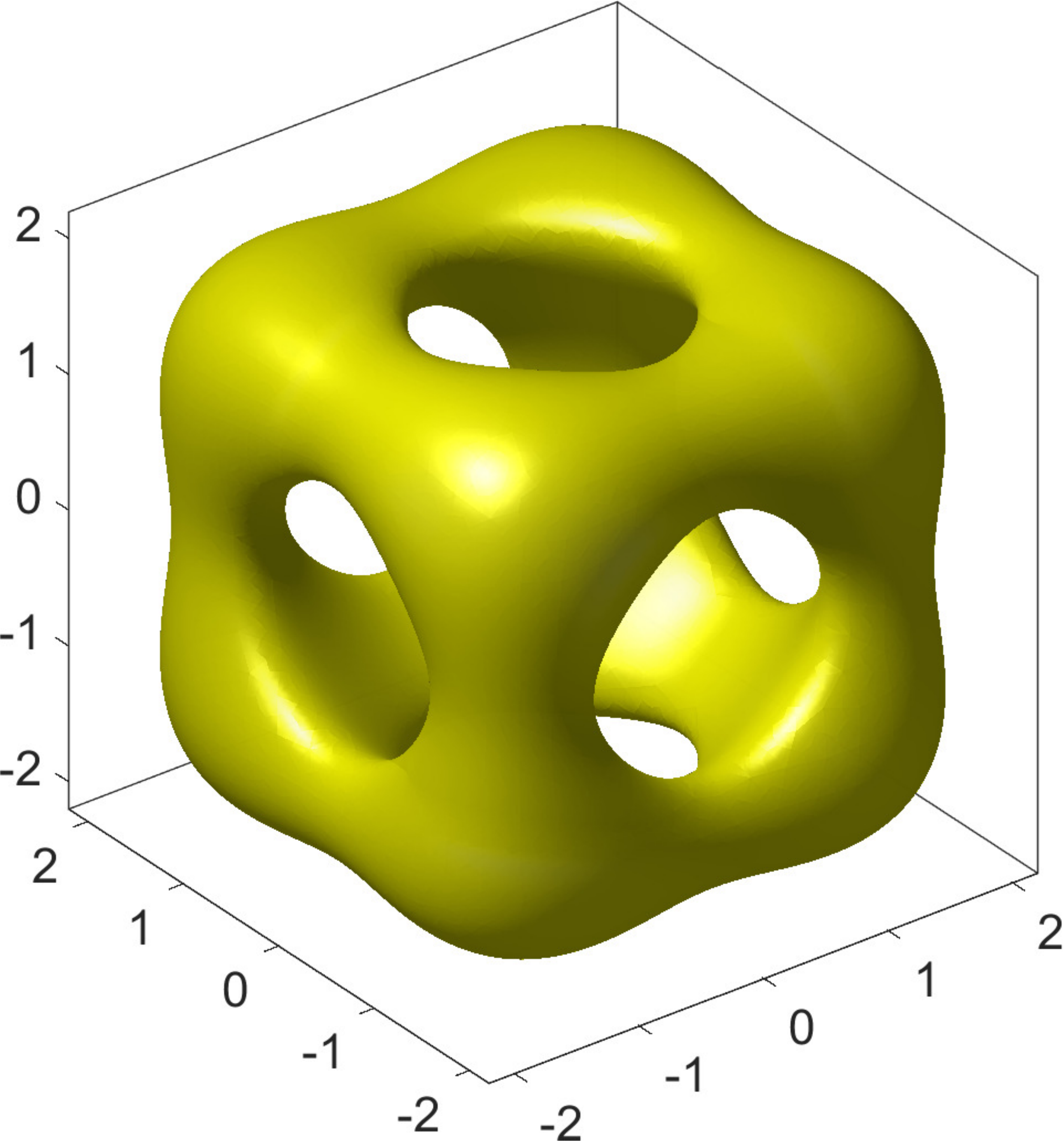}}\hfill\subfigure[mesh with $h=0.3$]{\includegraphics[width=0.25\textwidth]{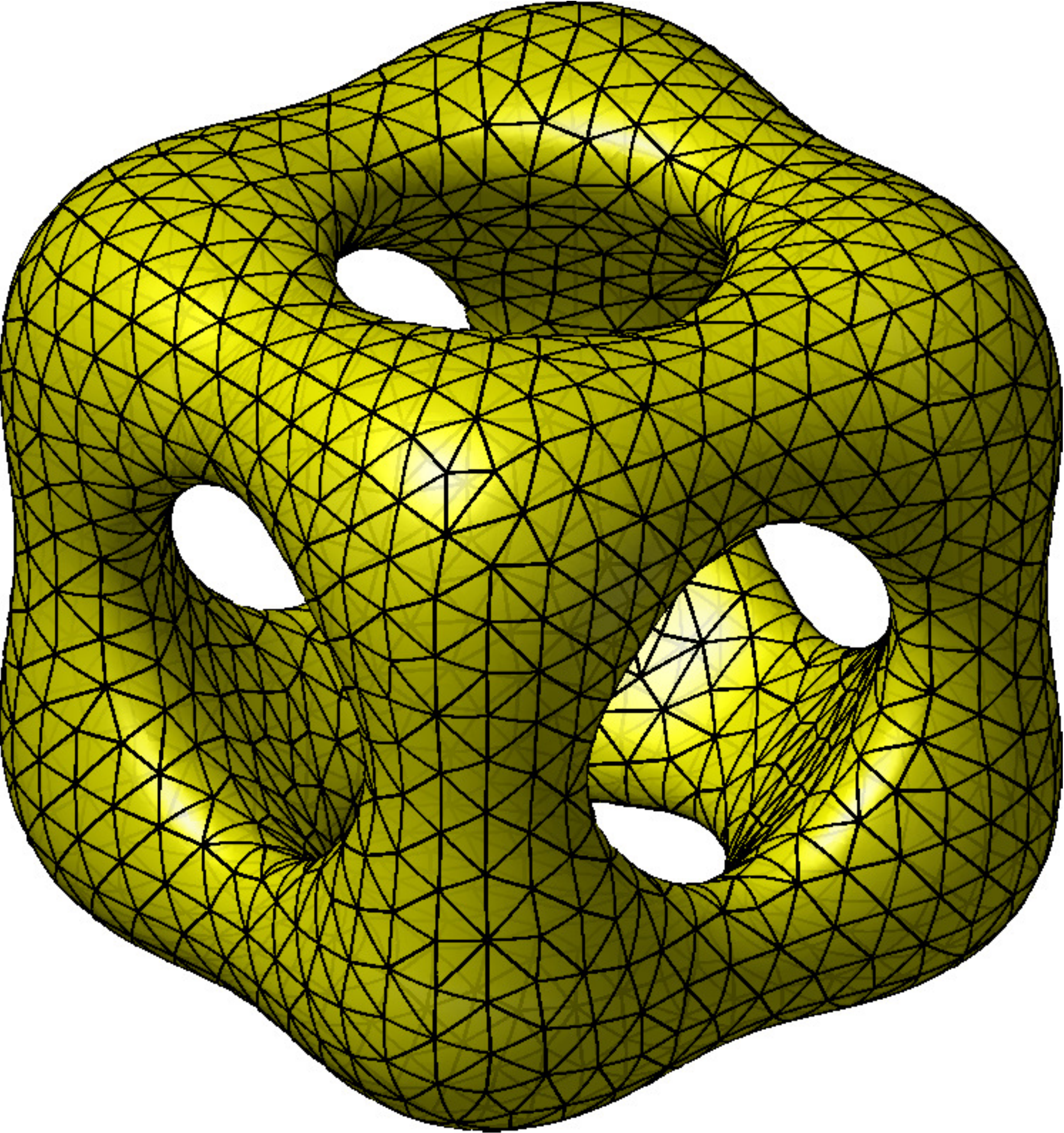}}\hfill\subfigure[mesh with $h=0.1$]{\includegraphics[width=0.25\textwidth]{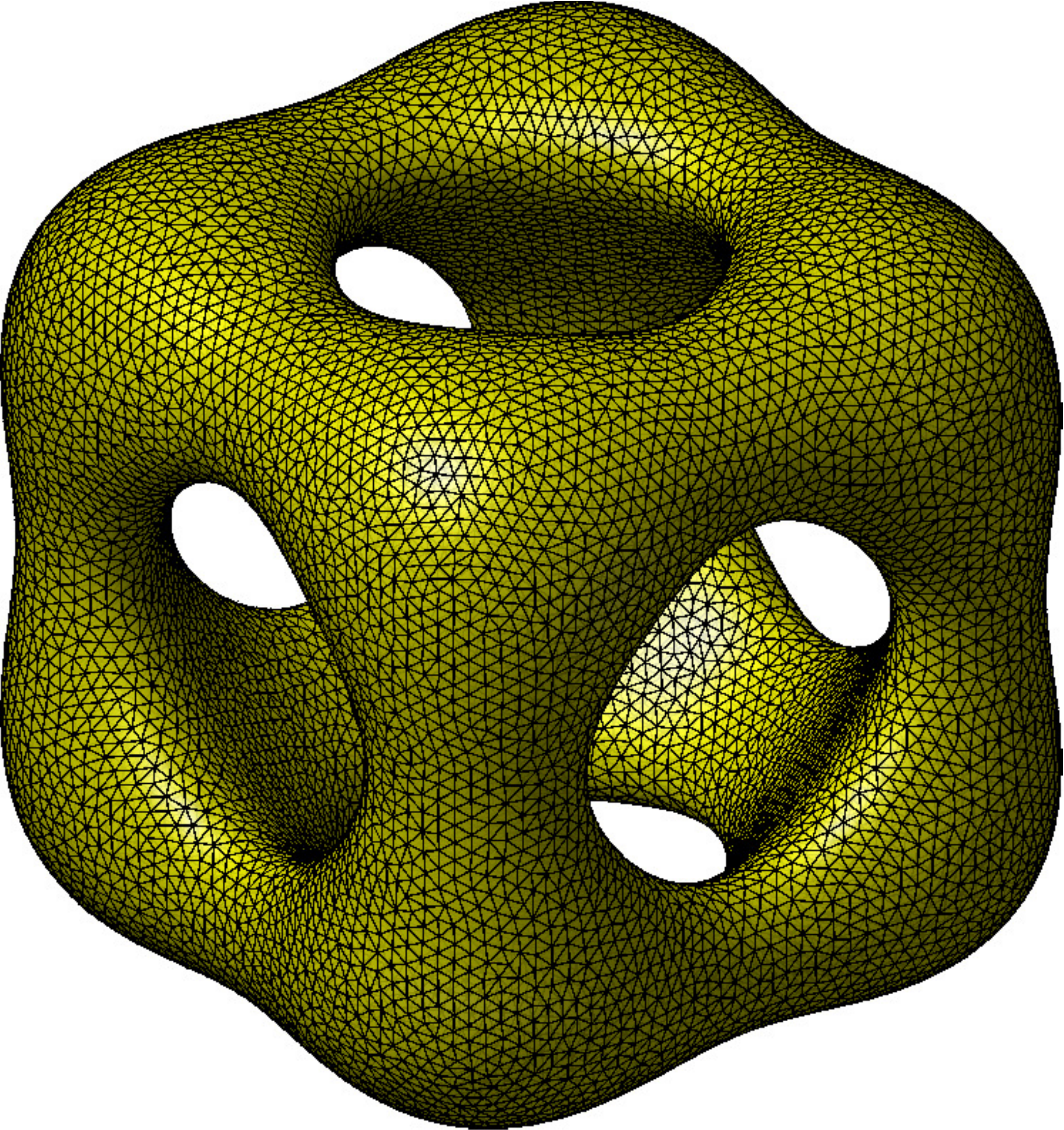}}

\caption{\label{fig:ZeroIsoSurfDomainAndMeshes}Manifold for the zero-isosurface
test case and meshes with different resolutions.}
\end{figure}

The next test case shows the potential to solve flows on zero-level
sets with the proposed models. Stationary Stokes and Navier-Stokes
flows are considered. The scalar function $\phi\left(\vek x\right):\mathbb{R}^{3}\rightarrow\mathbb{R}$
is based on \cite{Dziuk_2013a} and defined as
\begin{eqnarray*}
\phi\left(\vek x\right) & = & \left(x^{2}+y^{2}-4\right)^{2}+\left(x^{2}+z^{2}-4\right)^{2}+\left(y^{2}+z^{2}-4\right)^{2}+\\
 &  & \left(x^{2}-1\right)^{2}+\left(y^{2}-1\right)^{2}+\left(z^{2}-1\right)^{2}-15.
\end{eqnarray*}
The zero-isosurface of $\phi$ implies the compact manifold of interest,
$\Gamma=\left\{ \vek x:\phi\left(\vek x\right)=0\right\} $ and is
depicted in Fig.~\ref{fig:ZeroIsoSurfDomainAndMeshes}. In a first
step, meshes with linear triangular elements are generated using \texttt{distmesh}
\cite{Persson_2004a}. A scaling parameter $h$ may be chosen which
defines an average element length. In a second step, higher-order
elements are mapped to this linear surface mesh and their element
nodes are ``lifted'' \cite{Dziuk_2013a} such that they are on the
manifold $\Gamma$. Thereby, a higher-order accurate representation
$\Gamma^{h}$ is obtained.

\begin{figure}
\centering

\subfigure[$g_z(\vek x)$]{\includegraphics[width=0.35\textwidth]{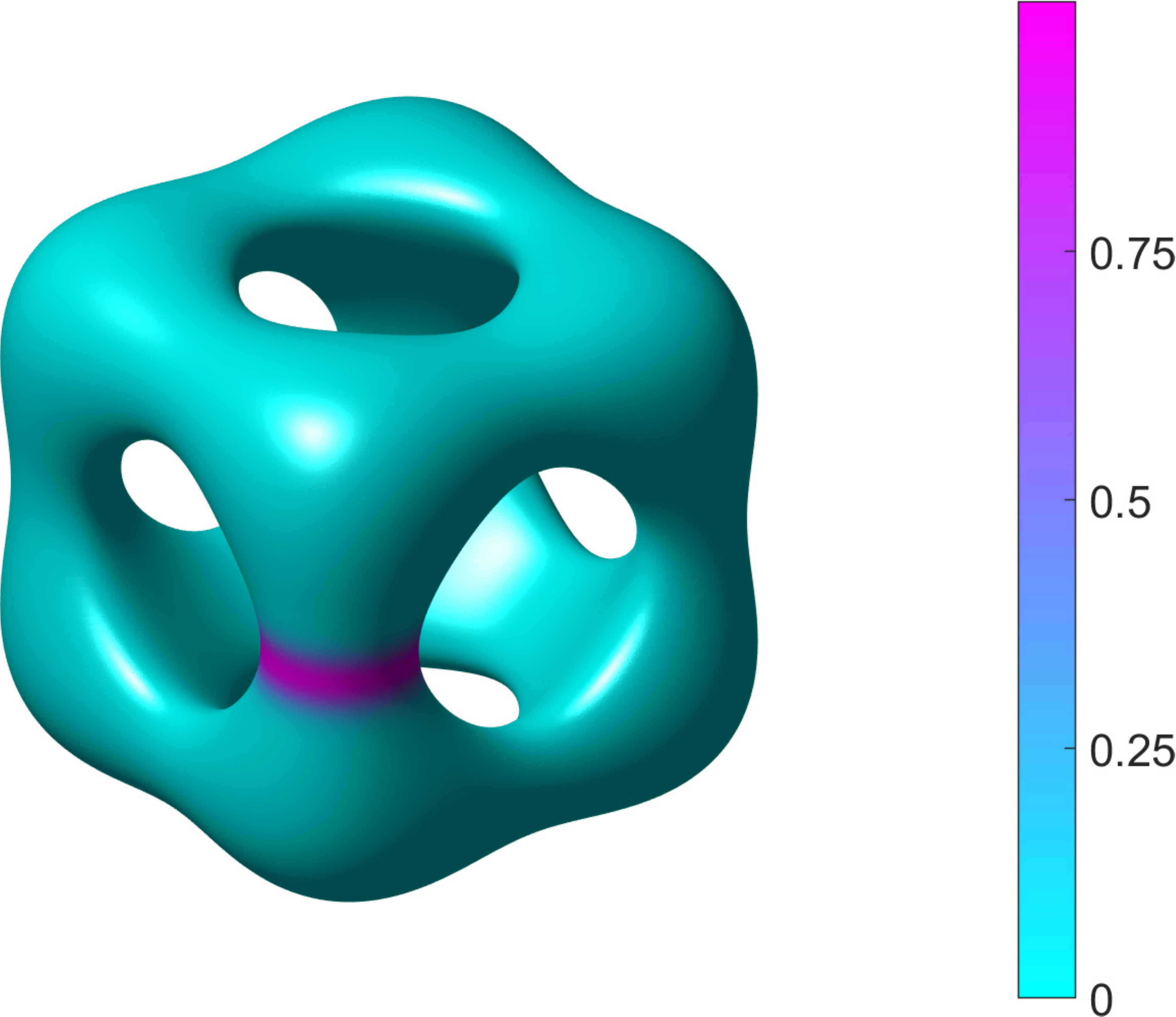}}\qquad\qquad\subfigure[$\left\Vert \vek u(\vek x)\right\Vert $]{\includegraphics[width=0.35\textwidth]{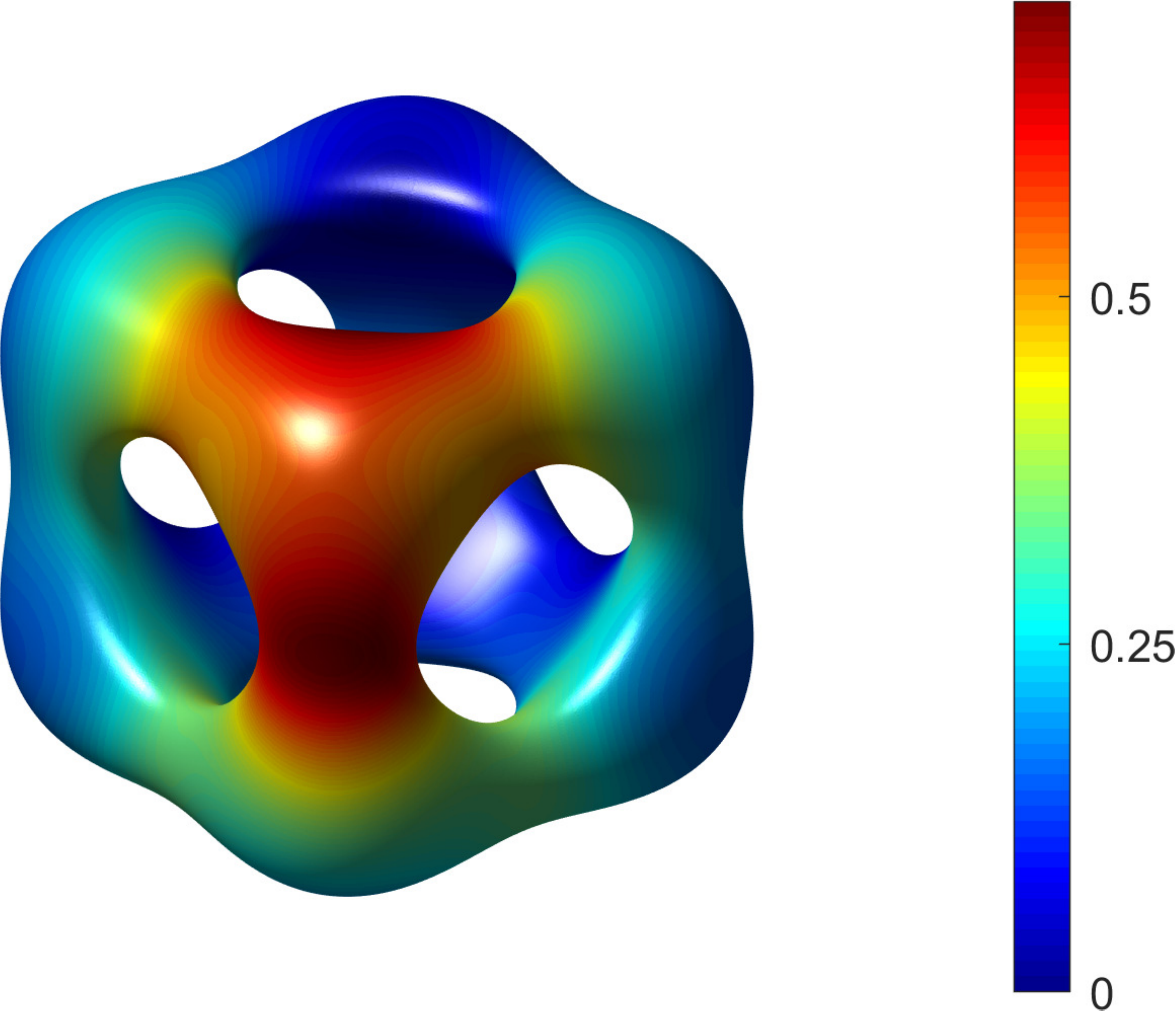}}

\subfigure[$p(\vek x)$]{\includegraphics[width=0.35\textwidth]{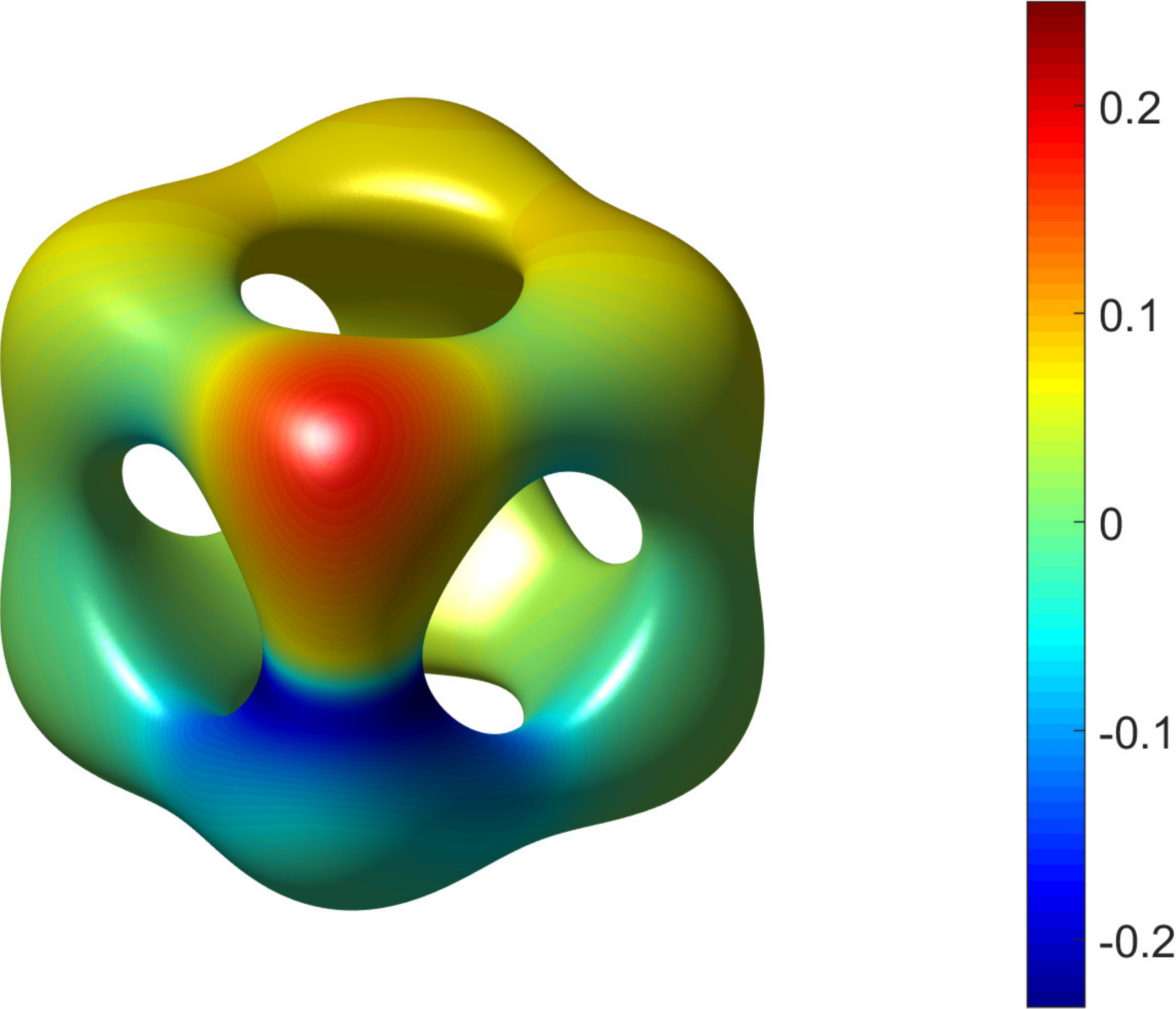}}\qquad\qquad\subfigure[$\omega^{\star}(\vek x)$]{\includegraphics[width=0.35\textwidth]{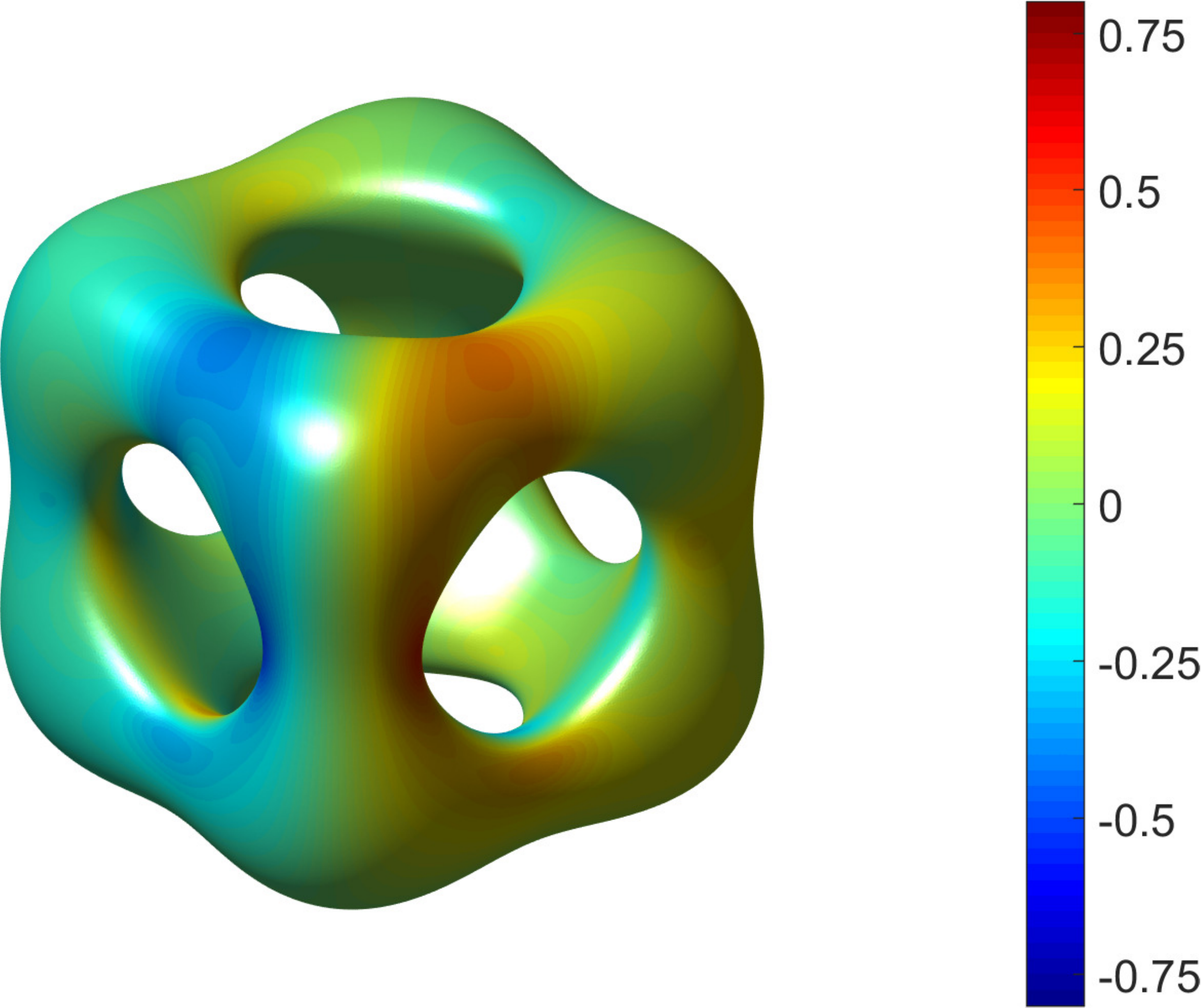}}

\caption{\label{fig:ZeroIsoSurfFields}Physical fields of the zero-isosurface
test case: (a) acceleration $g_{z}$, (b) velocity magnitude $\left\Vert \vek u\right\Vert $,
(c) pressure $p$, (d) vorticity $\omega^{\star}$.}
\end{figure}

As there are no boundaries present, an accelaration field in $z$-direction
drives the flow. That is, on the right hand side, $\vek g=\mat P\cdot\left[0,0,g_{z}\right]^{\mathrm{T}}$
where $g_{z}$ is determined by
\[
g_{z}\left(\vek x\right)=\begin{cases}
\exp\left(-\dfrac{z^{2}}{2\sigma_{0}^{2}}\right) & \text{with }\sigma_{0}=0.15\text{ for }x<0\text{ and }y<0,\\
0 & \text{else,}
\end{cases}
\]
and visualized in Fig.~\ref{fig:ZeroIsoSurfFields}(a). It is virtually
non-zero only for the left front ``pillar'' of the domain. The density
is $\varrho=1$ and the viscosity is $\mu=0.05$. For the case of
stationary Navier-Stokes flow, the corresponding velocity magnitude,
pressure fields and vorticity $\omega^{\star}$ according to Eq.~(\ref{eq:SpecialVorticity})
are seen in Figs.~\ref{fig:ZeroIsoSurfFields}(b) to (d), respectively.

\begin{figure}
\centering

\subfigure[$\varepsilon_{\mathrm{mom}}$]{\includegraphics[width=0.45\textwidth]{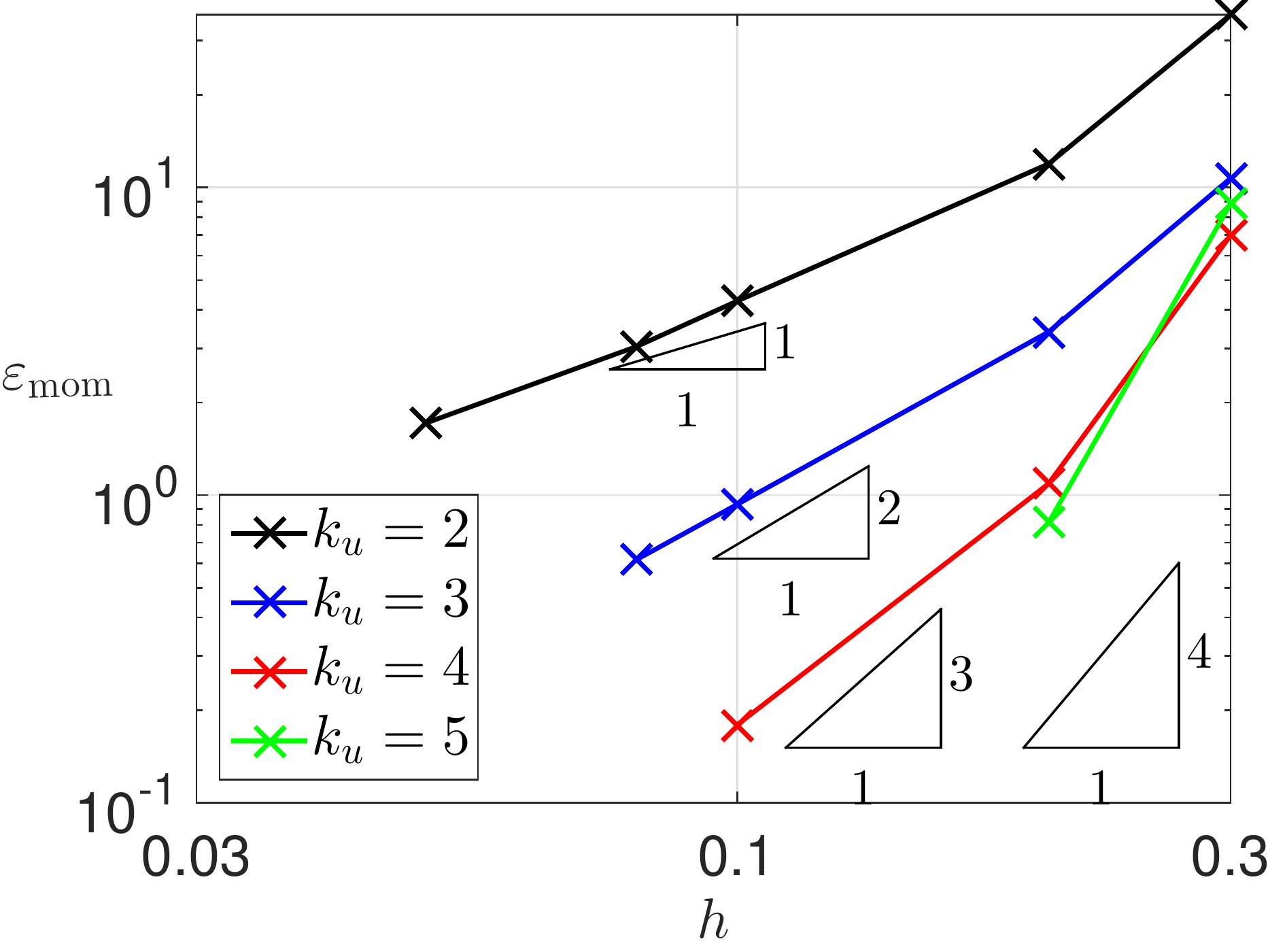}}\qquad\subfigure[$\varepsilon_{\mathrm{cont}}$]{\includegraphics[width=0.45\textwidth]{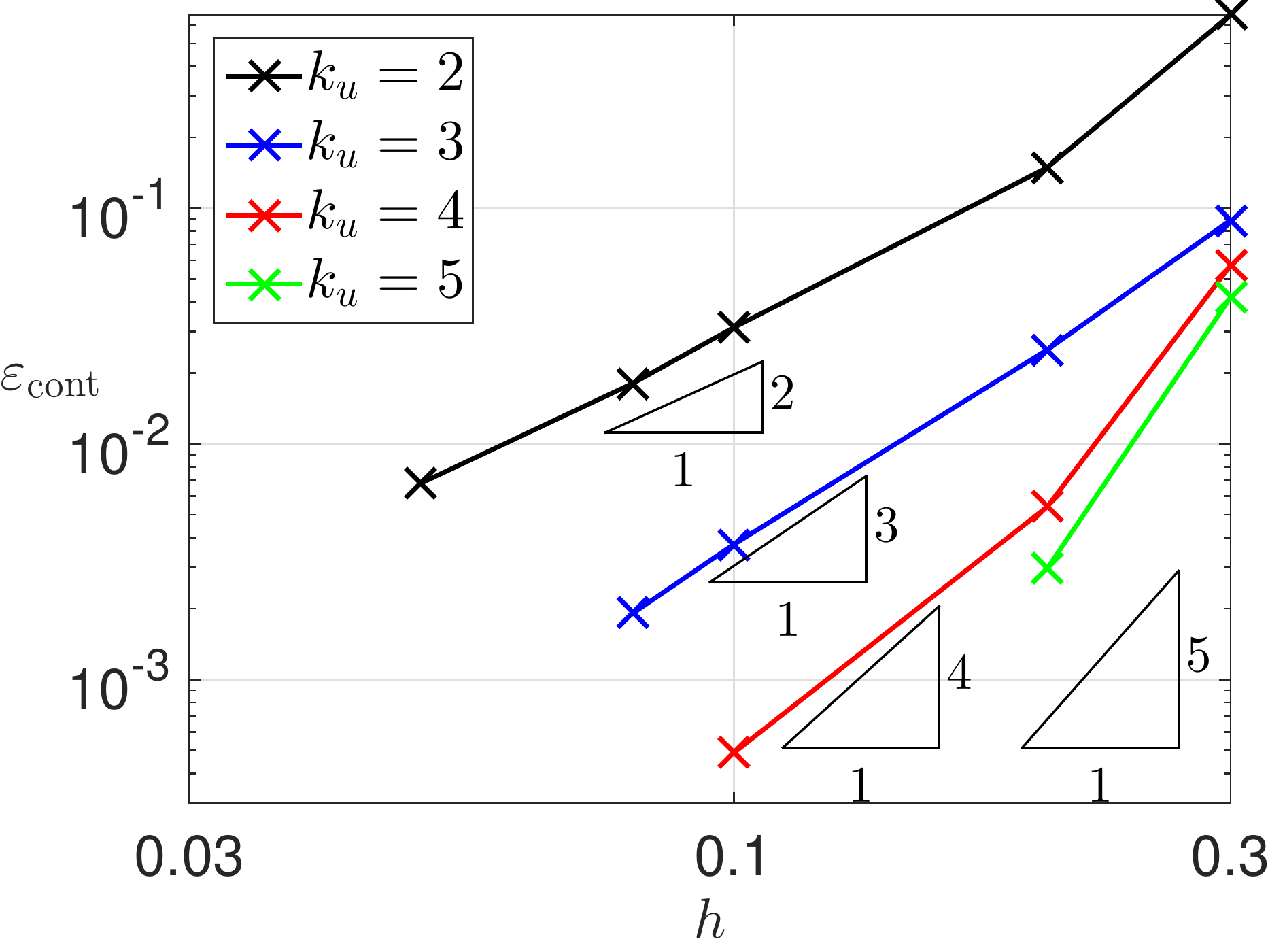}}

\caption{\label{fig:ZeroIsoSurfResNewA}Convergence results in (a) $\varepsilon_{\mathrm{mom}}$
and (b) $\varepsilon_{\mathrm{cont}}$ for the zero-isosurface test
case.}
\end{figure}

\begin{figure}
\centering

\subfigure[$w(\vek x)$]{\includegraphics[height=5cm]{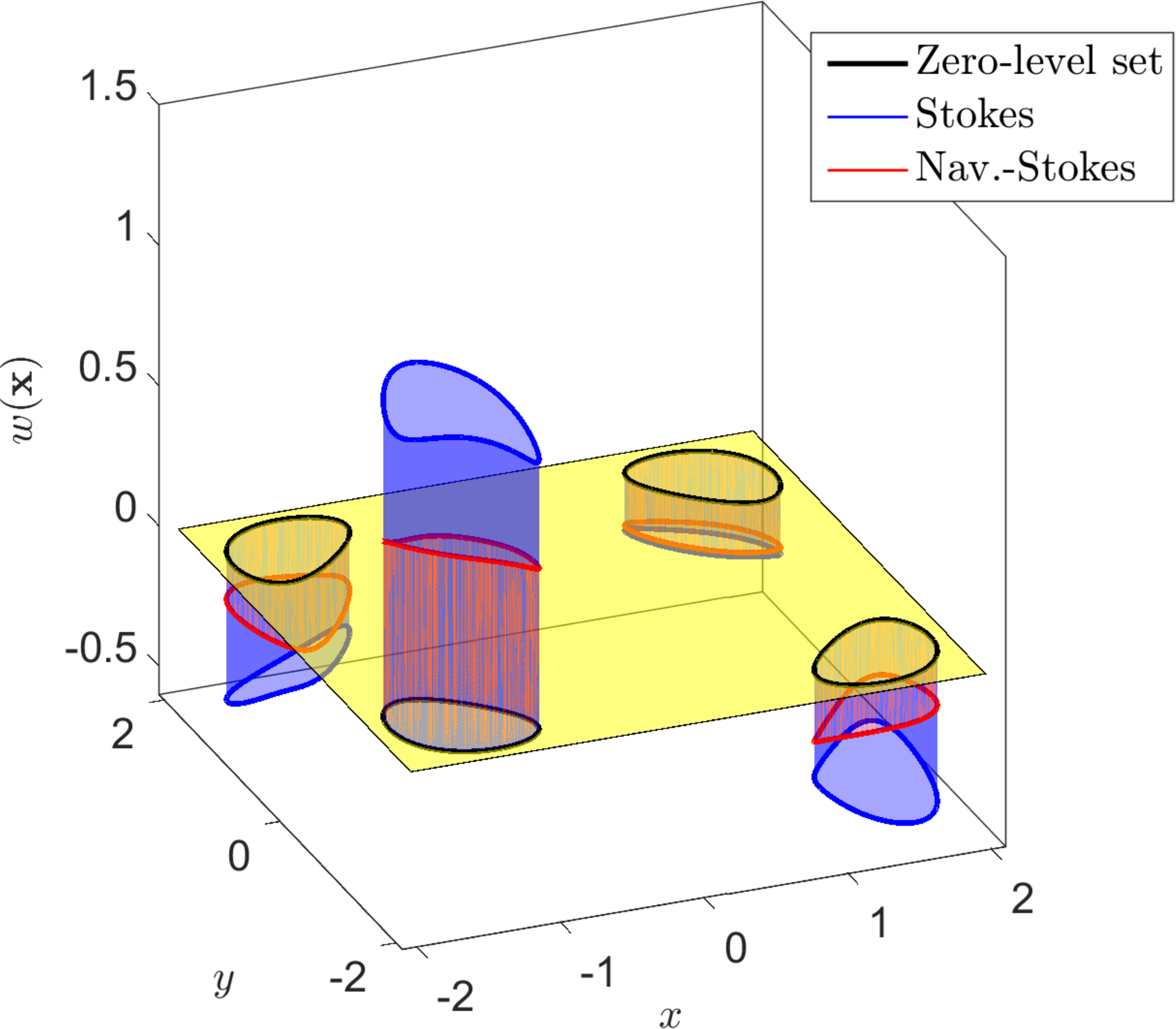}}\qquad\subfigure[$w(\vek x)$]{\includegraphics[height=5cm]{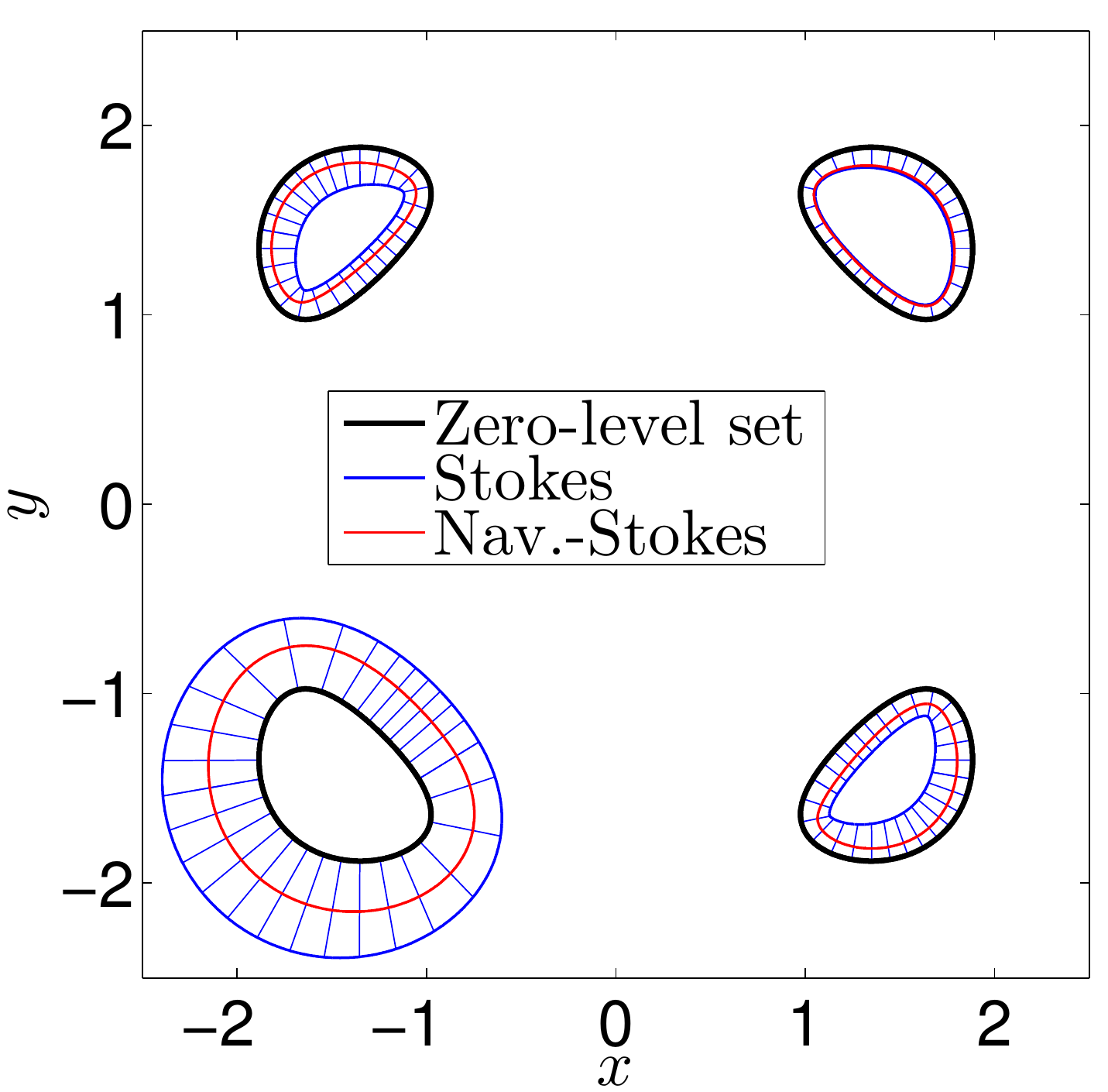}}

\caption{\label{fig:ZeroIsoSurfRes}Profiles for the vertical velocity $w\left(\vek x\right)$
in the plane with $z=0$, the scaling factor in (b) is $0.4$.}
\end{figure}

In the numerical studies, $2\leq k_{\vek u}\leq5$, $k_{p}=k_{\lambda}=k_{\vek u}-1$,
and $k_{\mathrm{geom}}=k_{\vek u}+1$ are used. As there is no analytical
solution available, convergence results are only shown in $\varepsilon_{\mathrm{mom}}$
and $\varepsilon_{\mathrm{cont}}$ in Fig.~\ref{fig:ZeroIsoSurfResNewA}.
Higher-order rates are clearly achieved. In order to make the solution
more quantitative, the velocity profiles for $w\left(\vek x\right)$
in the horizontal $xy$-plane (at $z=0$) are shown in Fig.~\ref{fig:ZeroIsoSurfRes}.
The four closed black lines represent the intersection of the plane
with the vertical ``pillars'' of the zero-isosurface. Fig.~\ref{fig:ZeroIsoSurfRes}(a)
shows $w(\vek x)$ as a third dimension, and (b) shows the same result
where $w(\vek x)$ is plotted in normal direction of the plane-pillar
intersections with a scaling factor of $0.4$. A clear convergence
to these profiles was observed when using meshes with different resolutions
and orders.

\subsection{Cylinder flows}

\begin{figure}
\centering

\subfigure[flat]{\includegraphics[width=0.3\textwidth]{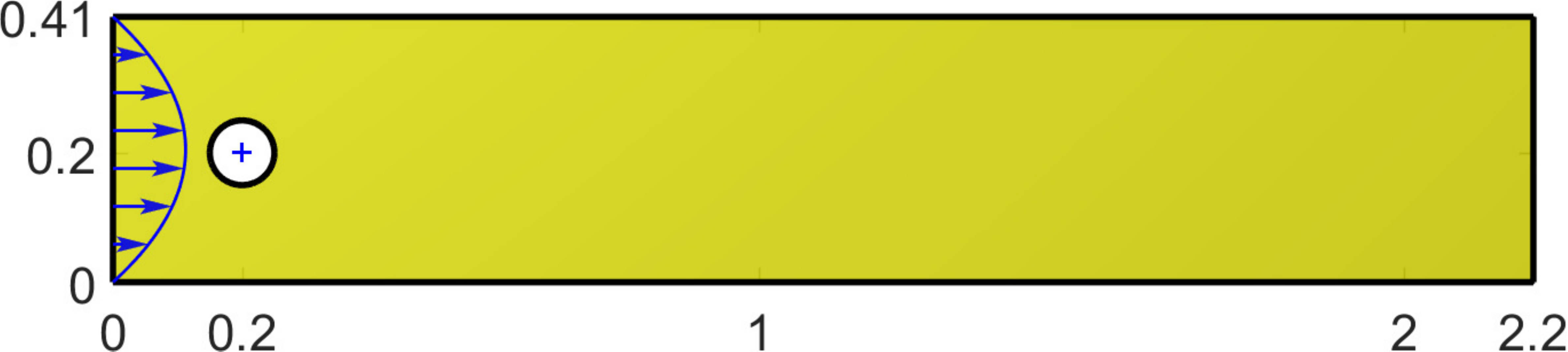}}\hfill\subfigure[map A]{\includegraphics[width=0.3\textwidth]{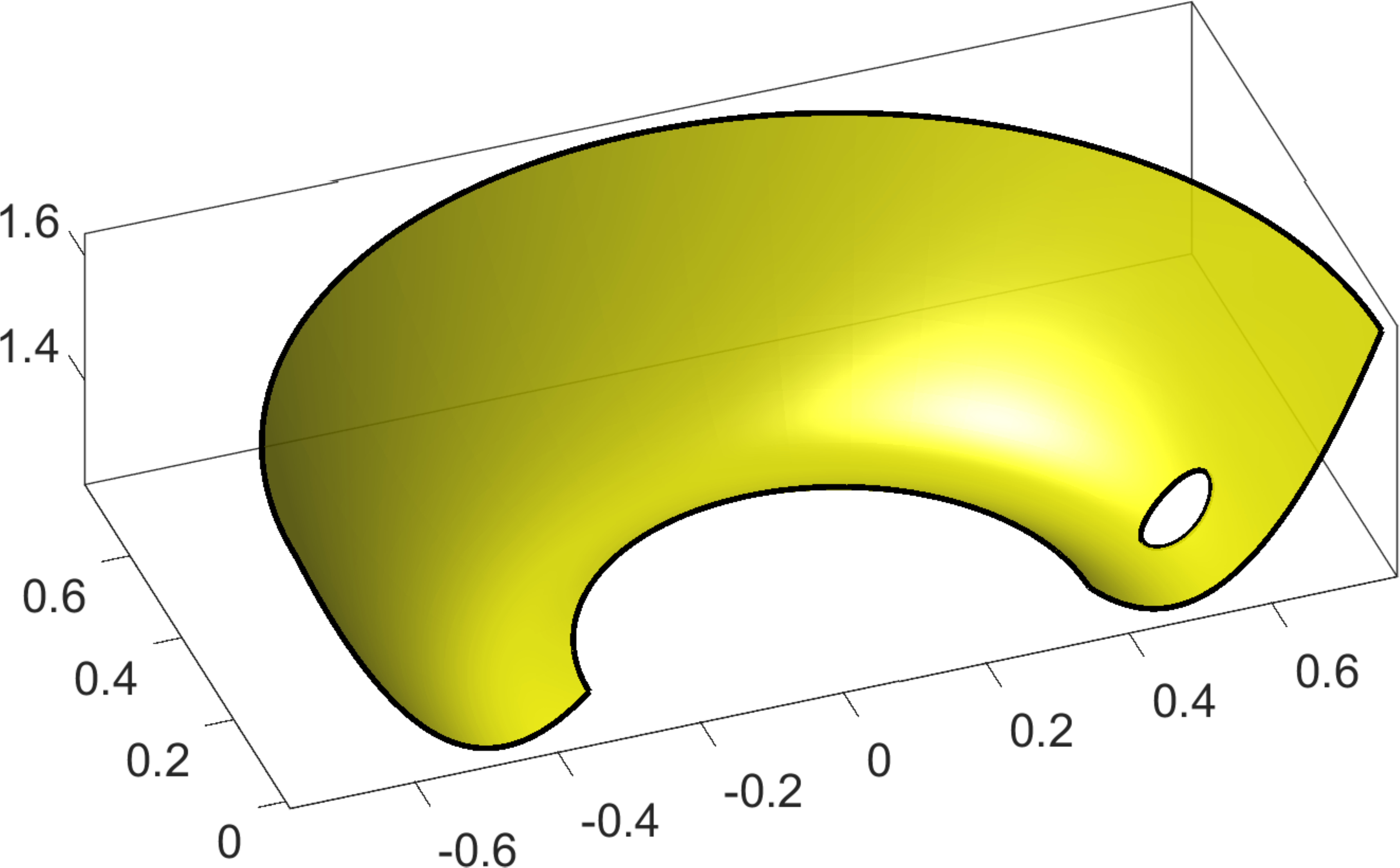}}\hfill\subfigure[map B]{\includegraphics[width=0.3\textwidth]{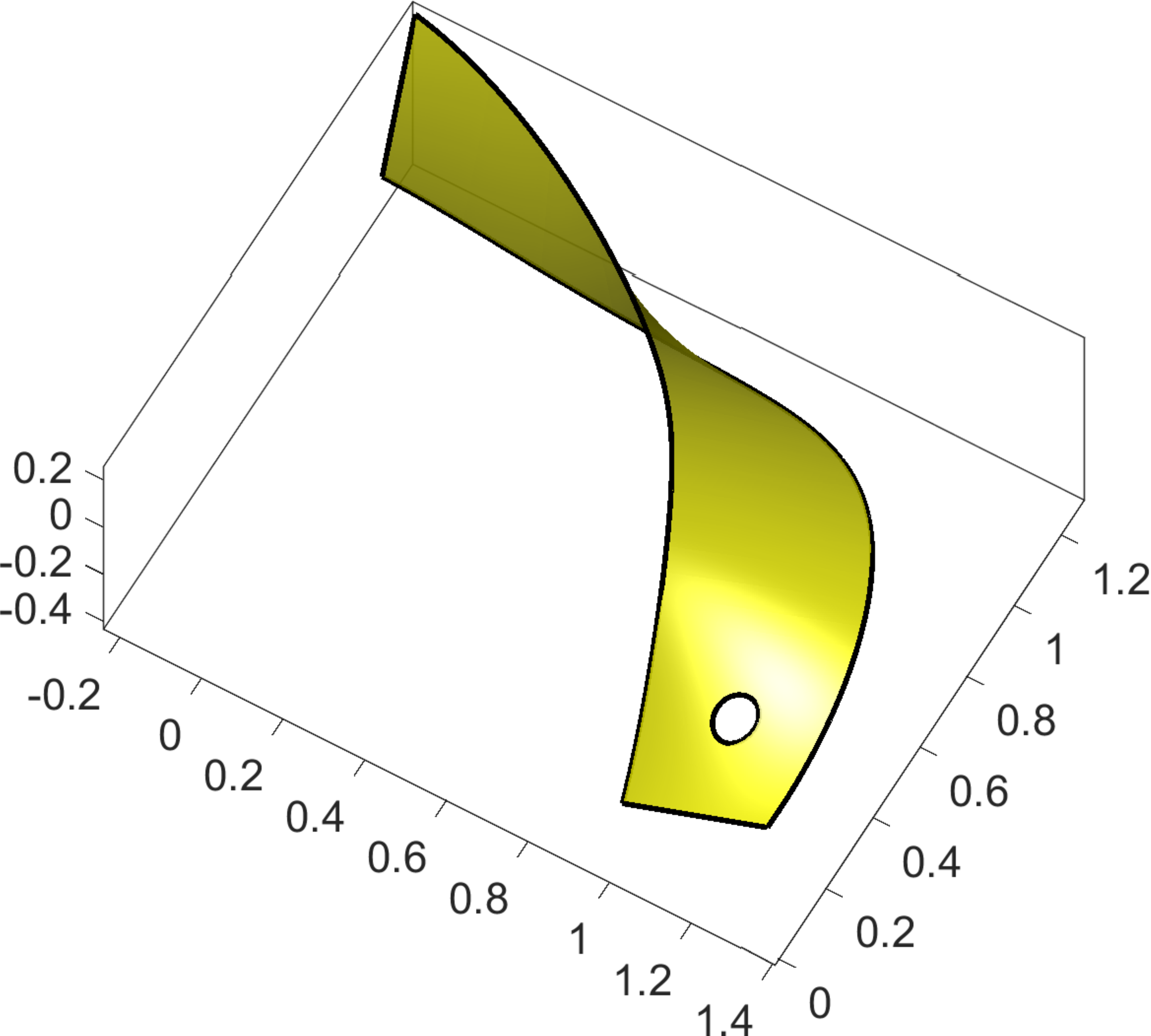}}

\caption{\label{fig:TurekDomain}Manifolds for the cylinder flow test case.}
\end{figure}

As an example for the instationary Navier-Stokes equations, the following
test case is based on a channel flow around a cylinder according to
\cite{Schaefer_1996a}. The geometry is first described in 2D, labelled
$\Omega_{\mathrm{2D}}$, and later on mapped to obtain curved surfaces
in 3D. In 2D, the cylinder with a diameter of $0.1$ is placed slightly
unsymmetrically in $y$-direction of the channel in $\left[0,2.20\right]\times\left[0,0.41\right]$,
see Fig.~\ref{fig:TurekDomain}(a). No-slip boundary conditions are
applied on the upper and lower wall and on the cylinder surface. A
quadratic velocity profile for $u$, with $u_{\mathrm{max}}=1.5$,
and $v=0$ is applied at the inflow on the left side of the domain.
At the outflow, traction-free boundary conditions are used. The density
and viscosity are prescribed as $\varrho=1.0$ and $\mu=0.001$. This
results in a Reynolds number of $\textrm{Re}=\varrho\cdot u_{m}\cdot L/\mu=100$
when taking the cylinder diameter as a length scale $L$ and the average
inflow velocity $u_{m}=1.0$ at the inflow. At this Reynolds number,
periodic flow patterns known as the K\'arm\'an vortex street are
observed behind the cylinder. Reference solutions are given for the
lift and drag coefficients $c_{L}$ and $c_{D}$ of the cylinder \cite{Schaefer_1996a}
and the current implementation confirms these numbers for the flat
case (i.e., in 2D or when the flat 2D domain is transformed by a rigid
body motion to 3D). The reference Strouhal number $\mathrm{St}=D/\left(u_{m}T\right)$,
with the diameter $D=0.1$ of the cylinder, and the time $T$ for
$2$ periods of the curve of $c_{D}$, is given as $0.295\leq\mathrm{St}\leq0.305$,
resulting in a frequency of about $f=3.33\,\nicefrac{1}{\mathrm{s}}$.

The 2D domain is mapped to three dimensions using two different maps.
Assume that the coordinates of the 2D domain $\Omega_{\mathrm{2D}}$,
as seen in Fig.~\ref{fig:TurekDomain}(a), are given in coordinates
$(a,b)$. Map A, $\vek x\left(\vek a\right):\mathbb{R}^{2}\rightarrow\mathbb{R}^{3}$,
is defined as
\begin{eqnarray*}
x(\vek a) & = & \cos\left(\frac{\pi\cdot a}{2.2}\right)\cdot(b+0.35),\\
y(\vek a) & = & \sin\left(\frac{\pi\cdot a}{2.2}\right)\cdot(b+0.35),\\
z(x(\vek a),y(\vek a)) & = & 2+\nicefrac{1}{2}\sqrt{x^{2}+y^{2}}-\sin\left(3\sqrt{x^{2}+y^{2}}\right).
\end{eqnarray*}
For map B, we first define an intermediate mapping $\vek r\left(\vek a\right):\mathbb{R}^{2}\rightarrow\mathbb{R}^{3}$
applying some twist to the domain,
\begin{eqnarray*}
r\left(\vek a\right) & = & a,\\
s\left(\vek a\right) & = & -\left(1+q\left(a\right)\right)\cdot\left(b-0.205\right)\cdot\cos\left(\nicefrac{\pi}{6}\left(1-\nicefrac{25}{11}\cdot a\right)\right),\\
t\left(\vek a\right) & = & -\left(1+q\left(a\right)\right)\cdot\left(b-0.205\right)\cdot\sin\left(\nicefrac{\pi}{6}\left(1-\nicefrac{25}{11}\cdot a\right)\right).
\end{eqnarray*}
with $q(a)=-0.2/2.42\cdot a^{2}+0.44/2.42\cdot a$. This is further
mapped by $\vek x\left(\vek r\right):\mathbb{R}^{3}\rightarrow\mathbb{R}^{3}$
defined as
\begin{eqnarray*}
x\left(\vek r\right) & = & \cos\left(\nicefrac{50}{198}\cdot\pi\cdot r\right)\cdot\left(s+\nicefrac{6}{5}\right),\\
y\left(\vek r\right) & = & \sin\left(\nicefrac{50}{198}\cdot\pi\cdot r\right)\cdot\left(s+\nicefrac{6}{5}\right),\\
z\left(\vek r\right) & = & t+\nicefrac{1}{5}\sin\left(3r\right).
\end{eqnarray*}
The resulting curved manifolds according to map A and B are visualized
in Figs.~\ref{fig:TurekDomain}(b) and (c), respectively. Note that
also the inflow velocities are mapped accordingly based on the Jacobians
of the respective mappings to ensure that they are in the tangent
space at $\partial\Gamma_{\mathrm{D}}$.

\begin{figure}
\centering

\subfigure[$\left\Vert \vek u(\vek x)\right\Vert $]{\includegraphics[width=0.3\textwidth]{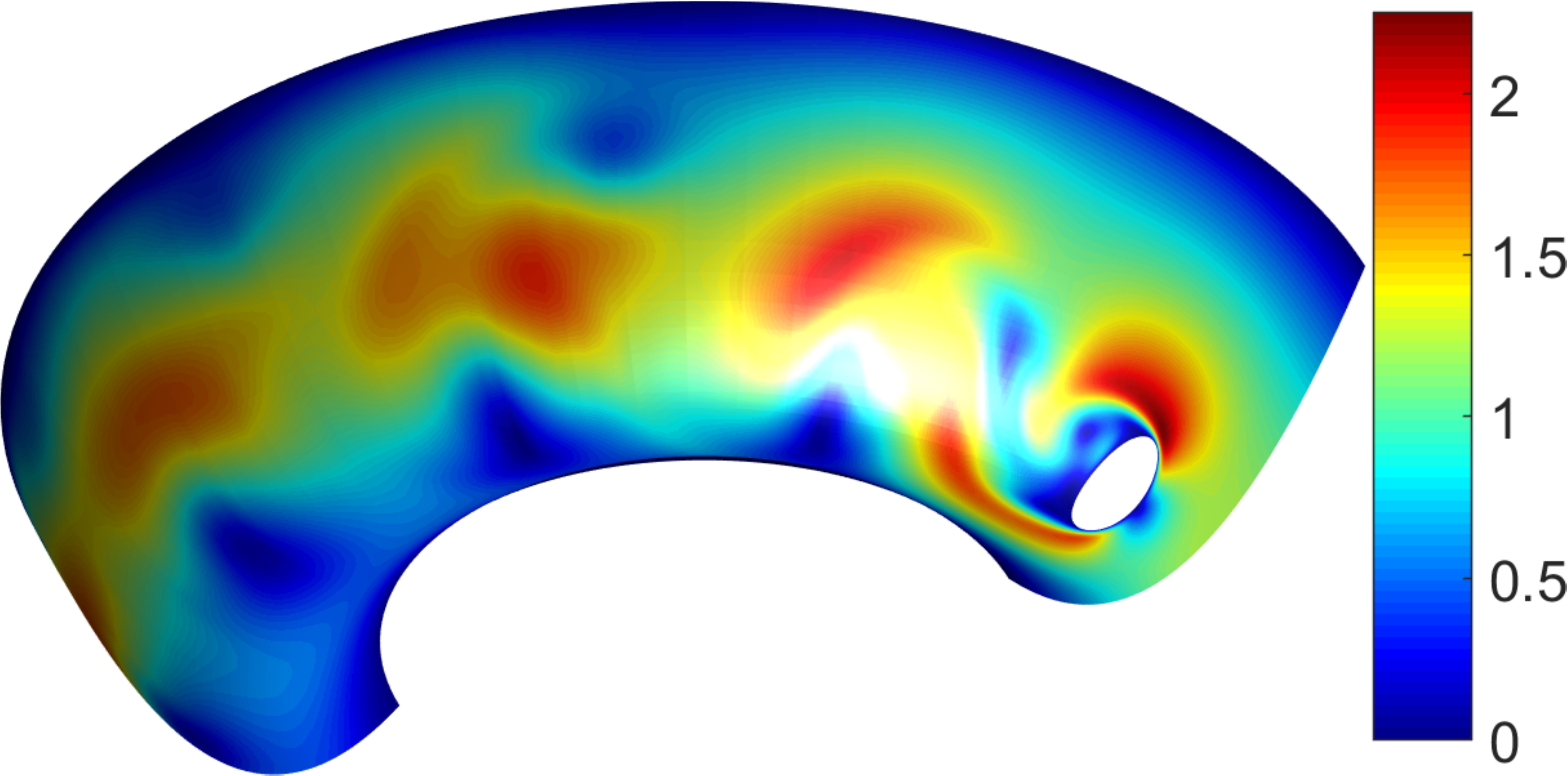}}\hfill\subfigure[$p(\vek x)$]{\includegraphics[width=0.3\textwidth]{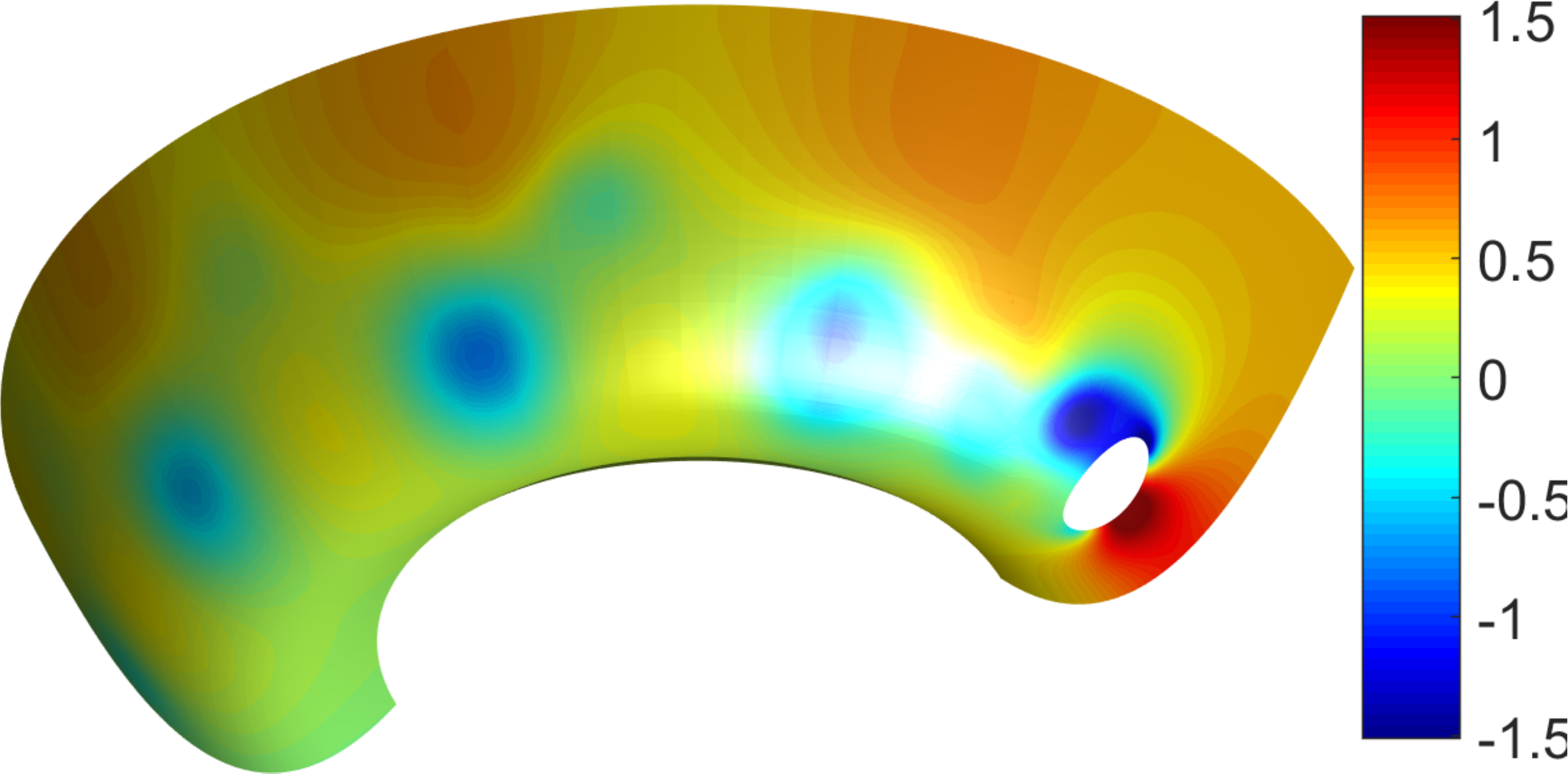}}\hfill\subfigure[$\omega^{\star}(\vek x)$]{\includegraphics[width=0.3\textwidth]{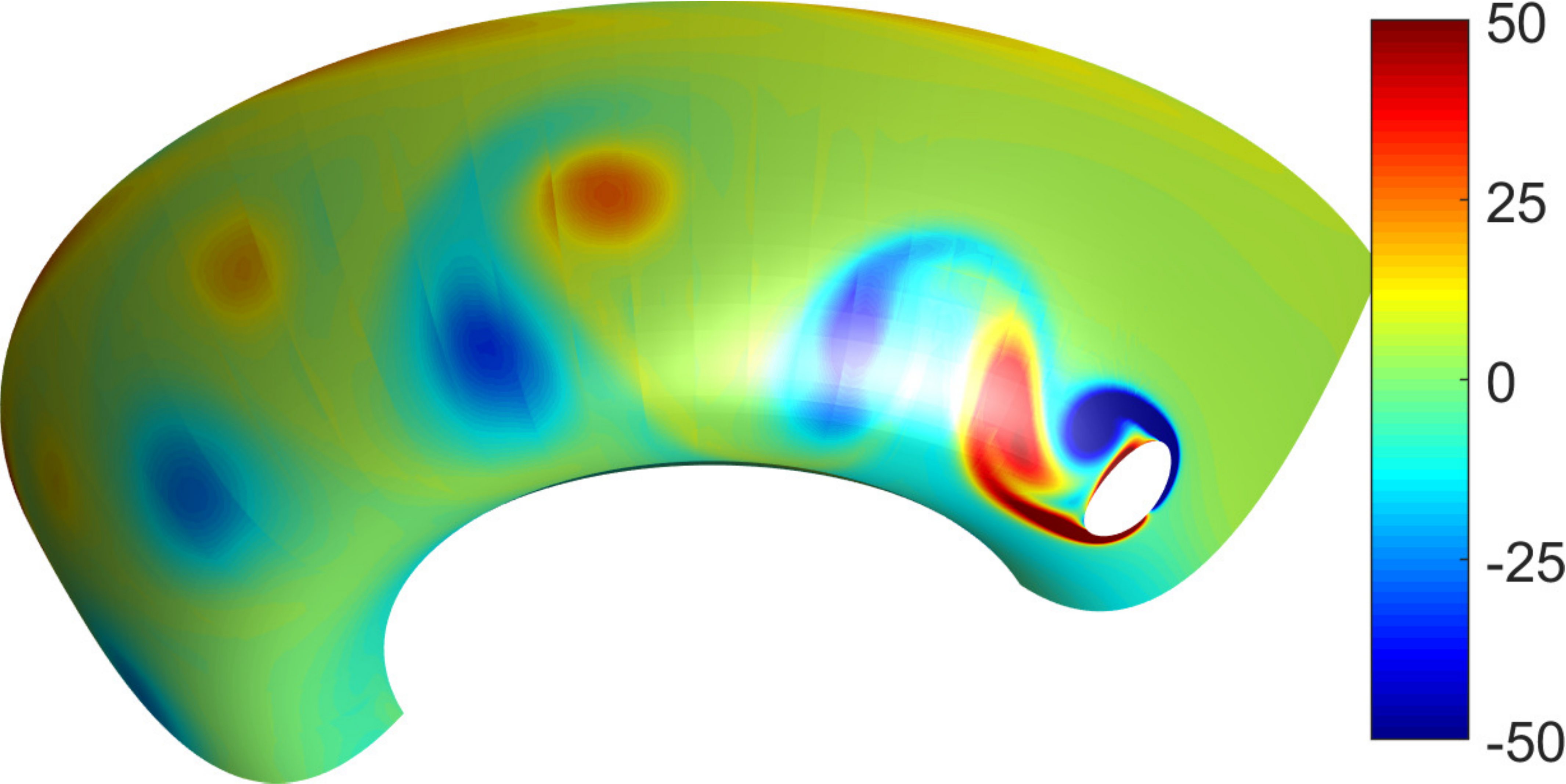}}

\caption{\label{fig:TurekFieldsA}Physical fields for the cylinder flow test
case according to map A: (a) velocity magnitude $\left\Vert \vek u\right\Vert $,
(b) pressure $p$, (c) vorticity $\omega^{\star}$.}
\end{figure}

\begin{figure}
\centering

\subfigure[$\left\Vert \vek u(\vek x)\right\Vert $]{\includegraphics[width=0.3\textwidth]{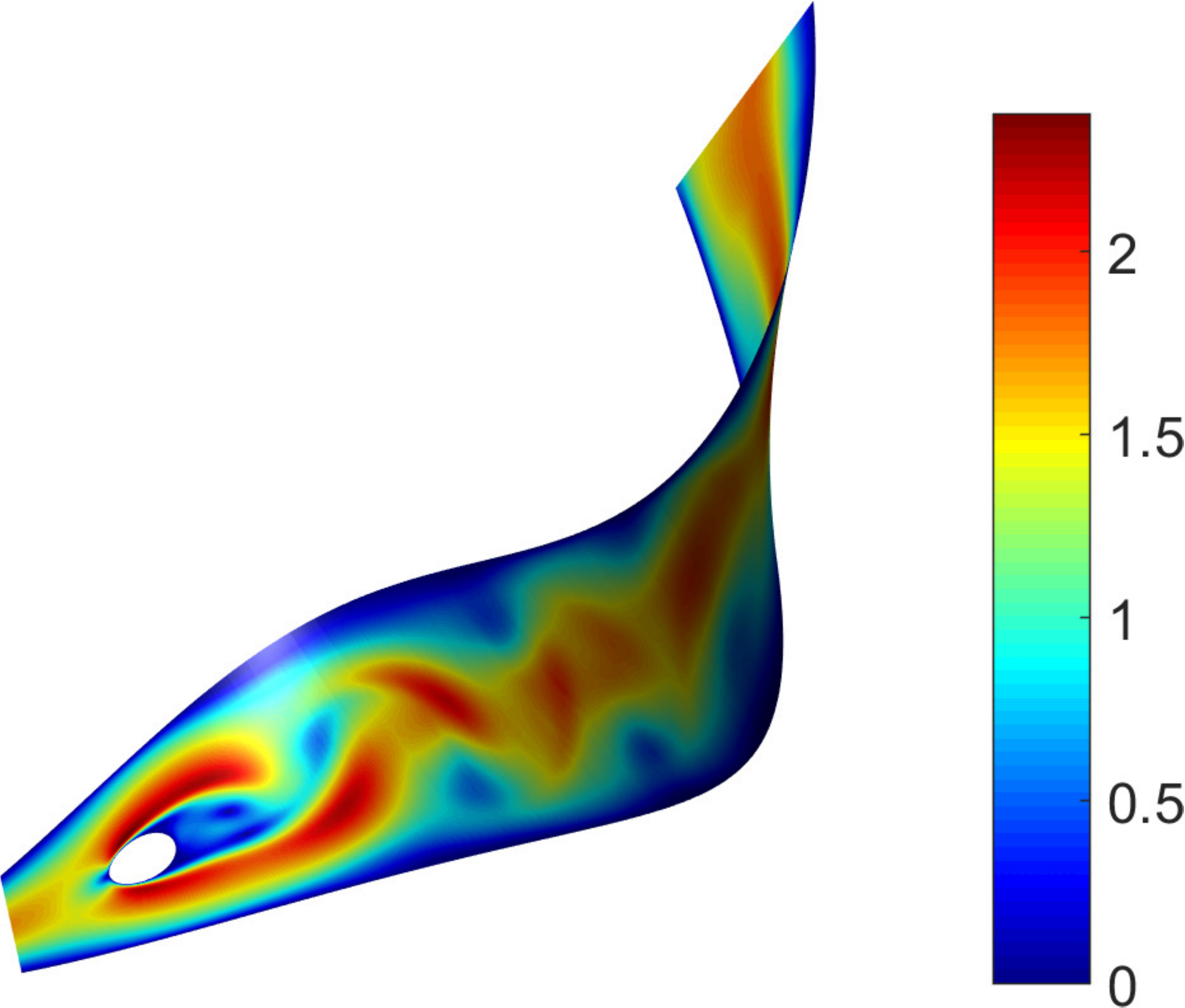}}\hfill\subfigure[$p(\vek x)$]{\includegraphics[width=0.3\textwidth]{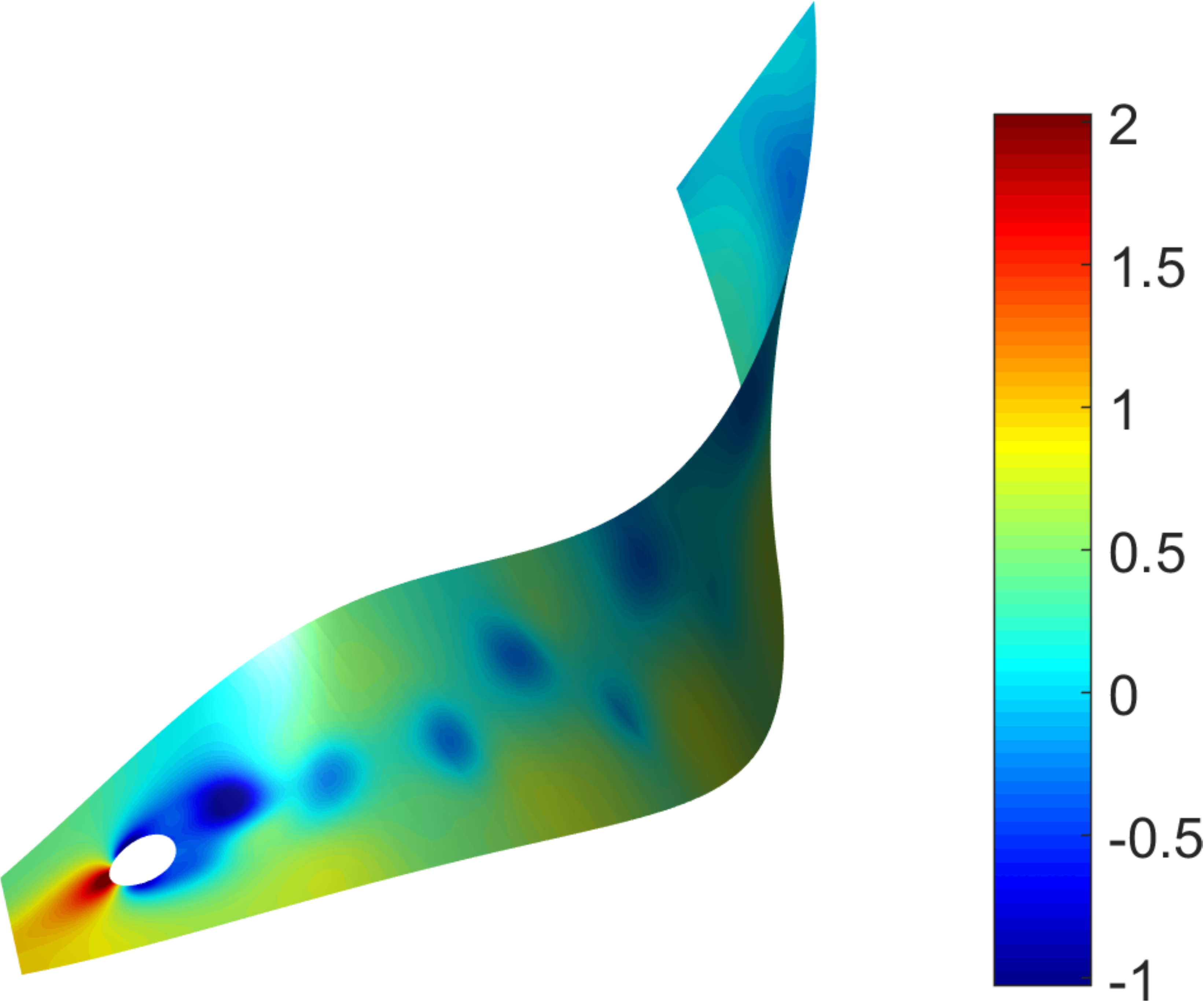}}\hfill\subfigure[$\omega^{\star}(\vek x)$]{\includegraphics[width=0.3\textwidth]{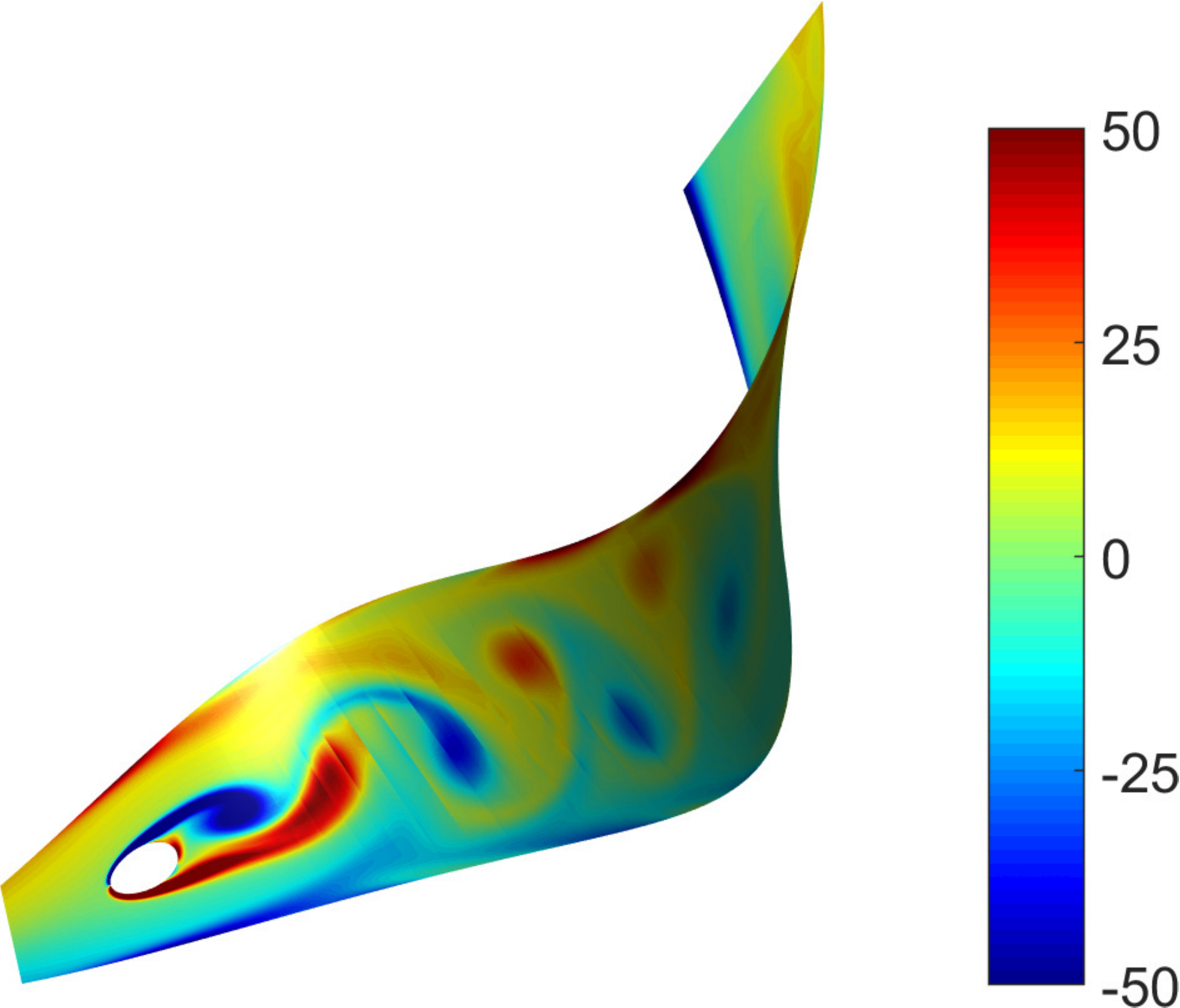}}

\caption{\label{fig:TurekFieldsB}Physical fields for the cylinder flow test
case according to map B: (a) velocity magnitude $\left\Vert \vek u\right\Vert $,
(b) pressure $p$, (c) vorticity $\omega^{\star}$.}
\end{figure}

The initial condition on the manifolds is $\vek u_{0}\left(\vek x\right)=\vek0$.
The observed time interval is $\tau=\left[0,6\right]$ and the inflow
velocities are ramped by a cubic function in time,
\[
R\left(t\right)=\begin{cases}
-2\cdot\left(\nicefrac{t}{t^{\star}}\right)^{3}+3\cdot\left(\nicefrac{t}{t^{\star}}\right)^{2} & \quad\text{for }t\leq t^{\star},\\
1 & \quad\text{else},
\end{cases}
\]
with $t^{\star}=0.96$. That is, after $t^{\star}$, the full velocity
profile is active at the inflow. Figs.~\ref{fig:TurekFieldsA} and
\ref{fig:TurekFieldsB} show the velocity magnitude, pressure field,
and vorticity $\omega^{\star}$ at time $t=6$ for the two mappings.
The expected vortex shedding can be clearly seen.

\begin{figure}
\centering

\subfigure[coarse, $972$ elements]{\includegraphics[width=0.45\textwidth]{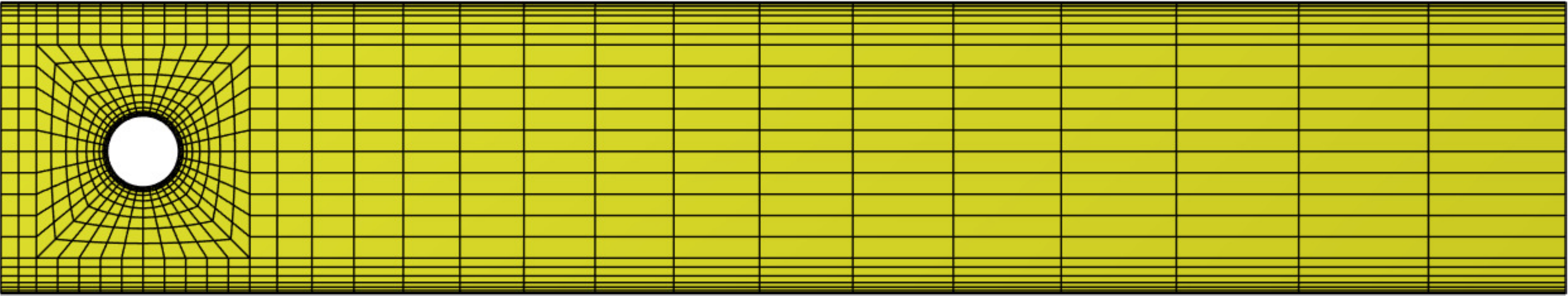}}\hfill\subfigure[fine, $1920$ elements]{\includegraphics[width=0.45\textwidth]{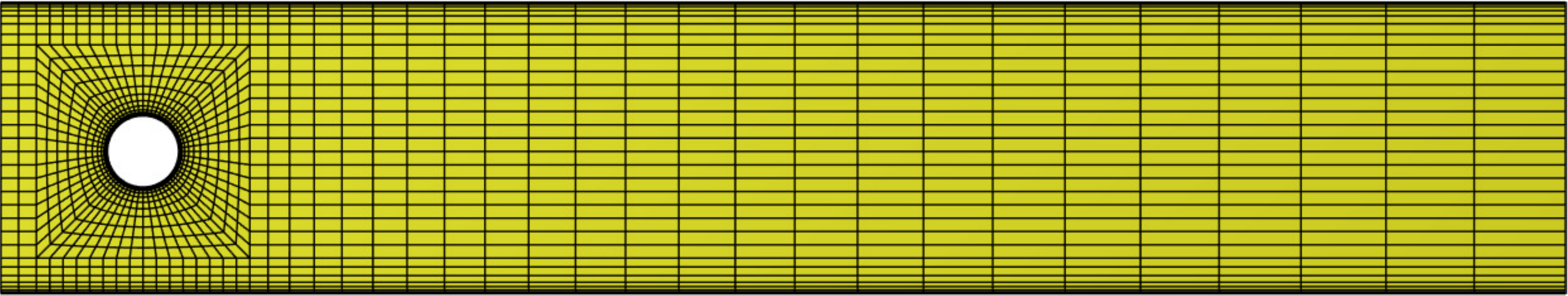}}

\caption{\label{fig:TurekMeshes}Coarse and fine mesh for the cylinder flow
test case; these meshes are later mapped according to map A or B.}
\end{figure}

Two different meshes with $972$ and $1920$ elements each are used
which are refined at the no-slip boundaries to resolve the boundary
layers. They are visualized for $\Omega_{\mathrm{2D}}$ in Fig.~\ref{fig:TurekMeshes}
and mapped to the manifolds accordingly. We use element orders of
$k_{\mathrm{geom}}=4$, $k_{\vek u}=3$, $k_{p}=2$ and $k_{\lambda}=2$
in the numerical studies shown here. Higher orders achieved virtually
indistinguishable results for the quantities shown below. It is also
noted that the Crank Nicolson method used for the time discretization
is only second-order accurate. For the time discretization $n_{\mathrm{step}}=\left\{ 150,300,600,1200,2400,4800\right\} $
time steps are used. To make the results more quantitative, the stresses
at the cylinder wall are summed up to obtain a force resultant $F\left(t\right)=\left\Vert \vek F\left(t\right)\right\Vert $
in 3D. This is the equivalent of the lift and drag coefficients for
the flat 2D case. Furthermore, the pressure difference between the
front and back position of the cylinder (in $\Omega_{\mathrm{2D}}$,
mapped to three dimensions) is computed, i.e., $\Delta p\left(t\right)=p_{\mathrm{front}}\left(t\right)-p_{\mathrm{back}}\left(t\right)$.

\begin{figure}
\centering

\subfigure[$F(t)$, $t\in(0,6)$]{\includegraphics[width=0.35\textwidth]{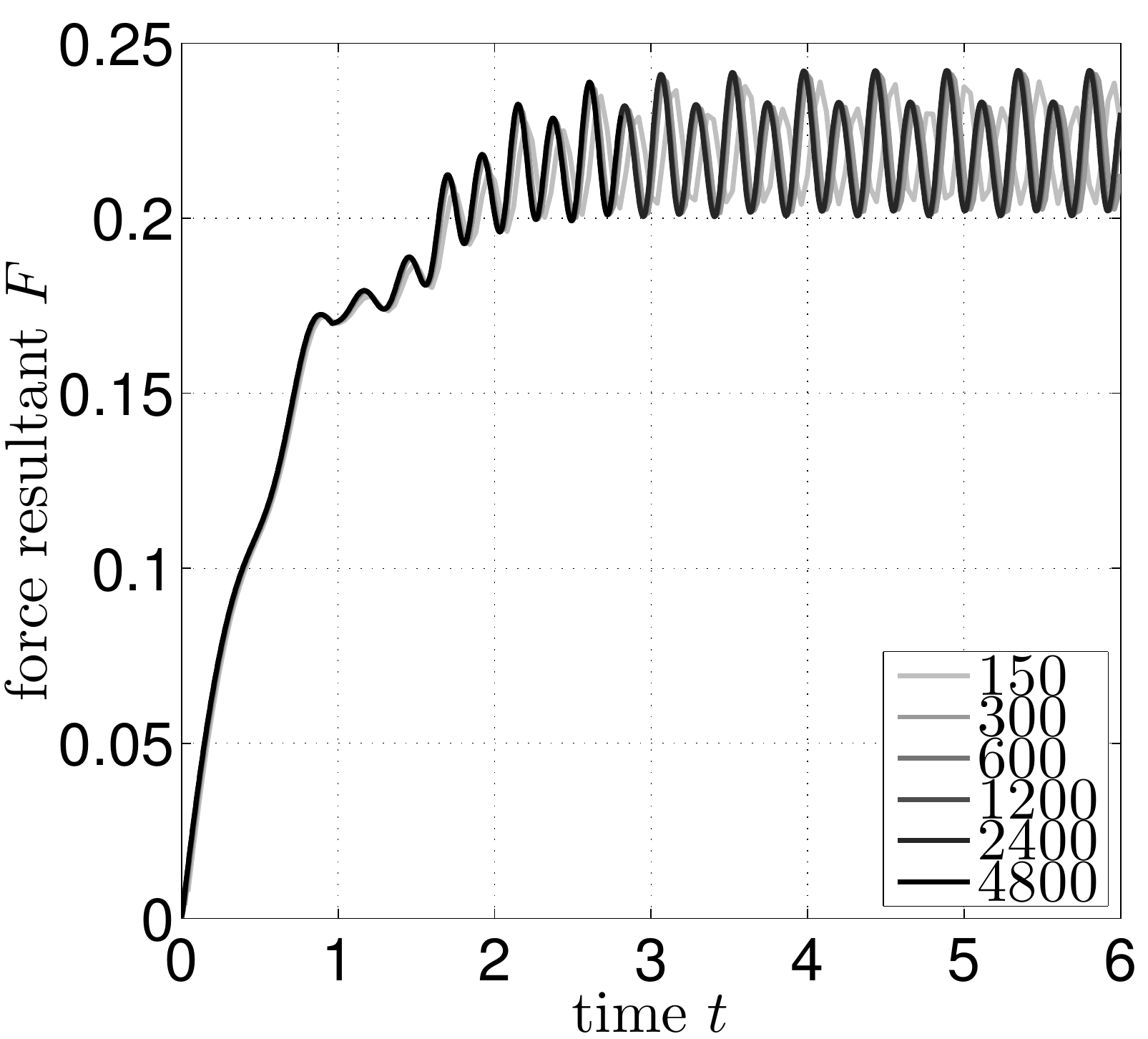}}\qquad\subfigure[$F(t)$, $t\in(5.2,6)$]{\includegraphics[width=0.35\textwidth]{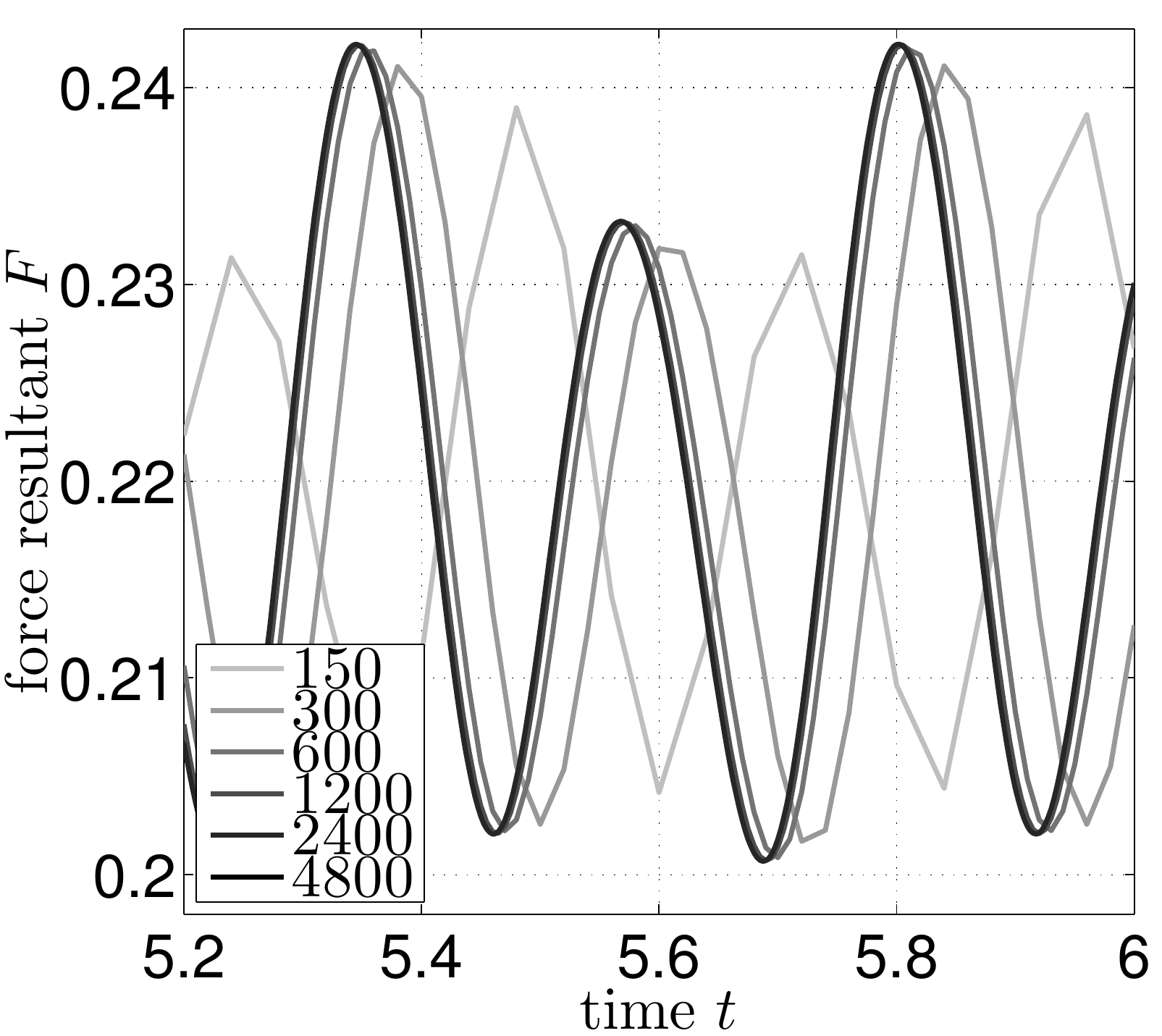}}

\subfigure[$\Delta p(t)$, $t\in(0,6)$]{\includegraphics[width=0.35\textwidth]{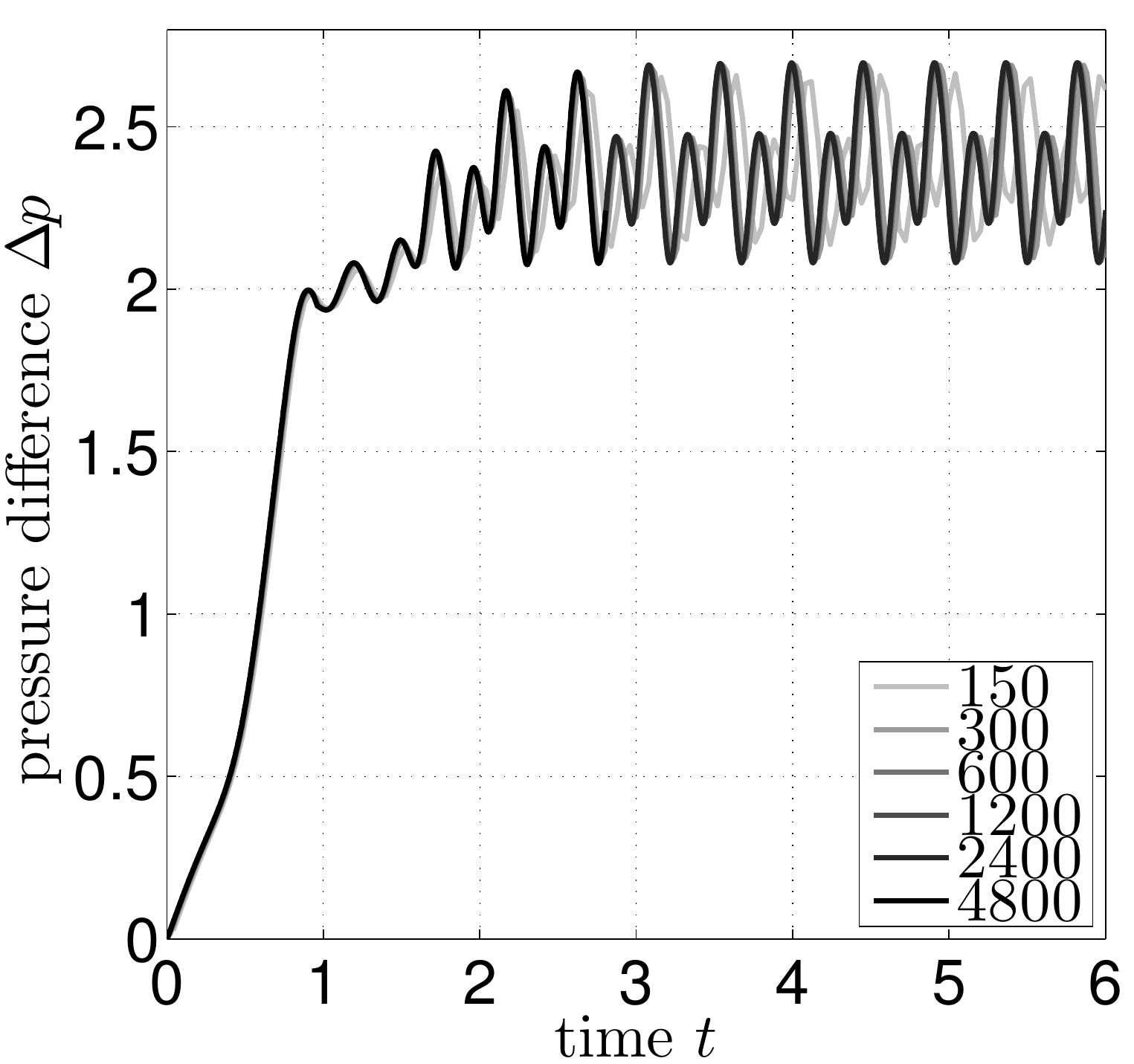}}\qquad\subfigure[$\Delta p(t)$, $t\in(5.2,6)$]{\includegraphics[width=0.35\textwidth]{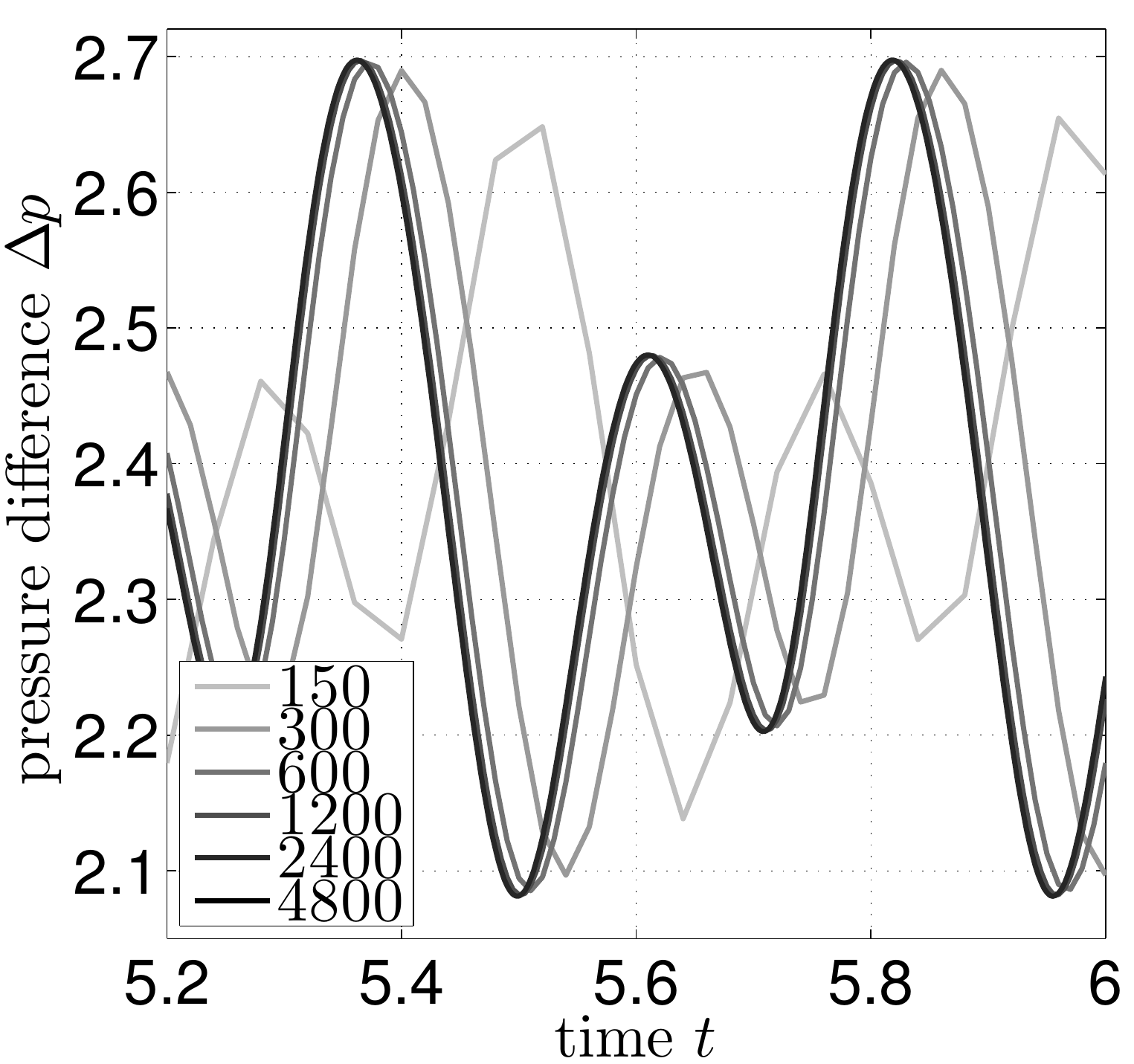}}

\caption{\label{fig:TurekResAMapA}Force resultant $F(t)$ and pressure difference
$\Delta p(t)$ obtained on the fine mesh of the cylinder flow test
case according to map A.}
\end{figure}

\begin{figure}
\centering

\subfigure[$F(t)$, $t\in(0,6)$]{\includegraphics[width=0.35\textwidth]{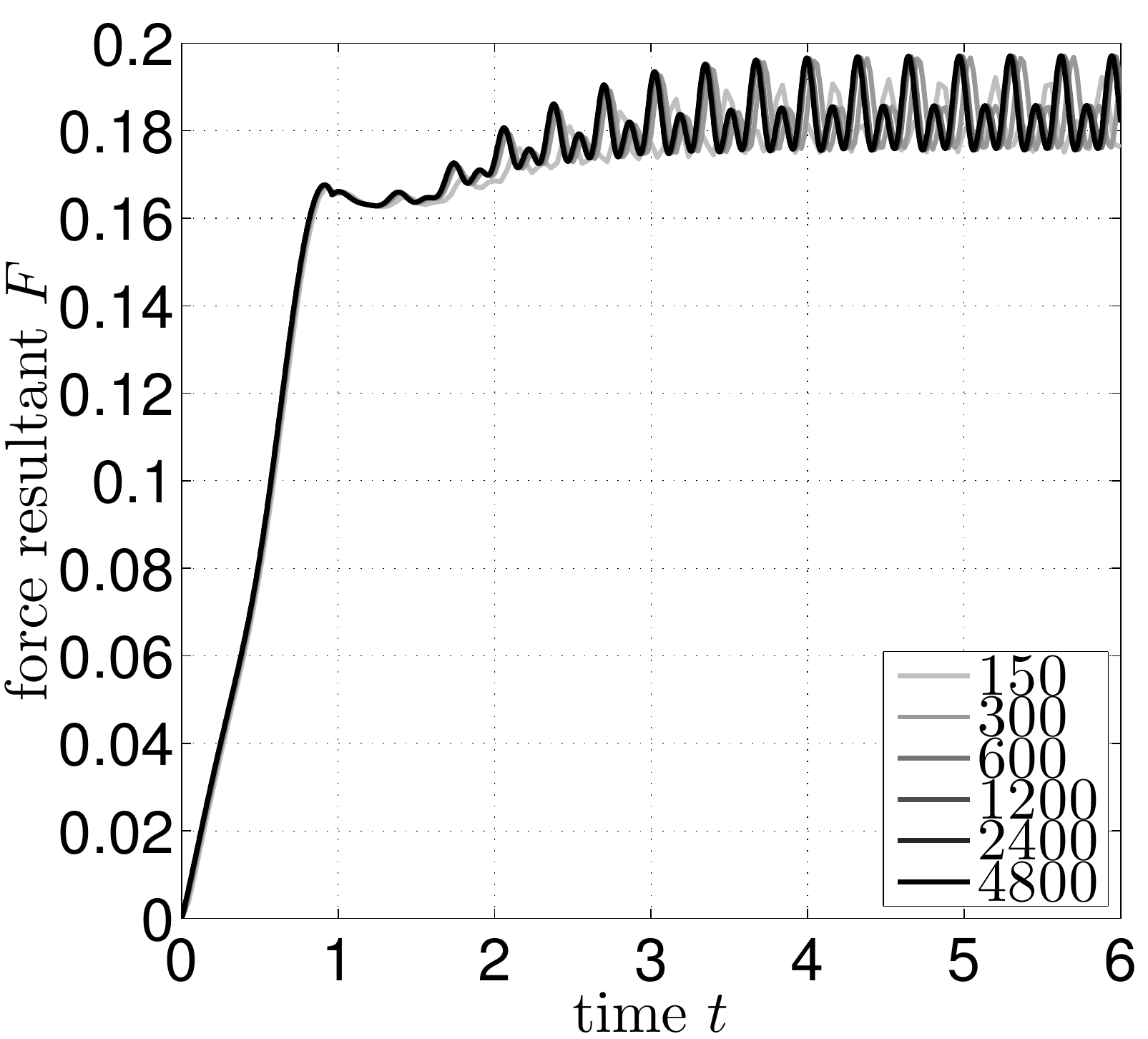}}\qquad\subfigure[$F(t)$, $t\in(5.2,6)$]{\includegraphics[width=0.35\textwidth]{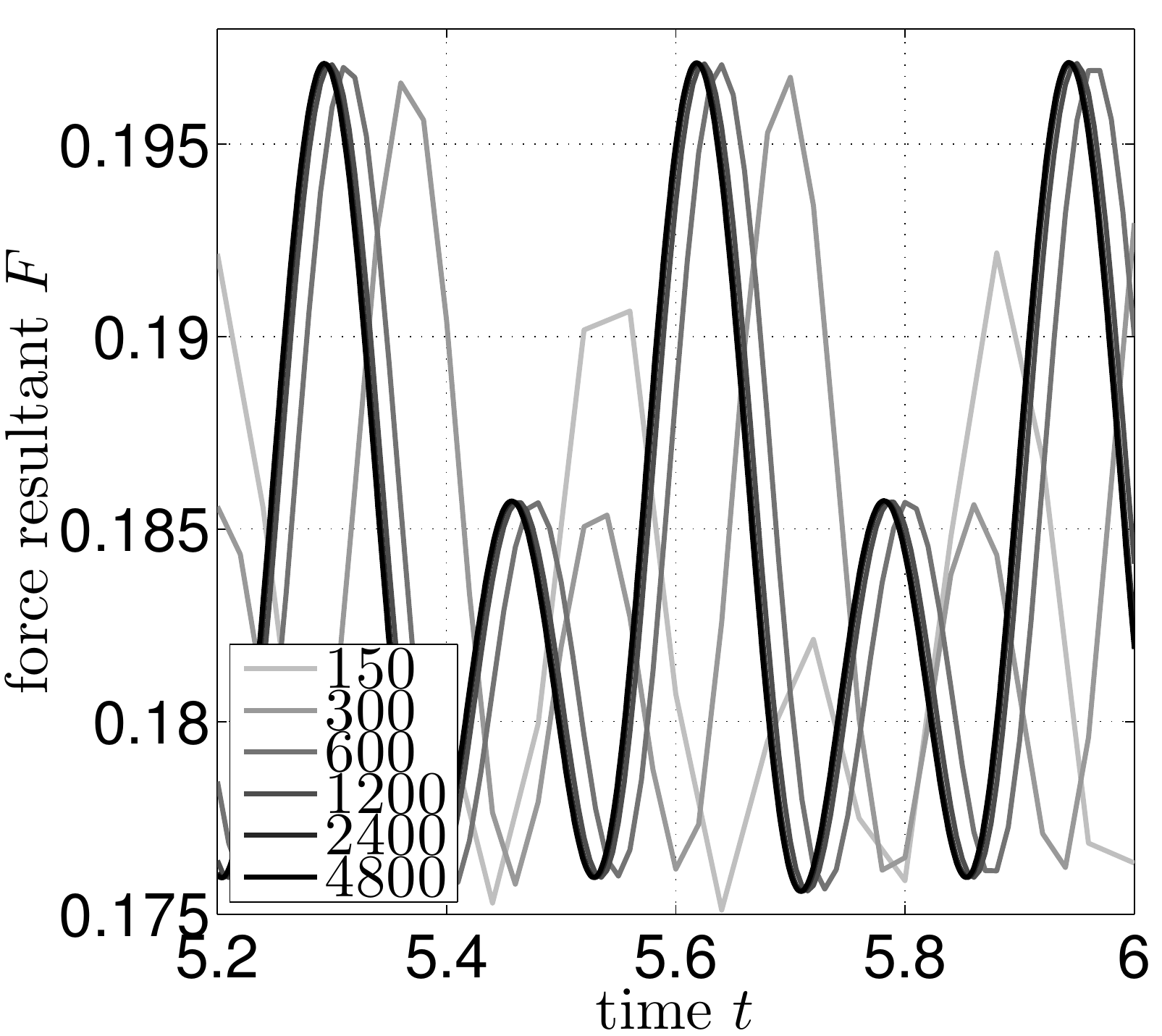}}

\subfigure[$\Delta p(t)$, $t\in(0,6)$]{\includegraphics[width=0.35\textwidth]{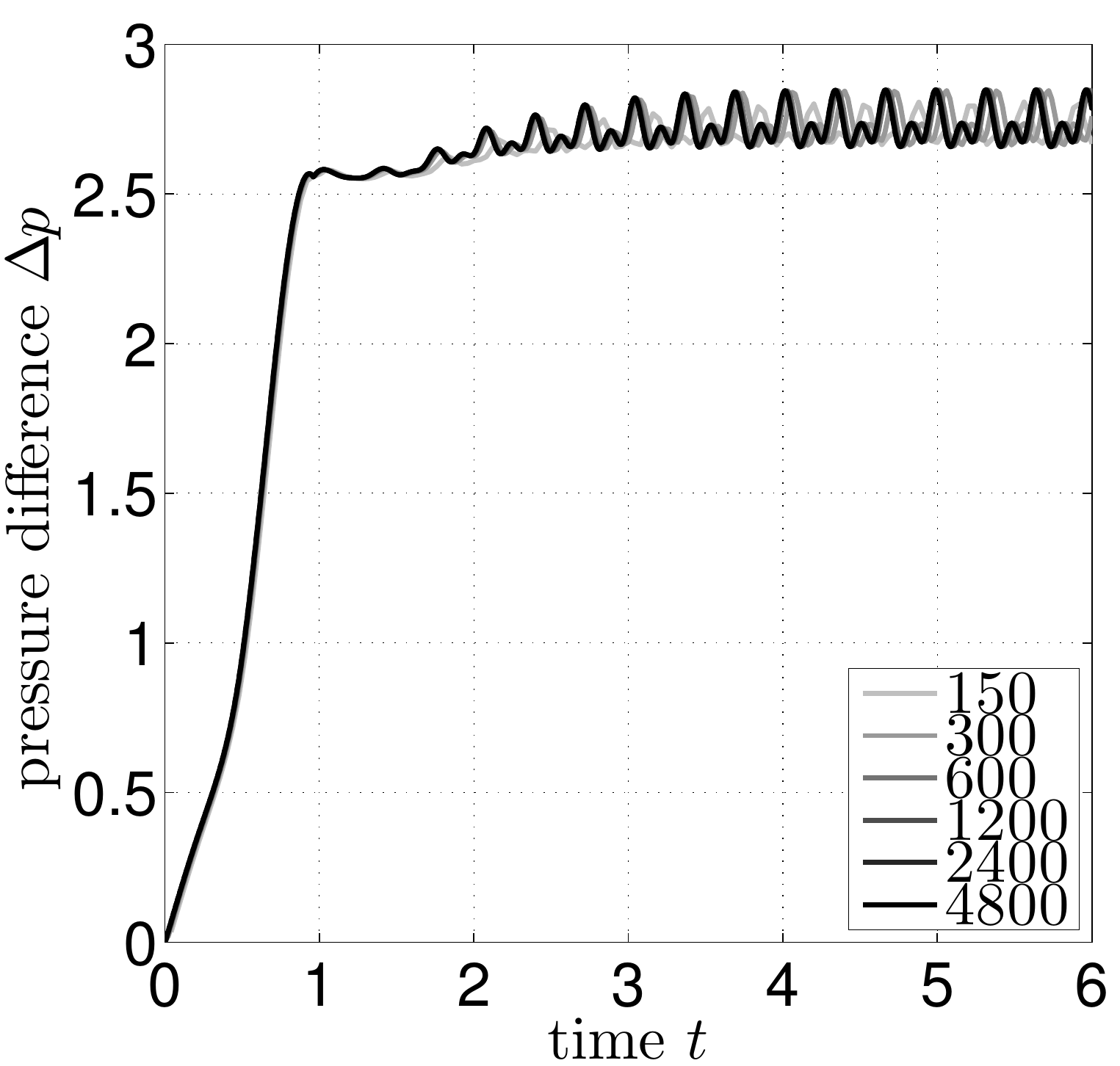}}\qquad\subfigure[$\Delta p(t)$, $t\in(5.2,6)$]{\includegraphics[width=0.35\textwidth]{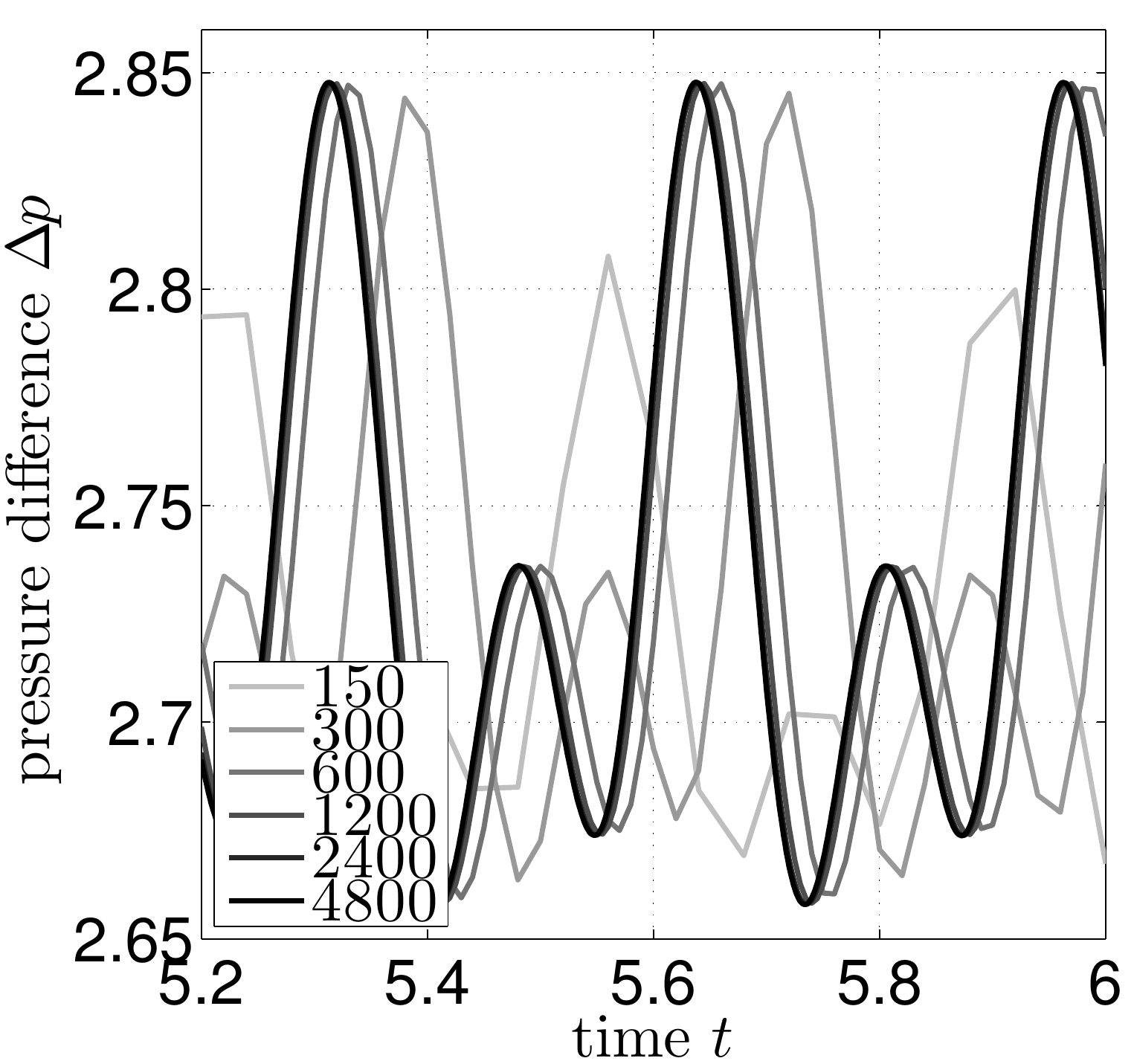}}

\caption{\label{fig:TurekResAMapB}Force resultant $F(t)$ and pressure difference
$\Delta p(t)$ obtained on the fine mesh of the cylinder flow test
case according to map B.}
\end{figure}

\begin{figure}
\centering

\subfigure[map A, $F(t)$, $t\in(5.2,6)$]{\includegraphics[width=0.35\textwidth]{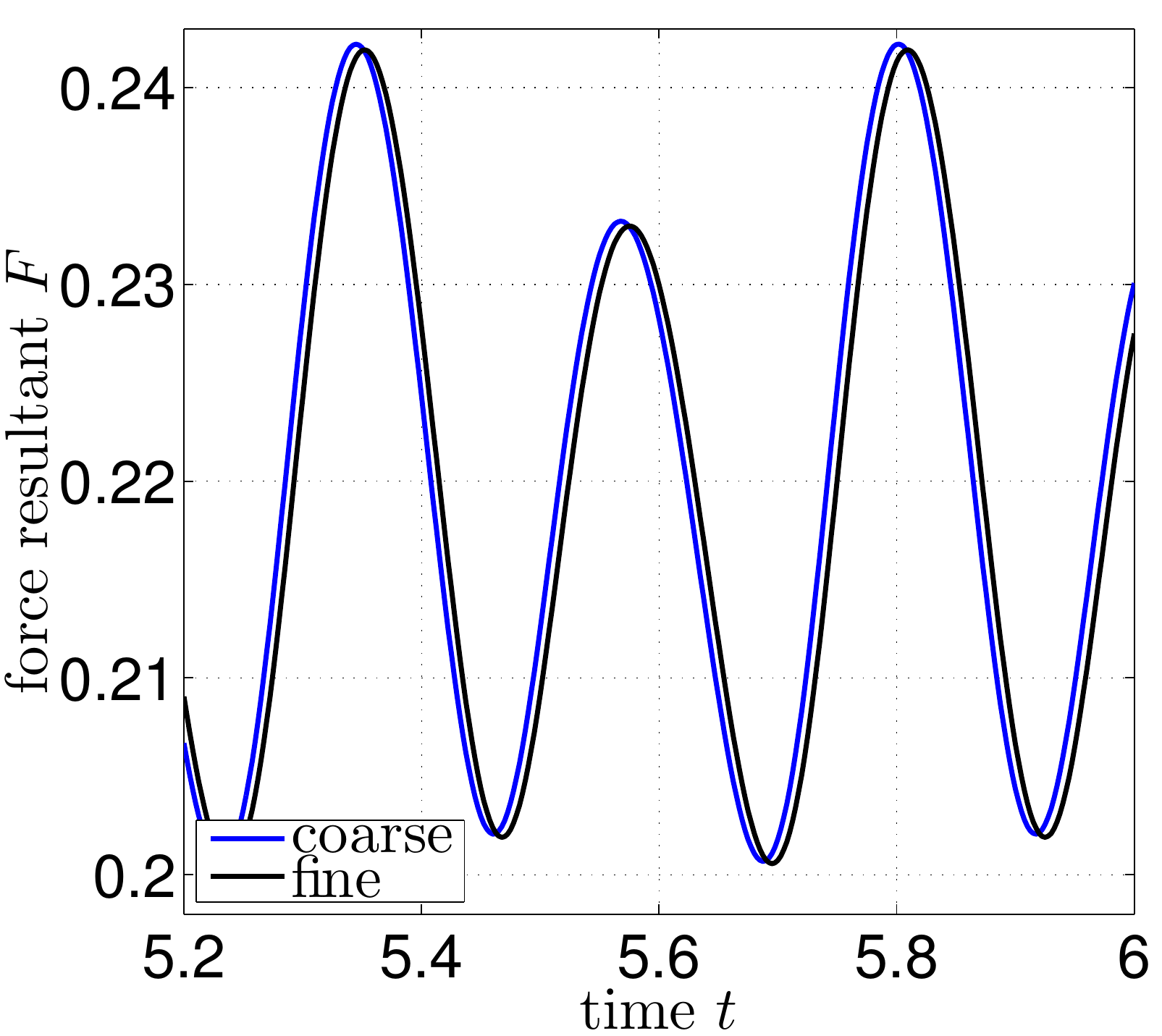}}\qquad\subfigure[map B, $F(t)$, $t\in(5.2,6)$]{\includegraphics[width=0.35\textwidth]{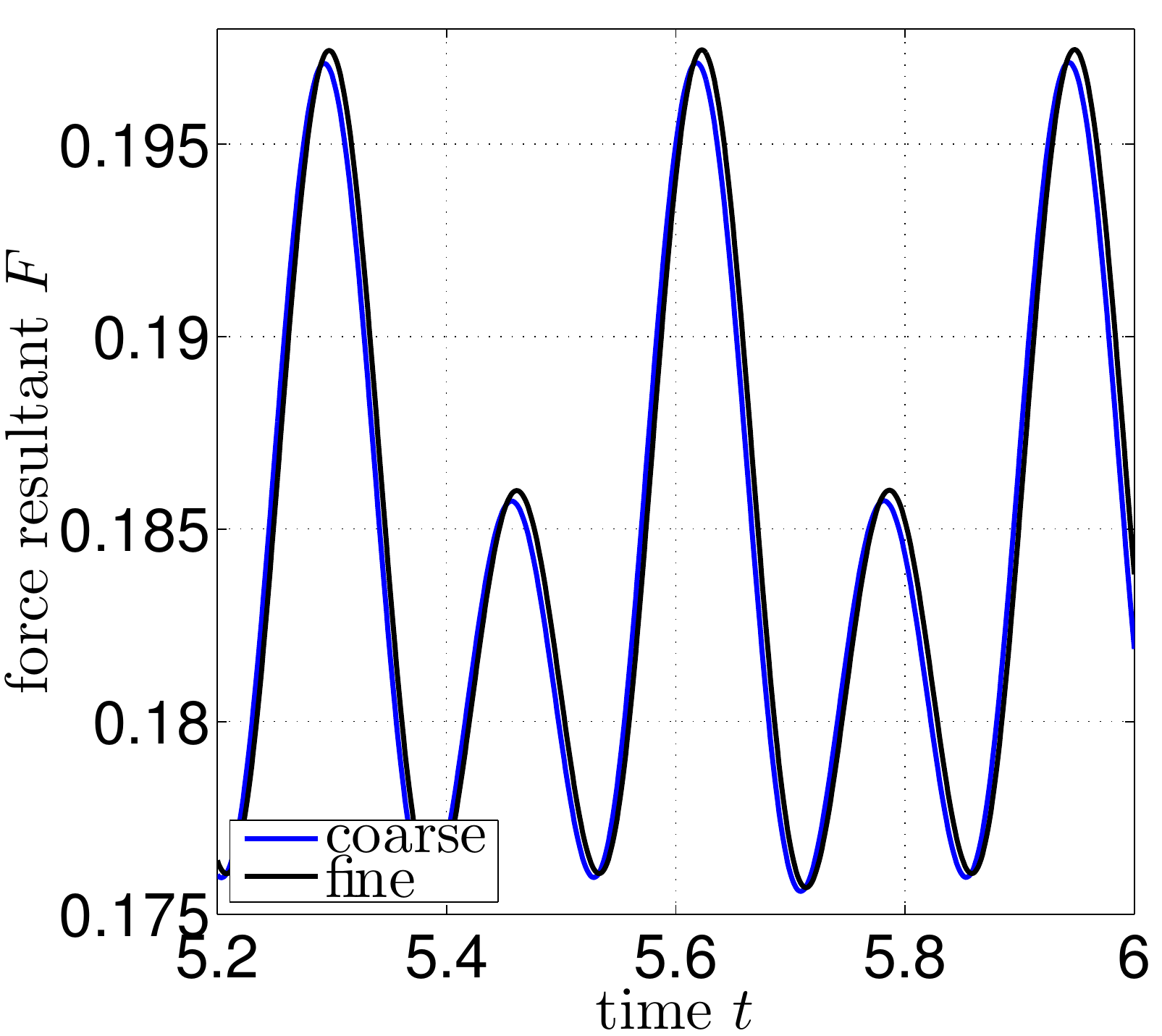}}

\caption{\label{fig:TurekResB}Force resultant $F(t)$ for the cylinder flow
test case according to map A and B for the fine and coarse meshes
with $4800$ time steps.}
\end{figure}

The results for map A are shown in Fig.~\ref{fig:TurekResAMapA}
for the different number of time steps. It can be seen that after
about $2\mathrm{s}$, the expected vortex shedding is almost established.
After $3\mathrm{s}$, the resulting oscillations remain virtually
unchanged. The time interval $\left[5.2,6\right]$ is shown in more
detail in Figs.~\ref{fig:TurekResAMapA}(b) and (d) for $F\left(t\right)$
and $\Delta p\left(t\right)$, respectively. The convergence with
increasing number of time steps is clearly demonstrated. Fig.~\ref{fig:TurekResAMapB}
shows the results in the same style for map B; the same conclusions
may be drawn. The spatial convergence is investigated in Fig.~\ref{fig:TurekResB}
where it is found that the coarse and fine mesh employed here obtain
very similar results for the chosen element orders. The frequency
of the oscillations for map A is $f_{A}=2.191\,\nicefrac{1}{\mathrm{s}}$
and for map B is $f_{B}=3.078\,\nicefrac{1}{\mathrm{s}}$; for the
flat case the frequency is $f=3.33\,\nicefrac{1}{\mathrm{s}}$.


\section{Conclusions\label{X_Conclusions}}

The surface FEM with higher-order elements is applied to solve Stokes
and Navier-Stokes flows on (fixed) manifolds. For the governing equations,
the classical gradient and divergence operators are replaced by their
tangential counterparts. An additional constraint is needed to ensure
that the velocities are in the tangent space of the manifold. Stabilization
is required for the case of Navier-Stokes flows at large Reynolds
numbers and the standard streamline-upwind Petrov-Galerkin (SUPG)
approach is used herein.

For the discretization, the surface FEM is employed with quadrilateral
or triangular elements. Element spaces of different orders are used
for (i) the geometric approximation of the manifold, $k_{\mathrm{geom}}$,
(ii) the approximation of the velocity fields, $k_{\vek u}$, (iii)
the pressure field, $k_{p}$, and (iv) the Lagrange multiplier field
for the enforcement of the tangential velocity constraint, $k_{\lambda}$.
The choice of these orders affects the properties of the resulting
FEM in terms of conditioning, accuracy, and stability. Particularly
useful combinations for a chosen order $k_{\vek u}$ are $k_{\mathrm{geom}}=k_{\vek u}+1$
and $k_{p}=k_{\lambda}=k_{\vek u}-1$. Some benchmark test cases for
flows on manifolds are proposed and higher-order convergence rates
are achieved. The notation used in this work is closely related to
the engineering literature for the FEM in fluid mechanics. Implementational
matters are outlined.

There is a large potential for future research related to this work:
One may investigate different stabilization methods such as Galerkin
least-squares stabilization and variational multiscale methods. Stabilization
may also be useful to circumvent the Babu\v ska-Brezzi condition
and enable equal-order shape functions for the velocities and pressure.
The tangential velocity constraint may be more efficiently enforced
based on penalty methods or other Lagrange multiplier approaches such
as the Uzawa method. We believe that flows on manifolds have a strong
potential for fundamental research in mathematics, physics, and engineering.

\section{Acknowledgements}

The fruitful discussions with Dr.~Sven Gro{\ss}, Dr.~Thomas R{\"{u}}berg
and Prof.~G{\"{u}}nther Of are gratefully acknowledged.

\bibliographystyle{schanz}
\addcontentsline{toc}{section}{\refname}\bibliography{FriesRefs}
 
\end{document}